%% file: Clementini-AA-2016-29925-accepted.tex
\documentclass{aa_gaia}
\usepackage{graphicx}
\usepackage{txfonts}
\usepackage{xcolor}
\usepackage{pdflscape}
\usepackage{siunitx}
\usepackage{longtable}
\definecolor{ao}{rgb}{0.0, 0.5, 0.0}
\begin{document} 

\title{\textit{\textbf{Gaia}} Data Release 1}

\subtitle{Testing the parallaxes with local Cepheids and RR Lyrae stars}

\titlerunning{Testing the parallaxes with local Cepheids and RR Lyrae stars}

\include{authors-c2}

\authorrunning{ Gaia Collaboration, Clementini et al.}

   \date{Received ... ; accepted .....}

% \abstract{}{}{}{}{} 
% 5 {} token are mandatory

  \abstract
{Parallaxes for 331 classical Cepheids, 31 Type~II Cepheids and 364 RR Lyrae stars 
in common between {\it  Gaia} and the Hipparcos and Tycho-2 catalogues are published in  {\it Gaia} Data Release 1 (DR1) as part of the Tycho-{\it Gaia} Astrometric Solution (TGAS, \citealt{lindegren16}).}
{In order to test these first parallax measurements of the primary standard candles of the cosmological distance ladder,  
that involve astrometry collected by {\it Gaia} during the initial 14 months of science operation,  
we compared them with literature estimates and derived new 
 period-luminosity ($PL$), period-Wesenheit ($PW$) relations for classical and Type~II Cepheids and infrared $PL$, $PL$-metallicity ($PLZ$) and  optical luminosity-metallicity  ($M_V$-{\rm [Fe/H]})  relations for the RR Lyrae stars,  with  zero points based on TGAS.}
 {Classical Cepheids were carefully selected in order to discard known or suspected binary systems. The final sample  comprises 102 fundamental mode pulsators, with periods ranging from 1.68 to 51.66~days  (of which 33 with $\sigma_{\varpi}/\varpi <$0.5). 
The Type II Cepheids  include a total of 26  W Virginis and BL Herculis stars spanning the period range from 1.16 to 30.00~days (of which only 7 with $\sigma_{\varpi}/\varpi <$0.5). The RR Lyrae stars include 200 sources with pulsation period ranging from  0.27 to 0.80~days (of which 112 with $\sigma_{\varpi}/\varpi <$0.5). 
 The  new relations were computed  using multi-band  ($V,I,J,K_{\mathrm{s}}, W_1$)  photometry and spectroscopic metal abundances available in the literature, and applying three alternative approaches: 
(i) by linear least squares fitting the absolute magnitudes 
  inferred from direct transformation of the TGAS parallaxes;  (ii) by adopting astrometric-based luminosities, 
   and (iii) using a 
  Bayesian fitting approach. The latter two methods work in parallax space where parallaxes are used directly, thus maintaining symmetrical errors and allowing negative parallaxes to be used.
The TGAS-based $PL, PW, PLZ$ and  $M_V-{\rm [Fe/H]}$ relations are confronted 
by comparing the distance to the Large Magellanic Cloud provided by different types of pulsating stars and alternative fitting methods.
}
{Good agreement is found from direct comparison of the parallaxes of RR Lyrae stars for which 
  both TGAS and HST measurements (\citealt{Benedict2011}) are available.
  Similarly, very good agreement is found between the TGAS  values and the parallaxes
   inferred from the absolute magnitudes of Cepheids and RR Lyrae stars analysed 
 with  the Baade-Wesselink 
  method. TGAS values also compare favourably with the parallaxes inferred by theoretical model fitting of 
  the multi-band light curves for two (out of three) classical Cepheids and one RR Lyrae star in our samples. 
The  $K$-band $PL$ relations show the significant 
  improvement  of the TGAS parallaxes for Cepheids and RR Lyrae stars with respect to the Hipparcos
  measurements.  This is particularly true for the RR Lyrae stars  
for which improvement in quality and statistics is impressive. 
}
 {TGAS parallaxes bring a significant added value to the previous Hipparcos estimates. The relations presented in this paper represent  first {\it Gaia}-calibrated relations and form a ``work-in-progress" milestone report 
in the wait for {\it Gaia}-only  parallaxes 
of which a first solution will become available with {\it Gaia}'s Data Release 2 (DR2) in 2018.}
\keywords{Astrometry  -- Parallaxes -- Stars: distances -- Stars: variables: Cepheids  -- Stars: variables: RR Lyrae --  Methods: data analysis} 
\maketitle
%
%________________________________________________________________

\section{Introduction}\label{sec:intro}
On 14 September 2016, photometry and astrometry data collected by the {\it Gaia} mission during the first 14 months of science operation have been released to the public with  {\it Gaia} first data release (hereinafter {\it Gaia} DR1; {\citealt{gaiacol-prusti, gaiacol-brown}). In particular, {\it Gaia}  DR1 catalogue includes positions, proper motions and parallaxes for about 2 million stars in common between {\it Gaia} and the Hipparcos and Tycho-2 catalogues,  computed as part of the Tycho-{\it Gaia} Astrometric Solution (TGAS), of which the principles are discussed in  \citet{michalik15} and results published in {\it Gaia} DR1 are described in detail in \citet{lindegren16}. 
Among the TGAS sources is a sample of Galactic pulsating stars that includes 331 classical Cepheids, 31 Type~II Cepheids and 364 RR Lyrae stars. 
 As part of a number of checks 
 performed within the {\it Gaia} Data Processing and Analysis Consortium (DPAC) we have tested TGAS parallaxes for Cepheids and RR Lyrae stars 
by building canonical relations followed by  these variable stars, such as the period-luminosity ($PL$) and period-Wesenheit ($PW$) relations for classical and Type~II Cepheids and the infrared $PL$, $PL$-metallicity ($PLZ$) and  optical luminosity-metallicity  ($M_V$-{\rm [Fe/H]})  relations for RR Lyrae stars,   with zero points based on TGAS parallaxes. Results of these tests are presented in this paper. 

 Thanks to the characteristic $PL$ relation
discovered at the beginning of the last century by Mrs Henrietta Swan Leavitt (1868 - 1921), classical
Cepheids have become the basis of an absolute calibration of the
extragalactic distance scale \citep[see e.g.][and references
therein]{free01,saha06, fio13, riess11,riess16}. 
The $PL$ is a statistical relation with an
 intrinsic dispersion caused by the finite width of the instability
strip for pulsating stars. This dispersion is particularly significant in the optical bands (e.g. $B,V$), where
it is of the order of $\pm$ 0.25~mag, but decreases moving towards
longer wavelengths becoming less than $\sim \pm$ 0.1~mag in the near and
mid-infrared (NIR and MIR) filters \citep[see e.g.][and references
therein]{madore91,cmm00,marconi05,ngeow12,ripe12,inno13,Gieren2013}.
Main open issues concerning the use of the Cepheid $PL$ 
  for extragalactic distance determinations are: (i)  the dependence
of the $PL$ relation on chemical composition, on which
no general consensus has been reached yet   in the literature and,  (ii) 
the possible nonlinearity of the Cepheid $PL$ relations
at the longest periods,  for which some authors find evidence in the form of a break around 
10~days, with a clear corresponding change of the $PL$ slope in 
$B,V,R$ and $I$  \citep[see e.g.][]{nk06,ta03}. 
 Both metallicity (and helium) dependence and  nonlinearity effect, besides the 
 effect of 
 the finite intrinsic width of the instability strip mentioned above,  
 decrease when moving from
optical to NIR and MIR passbands \citep[see e.g.][and references therein]{madore91,cmm00,marconi05,ripe12,ripe16,inno13,Gieren2013}.

When optical bands are used great advantages are obtained by adopting
reddening-free formulations of the $PL$ relation, called Wesenheit functions ($PWs$) 
\citep[see][]{mdr82,cmm00,ripe12}. These relations include a colour
term,  thus partially correcting for the intrinsic width of the
instability strip, whose coefficient is given by the ratio of total
to selective extinction. The Wesenheit relation in the $V,I$ bands, $PW(V,I)$, is often adopted to derive accurate extragalactic
distances as it is  widely recognized to be little
dependent on metallicity \citep[see e.g.][and references
therein]{bono10}. Other filter combinations, extending to the
NIR, are also commonly used in the literature
\citep[see e.g.][]{riess11,riess16,fio13,ripe12,ripe16}. 
However, all these relations need an accurate calibration of their  zero points and a quantitative assessment 
of the dependence of slope and  zero
point on the chemical composition, as any systematic effects on the
coefficients of both $PL$ and $PW$ relations directly propagates in the
calibration of the secondary distance indicators and the estimate of the Hubble constant, H$_0$.
{\it Gaia} will play a crucial role to definitely address all these issues of the Cepheid-based 
distance ladder. 

On the other hand, an alternative and independent route to H$_0$ using the cosmic ``distance ladder" method is provided by
Population II pulsating stars such as the RR Lyrae stars, (see, e.g. \citealt{beaton2016}, and references therein), 
the Type II Cepheids and
the SX Phoenicis variables, old ($t \gtrsim 10$ Gyr), sub-solar mass variables, that typically populate globular clusters and galactic halos.
While Type II Cepheids and
SX Phoenicis stars follow $PL$ relations, the  standard candle commonly 
associated with  RR Lyrae stars is  the relation existing between the mean absolute visual magnitude $\langle M_V{(RR)} \rangle$
and the iron content [Fe/H], usually assumed in a  linear form: 
$M_V{(RR)} = a{{\rm [Fe/H]}} + b$.  Current determinations of the slope $a$
and zero point $b$ of this relation span a wide range of values \citep[see e.g.][and references therein]{Clementini2003,cacciari03,marconi15} and  theoretical investigations
based on evolutionary and pulsation models also suggest
a change in the slope at [Fe/H]$\approx-1.5$~dex \citep[see
e.g.][]{capu00,cassisi98,lee90}.
The other characteristic relation that makes RR Lyrae stars  fundamental 
primary distance indicators for systems mainly composed by Population II stars is the $PL$ relation they conform to at infrared wavelengths and in the 
 $K$ (2.2~$\mu$m) band in particular, as first pointed out in the
pioneering investigations of \citet{long86,long90}. 
Due to the strict relation between the $V-K$ colour and the effective
temperature and between the latter quantity and the pulsation period
the nearly horizontal distribution of the RR Lyrae stars in the $M_V$ {\it versus}
$\log{P}$ plane evolves into a strict $PL$ relation in the $M_K$ {\it versus}
$\log{P}$ plane  (see e.g. fig. 2 of \citealt{Cat2004}), by which 
 longer periods correspond to brighter pulsators
in the $K$ band.  
It has also been demonstrated \citep{bono01}
 that the intrinsic dispersion of the $PL(K)$ relation
drastically decreases when metallicity
differences and evolutionary effects are taken into account.
However, coefficients and zero point of the $M_K - \log{P} -{\rm [Fe/H]}$ relation (hereinafter, $PM_{K}Z$) are still matter of 
debate in the literature and may differ significantly from one study to the other \citep[see e.g.][]{marconi15}.
\citet{Bono2003} and \citet{Cat2004} analysed the $PM_{K}Z$ from the semi-theoretical and theoretical point of view and found a non-negligible 
dependence of the RR Lyrae $K$-band absolute magnitude, $M_K$, on metallicity: $b=0.231 \pm 0.012$ and $b=0.175$, respectively. 
Conversely, the dependence of the $K$-band luminosity on metallicity derived in empirical studies is generally much shallower 
  \citep{DelP2006} 
or even negligible (\citealt{Sol2006,Sol2008}, \citealt{Bor2009}, \citealt{Muraveva2015}). 
As for the dependence of  $M_K$ on period, values in the literature vary  from $-2.101$ \citep{Bono2003} to $-2.73$ \citep{Muraveva2015}.

In this paper we use TGAS parallaxes of local Cepheids and RR Lyrae stars along with literature $V, I, J, K_{\mathrm{s}}, W_1$ photometry to compute new $PL$, $PW$ and $M_V$ - [Fe/H] relations through a variety of methods and compare their results. This enables us to test TGAS parallaxes for these primary standard candles. 
  Estimation of distances from trigonometric parallaxes is not straightforward and still matter of debate. The direct transformation to distance (and then absolute magnitude) by parallax inversion is not  often advisable if errors are large, since it causes asymmetric errors in the magnitudes and does not allow the use of negative parallaxes.  
Methods that operate in parallax space such as the Astrometric Based Luminosity (ABL, \citealt{Arenou1999}) and Bayesian approaches are to be preferred.  In this paper we adopt the least squares fit of the absolute magnitudes obtained from direct transformation of the parallaxes, the ABL method and a Bayesian approach to fit the various relations that Cepheids and RR Lyrae stars conform to, then compare the results that different types of variables and different fitting methods provide for the distance to the Large Magellanic Cloud (LMC).
Far from seeking  
 results on the cosmic distance ladder as re-designed by these first {\it Gaia} measurements, the exercise presented in this paper is meant to assess limitations and  potential of this first astrometry solution and to compare different methods of handling parallaxes.  
The present approach partially differs from the photometric parallax approach adopted in 
\citet{lindegren16} and \citet{cas17},  where literature Cepheid $PL$ relations (whether in the visual or the near-infrared)  are assumed to probe TGAS parallaxes of classical Cepheids, and, hopefully, it is less prone to shortcomings arising from the intrinsic 
width of the Cepheid instability strip and the poor knowledge about universality,  linearity and metallicity-dependence of the reference relations used in the aforementioned studies.

The paper is organized as follows:  in Section~\ref{sec:samples} we present the samples of Cepheids (both classical and Type II) and RR Lyrae stars we have analysed, describe how they were selected and compare their TGAS parallaxes with parallax values (Hipparcos and/or HST) available in the literature for some of them, with the parallaxes inferred from the theoretical modelling 
of the light curves and from Baade-Wesselink studies. In Section~\ref{sec:bias_meth} we analyse possible biases that affect  the Cepheid and RR Lyrae samples and describe the methods we used  to fit the various relations of these variable stars.
In Section~\ref{sec:dceps}  we present the photometric dataset used for the classical Cepheids and the derivation of the corresponding $PL$ and $PW$ relations.  Section~\ref{sec:t2ceps} is devoted to the Type II Cepheids, and Section~\ref{sec:rrls} to the RR Lyrae stars.  
 In Section~\ref{sec:lmc}, we discuss the TGAS-based 
relations derived in the previous sections 
by comparing the distance to the Large Magellanic Cloud they provided 
and present a few concluding remarks.

\section{Cepheid and RR Lyrae samples}\label{sec:samples}
\subsection{Sample selection}
The magnitude distribution of the sources for which a parallax measurement is available in {\it Gaia} DR1 is shown in  fig.~1 of
 \citet{gaiacol-brown} and includes  sources with a mean $G$-band  apparent magnitude between $\sim$ 5 and $\sim$ 13.5~mag (but only very few with $G \lesssim$ 7 mag).  The typical uncertainty of the TGAS parallaxes is 
 0.3 milliarcsecond (mas), to which a systematic component of 0.3~mas should be added, arising from model assumptions and simplifications of data processing for DR1, among 
 which, mainly,  position and colour of the sources, as widely discussed in  \citet{lindegren16} and  also summarised in section~6 of \citet{gaiacol-brown}. 
Since TGAS parallaxes are available for sources observed by Tycho-2 \citep{hog00}, of which  only a fraction are also in the Hipparcos catalogue (\citealt{esa97, van07a}),   
to  build the largest possible samples  
we used the list of  Cepheids and RR Lyrae stars in the  Tycho-2  catalogue as reference. To create this list 
we cross-identified the Tycho-2 whole catalogue  with  the 
  General Catalog of Variable Stars (GCVS database; \citealt{gcvs}) that contains a total of 1100 between classical and Type II Cepheids  
   and with the David Dunlap Observatory Database of Galactic Classical Cepheids (DDO\footnote{ http://www.astro.utoronto.ca/DDO/research/cepheids/}; \citealt{ddo}) that contains over 500  classical Cepheids. In particular, according to the variability types in the GCVS,  in these selections we included, under the definition of \textit{Classical Cepheids}, the following types: Cepheids and classical Cepheids or Delta Cephei-type variables (CEP and CEP(B), DCEP,  DCEPS and DCEPS(B), as labelled in the GCVS) and under 
  \textit{Type II Cepheids}, the following types:  CW, CWA, CWB, RV, RVA and RVB. 
 Crossmatching these databases with the Tycho-2 general catalogue ($\gtrsim$ 2.5 billion sources) and following supplements ($\gtrsim$ 18 thousand sources) we found  final samples of 388 classical and 33 Type II Cepheids\footnote{The cross-match between Tycho-2 and GCVS sources was done using  equatorial J2000 RA, DEC coordinates and assuming an astrometric error of 1 arcsec between catalogues. Conversely, we converted the DDO database equatorial B1950 coordinates to J2000 before matching the Tycho-2 and DDO catalogues and assumed 1-5 arcsec as maximum difference of the two sets of coordinates.}.   
 We then queried the  \texttt{tgas\_source} table in the {\it Gaia} Archive Core Systems (GACS)\footnote{\url{http://Gaia.esac.esa.int/archive/}} to retrieve TGAS parallaxes and  {\it {\it Gaia}} $G$-band apparent magnitudes for the samples of 388 classical and 33 Type II Cepheids. Only for 331 of the  classical Cepheids in our list TGAS parallaxes and  {\it {\it Gaia}} $G$ magnitudes are actually available in GACS. They span $G$-band apparent magnitudes  in the range 4.68 $\leq G \leq$ 12.54~mag.  Their parallaxes span  the range from  $-$1.610 to 6.214~mas, with parallax errors in the range from 0.215 and 0.958~mas, and with 29 sources having TGAS negative parallax. %, where extreme negative values are given in brackets.
 The error distribution of TGAS parallaxes for the 331 classical Cepheids is shown by the pink histogram in Fig.~\ref{fig:CCcumul}.  
Of the 33 Type II Cepheids only 31 have $G$ magnitude and TGAS parallax available. They  span  $G$-band apparent magnitudes  in the range 6.89  $\leq G \leq$ 12.10~mag. Their parallaxes
span the  range from $-$0.234 to 3.847~mas, with parallax errors from 0.219~mas to 0.808~mas, and with negative parallax for 5 of them. The error distribution of the TGAS parallax for the 31 Type II Cepheids is shown by the green histogram in Fig.~\ref{fig:T2Ccumul}.
  
  Concerning the RR Lyrae stars, the  GCVS \citep{gcvs} contains information on 7954 such variables which are  labelled as RR, RR(B), RR:, RRA, RRAB, RRAB:, RRC, RRC:, where ":" means uncertain classification. We crossmatched the  GCVS  RR Lyrae star sample against  the Tycho-2 general catalogue  and its supplements, and found 421 sources in common.  
  Three sources, (namely, S Eri, V2121 Cyg and NZ Peg) have uncertain classification according to  ``The SIMBAD astronomical database" (\citealt{Wenger2000}) and were removed.   We then crossmatched  the remaining 418 sources against the \texttt{tgas\_source} table in  GACS  and found a TGAS parallax for 364 of them. Values of the $G$-band apparent magnitude for these 364 RR Lyrae stars are in the range  7.03 $\leq G \leq$~13.56~mag. Their  parallaxes are in the range from $-$0.837 to 13.131~mas with parallax errors spanning the range from 0.209 to 0.967~mas and  six stars  having negative parallaxes. 
  The error distribution of TGAS parallax for the  364 RR Lyrae stars is shown by the cyan histogram in Fig.~\ref{fig:RRcumul}. 
 
 \begin{figure}[t!]
\centering
\includegraphics[trim=50 300 0 300 clip, width=1.05\linewidth]{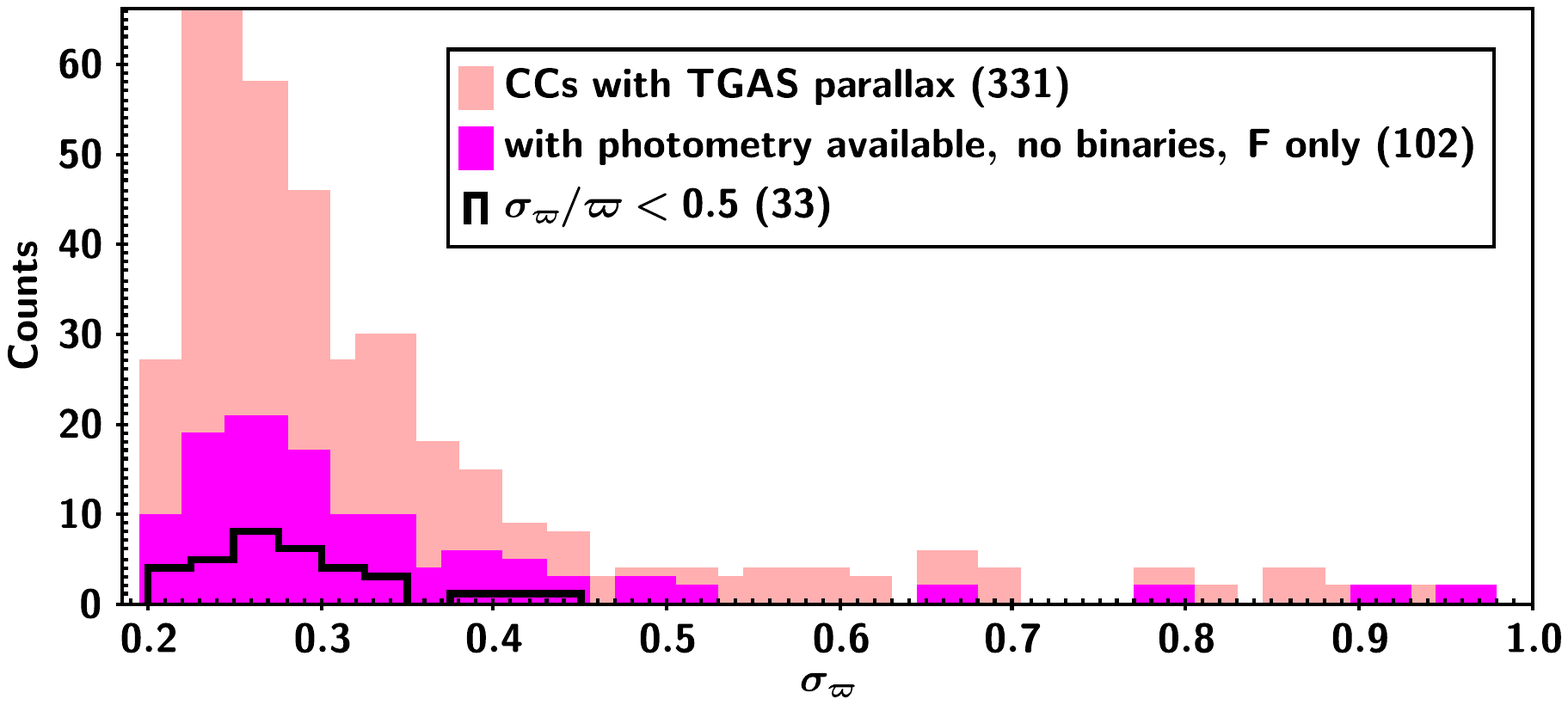}
\caption[]{Error distribution of TGAS parallaxes for classical Cepheids (CCs, in the label): whole sample (331 stars, pink), subsample with literature photometry after removing 
binaries and retaining only fundamental-mode (F) pulsators (102 stars, magenta), subsample of the previous 102 sources retaining only stars with positive parallax and 
parallax errors $\sigma_{\varpi}/\varpi<0.5$ (33 stars, black contour). The bin size is 0.025~mas.}
\label{fig:CCcumul}
\end{figure}

 \begin{figure}[t!]
\centering
\includegraphics[trim=50 300 0 300 clip, width=1.05\linewidth]{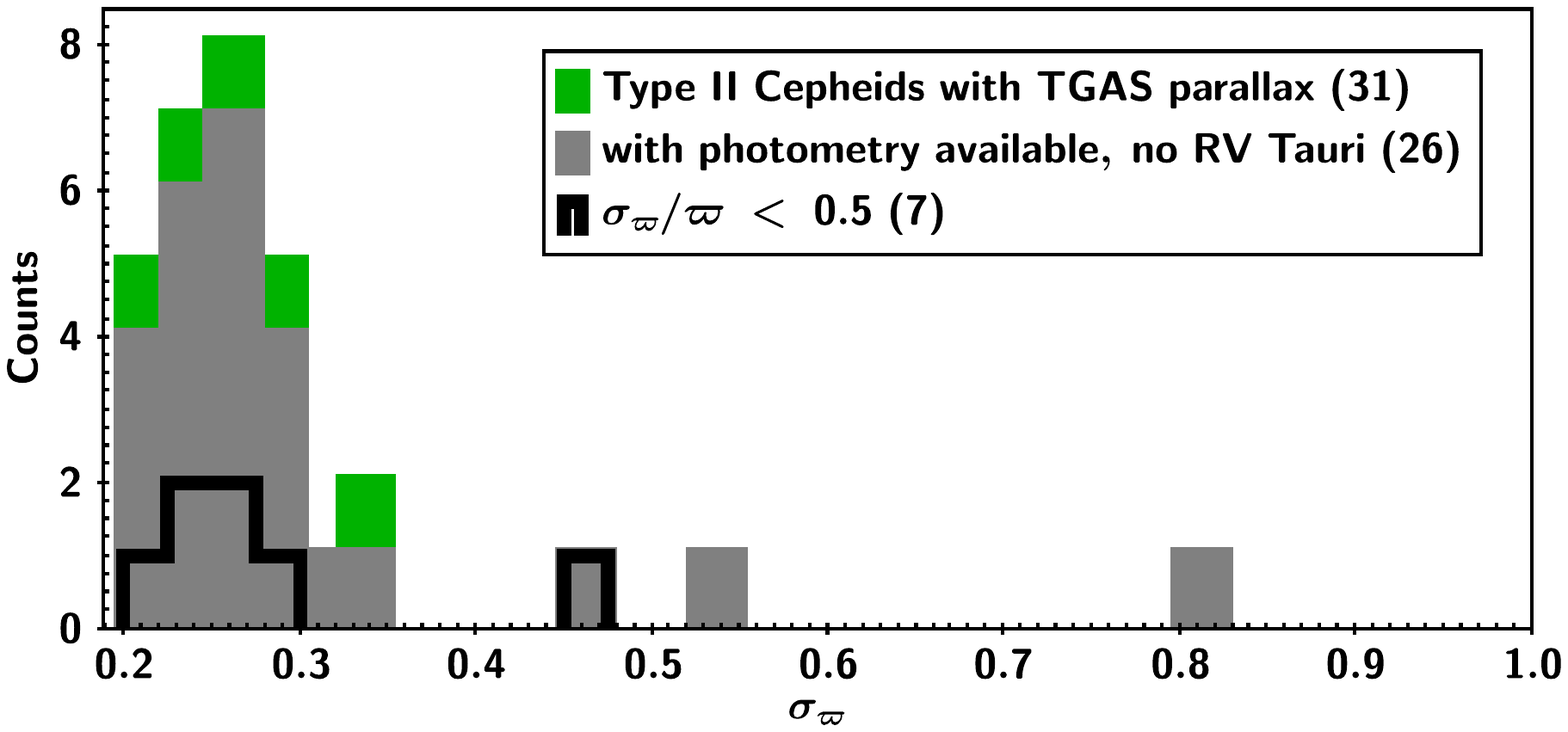}
\caption[]{Error distribution of TGAS parallaxes for Type II Cepheids: whole sample (31 stars, green), subsample with literature photometry and removing 
variables of RV Tauri type (26 stars, gray), subsample of the previous 26 sources retaining only stars with positive parallax and 
parallax errors $\sigma_{\varpi}/\varpi<0.5$ (7 stars, black contour).
The bin size is 0.025~mas.}
\label{fig:T2Ccumul}
\end{figure}

\begin{figure}[t!]
\centering
\includegraphics[trim=50 300 0 300 clip, width=1.05\linewidth]{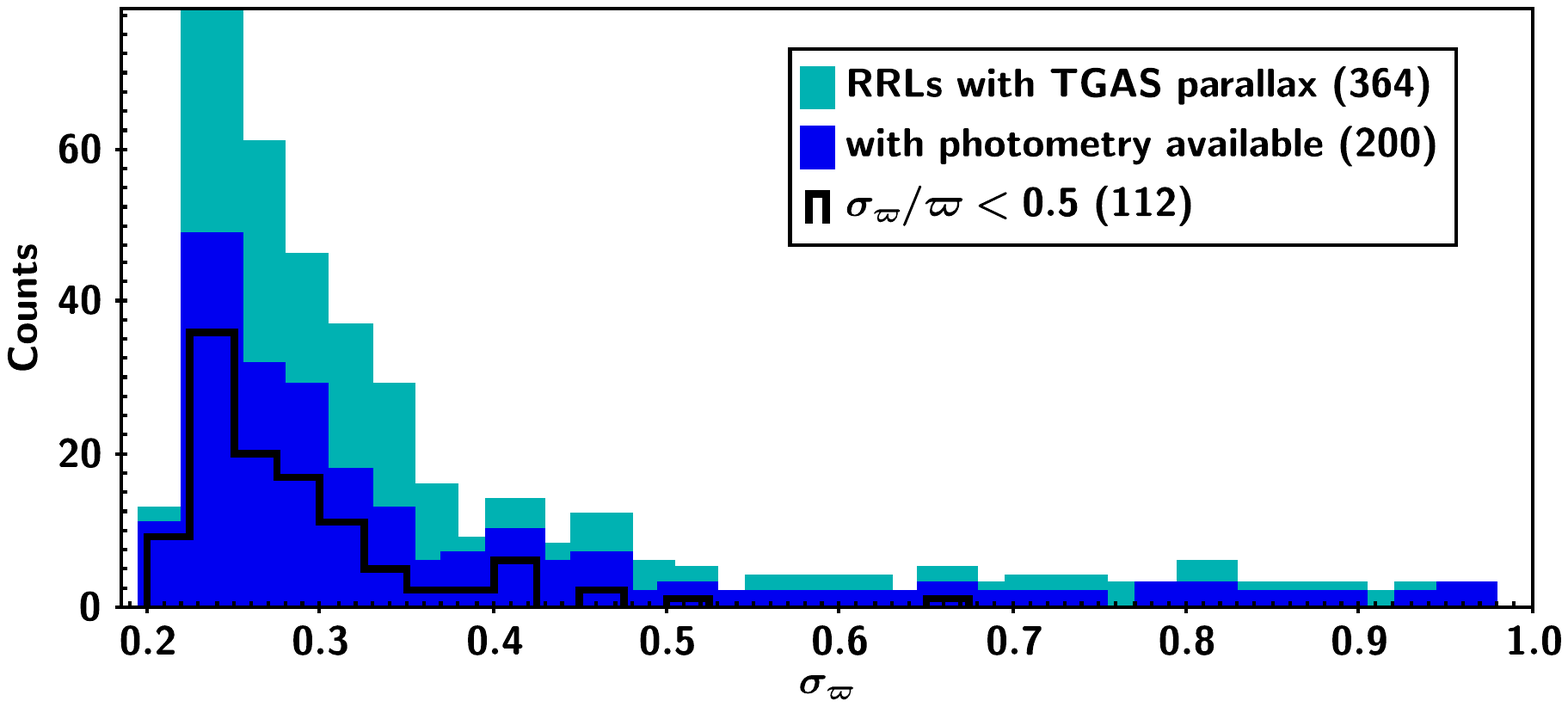}
\caption[]{Error distribution of the TGAS parallaxes for RR Lyrae stars (RRLs in the label): whole sample  (364 stars, cyan),  subsample with literature photometry (200 stars, blue), subsample of the previous 200 sources retaining only stars with positive parallax and 
parallax errors $\sigma_{\varpi}/\varpi<0.5$ (112 stars, black contour).  
The bin size is 0.025~mas.}
\label{fig:RRcumul}
\end{figure}

Finally, the  distribution on sky of the 331 classical Cepheids, 31 Type II Cepheids and 364 RR Lyrae stars considered in this paper is shown in Fig.~\ref{fig:skydistr},  where red filled circles mark the classical Cepheids that, as expected, mainly concentrate in the Milky Way (MW)  disk. Type II Cepheids and RR Lyrae stars are shown by green filled triangles and blue filled circles, respectively,  and nicely outline the MW halo. We note that by combining results from these three different standard candles  and the improved census of such variables that {\it Gaia} is expected 
to provide, it will be possible to further probe the MW 3D structure and  the all sky extension of the Galactic halo, a topic in which  {\it Gaia} already demonstrated its potential through 
discovery of over 300 new RR Lyrae stars in the far, still unexplored outskirts of one of our closest neighbours, the LMC (\citealt{clementini16}).

\begin{figure}
   \centering
   \includegraphics[trim=180 280 200 300 clip,  width=0.4\linewidth]{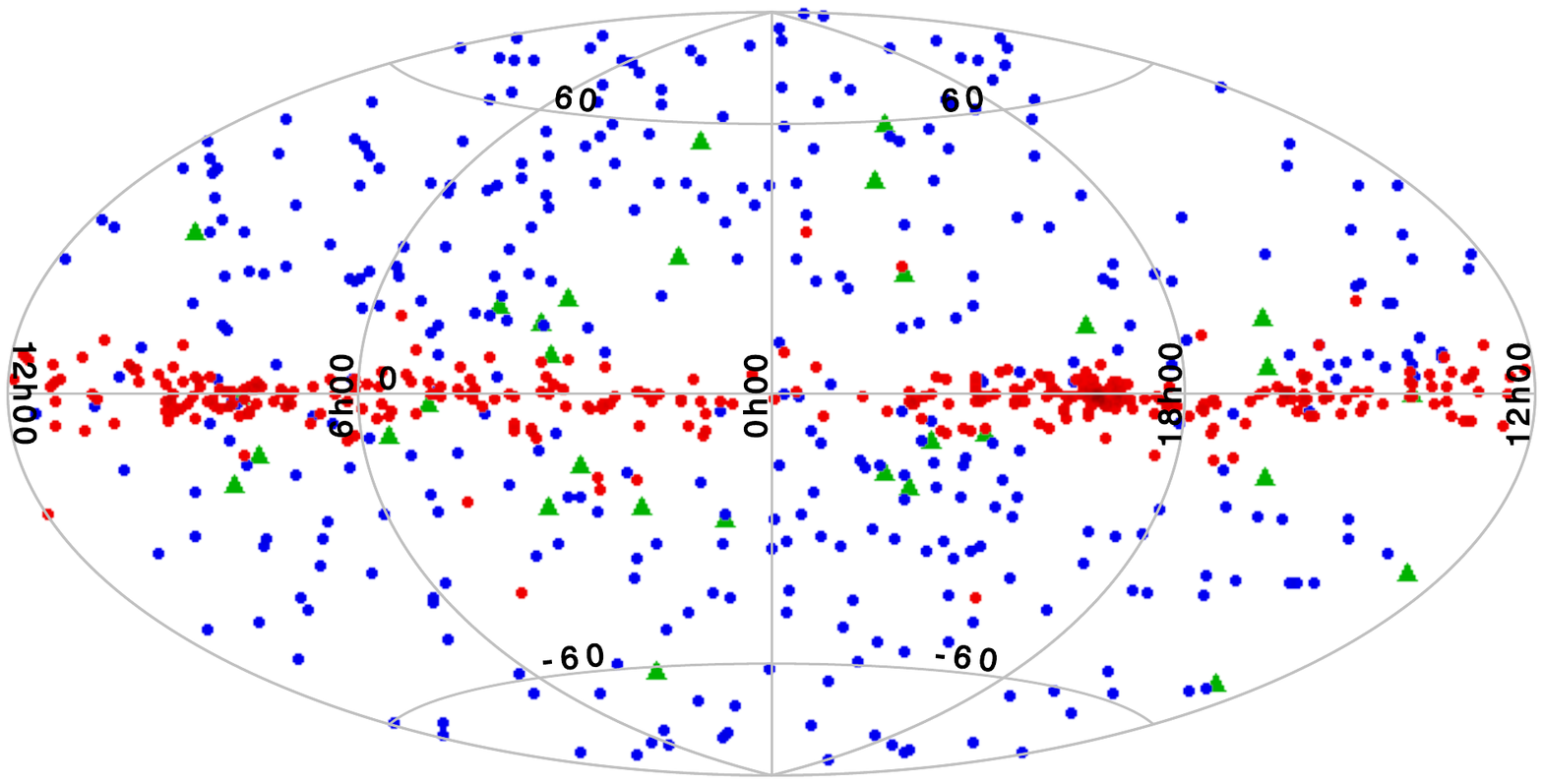}
   \caption{Sky distribution, in Galactic coordinates, of the 331 classical Cepheids  (red filled circles), 31 Type II Cepheids 
   (green filled triangles)  and 364 RR Lyrae stars (blue filled circles) discussed in this paper.} 
\label{fig:skydistr}
    \end{figure}

\subsection{Comparison with other parallax measurements}\label{comp_par}
Parallaxes obtained with the TGAS for classical  and Type~II Cepheids and for the RR Lyrae stars  published in {\it Gaia} DR1 are listed in Tables~\ref{tab:cephall}, ~\ref{tab:type2all} and ~\ref{tab:rrlsall},  where we also provide $G$-band magnitudes and other relevant photometric and spectroscopic information for these stars. 
In order to assess qualitatively the goodness of the TGAS parallaxes for Cepheids and RR Lyrae stars we compared the parallax values for variables having both TGAS and Hipparcos measurements (248 classical Cepheids, 31 Type~II Cepheids and 188 RR Lyrae stars).  The comparison TGAS {\it versus} Hipparcos for the classical Cepheids is shown in Fig.~\ref{fig:compCCs}, using black filled circles to mark the whole sample.
%In the figure, red and magenta filled circles highlight stars with $(\sigma_{\varpi})_{\rm Hipparcos} <$ 0.50 and 0.30~mas, respectively, whereas cyan filled circles are two stars 
%with $(\sigma_{\varpi}/\varpi)_{\rm Hipparcos} <0.20$, namely, 
We have labelled in the figure two stars, RW Cam and SY Nor, for which a significant discrepancy exists between Hipparcos and TGAS 
parallax values. Both stars are known to have very bright close-by companions (\citealt{evans1994, fernie2000}). We also do not plot the three sources with largest differences, namely, V1477 Aql, UX Per and AQ Pup. 
 The TGAS-Hipparcos comparison for classical Cepheids shows comforting results, the number of negative parallaxes  has reduced from 32\% in Hipparcos to only 4\% in TGAS:  of the 248 classical Cepheids 79 have a negative Hipparcos parallax to compare with only 5 of them still having  negative parallax and an additional 6 stars for a total of 11 sources in  TGAS. This is not surprising, since the fraction of negative parallaxes is expected to decrease when uncertainties get smaller.
We have created different  sub-samples based on absolute and relative errors of the Hipparcos parallaxes, in order to highlight the samples with most  reliable parallaxes. 
Classical Cepheids with $(\sigma_{\varpi}/\varpi)_{\rm Hipparcos} <0.20$ are marked in Fig.~\ref{fig:compCCs} by cyan filled circles, they are V2081 Cyg and PR Peg\footnote{Following the referee's comment that considering their periods and absolute magnitudes these two stars are maybe not classical Cepheids, we double-checked the literature and found that both stars are still classified as Cepheids in the latest version of the General catalogue of variable stars: Version GCVS 5.1 (\citealt{samus17}).}.  
Red and magenta filled circles highlight stars with $(\sigma_{\varpi})_{\rm Hipparcos} <$ 0.50 and 0.30~mas, respectively.
%Sources with $(\sigma_{\varpi})_{\rm Hipparcos}<$ 0.50 and 0.30~mas, respectively by red and magenta filled circles.
 Increasing agreement between the TGAS and Hipparcos results is found if  we consider only sources with precise Hipparcos values, suggesting that more precise Hipparcos measures correspond to more precise TGAS measures. 
Figures~\ref{fig:compT2Cs} and ~\ref{fig:compRRLs} show the same test but for Type~II Cepheids and RR Lyrae stars, respectively. Red filled circles in 
Fig.~\ref{fig:compT2Cs} indicate Type~II Cepheids with  $(\sigma_{\varpi})_{\rm Hipparcos} <0.50$.  Of the 31 Type~II Cepheids with both Hipparcos and TGAS parallaxes, 13 had a negative Hipparcos parallax (42\% of the sample), to compare with only 4 of them still having  negative parallax and an additional 1 for a total of 5 sources (16\%) in TGAS.  MZ Cyg is the source with largest discrepancy between
Hipparcos and TGAS among the Type~II Cepheids with a positive parallax value.
Red filled circles in Fig.~\ref{fig:compRRLs} are RR Lyrae stars with $(\sigma_{\varpi})_{\rm Hipparcos} <0.70$  
 while cyan filled circles are a few RR Lyrae stars with $(\sigma_{\varpi}/\varpi)_{\rm Hipparcos} <0.20$.  Of the 188 RR Lyrae stars with both Hipparcos and TGAS parallax, 59 had a negative Hipparcos parallax (31\% of the sample) to compare with only 2 of them still having  a negative parallax (1\%) in TGAS. CH Aql is the source with largest discrepancy between
Hipparcos and TGAS among the RR Lyrae stars with positive parallax values.
From these first global comparisons the improvement of {\it Gaia} with respect to Hipparcos is straighforward and is even more so for the Population~II standard candles, that is for RR Lyrae stars and Type~II Cepheids.

Considering now the most accurate astrometric parallaxes available in the literature,  we note that 3 classical Cepheids in our sample, namely, FF Aquilae (FF Aql), SY Aurigae (SY Aur) and  SS Canis Majoris (SS CMa), have their parallax measured with the Hubble Space Telescope (HST) by \citet{ben07}, \citet{riess14} and \citet{cas16}, respectively. The parallax of  FF Aql  was determined with the HST Fine Guidance Sensor, reaching  a precision of $\sigma_{\varpi}/$ $\varpi\sim$ 6$\%$. The astrometric measurements of SY Aur and SS CMa were obtained with the Wide Field Camera 3 (WFC3) by spatial scanning that improved the precision of the source position determination 
 allowing to derive parallaxes with uncertainties in the range of $\sim$ 0.3-0.5~mas ($\sigma_{\varpi}/$ $\varpi\sim$ 11-12$\%$). Parallax measurements available for these three stars are summarised 
 in the upper portion of Table~\ref{tab:varhst}.
   Taking into account the rather small sample and the much larger errors, as expected for these first {\it Gaia} parallaxes,  agreement between TGAS and HST is within 2$\sigma$ for FF Aql and SS CMa, and within 1$\sigma$ for SY Aur. 
 We also note that FF Aql is known to be in a binary system and this may have affected the measure of its parallax (see Section~\ref{sec:binary}).
Figure~\ref{fig:hhtccs} shows for these 3 classical Cepheids  the comparison between the TGAS and  HST parallax values (lower panel), between TGAS and Hipparcos (middle panel) and 
between Hipparcos and the HST (upper panel).
Going from top to bottom the agreement between different parallax values increases, the best agreement existing between the TGAS and HST values, thus confirming that TGAS, although less precise  than HST,  provides more reliable parallax measurements and an  improvement with respect to Hipparcos.

The parallax  has been measured  with the HST  only  for one of the Type~II Cepheids in our sample, VY~Pyx (\citealt{Benedict2011}). Results of the comparison between the TGAS, Hipparcos and HST parallaxes for this star are summarised in the mid portion of Table~\ref{tab:varhst} and shown in Fig.~\ref{fig:hhtt2}. The TGAS parallax for VY~Pyx differs significantly from the  HST and Hipparcos  values, which, on the other hand seem to be in reasonable agreement to each other. However, as discussed in \citet{Benedict2011} the  $K$-band absolute magnitude of VY~Pyx  inferred from the HST parallax places the star  1.19~mag below the  $PM_K$ relation defined by five RR Lyrae stars with parallax also measured by the HST (see  fig.~6 in \citealt{Benedict2011} and the discussion below), in contrast with the Type~II Cepheids being expected to lay  on the extrapolation to longer periods of the  RR Lyrae star $PM_K$ relation  (see e.g. \citealt{Ripepi2015} and references therein). \citet{Benedict2011} explain this discrepancy either as due to the wide range in absolute magnitude spanned by the short-period Type~II Cepheids or as caused by some anomaly in VY~Pyx itself.
 We have reproduced \citet{Benedict2011}'s fig.~6 in our Fig.~\ref{fig:Ben}  using for the 5 RR Lyrae stars in Benedict et al.'s sample the $M_K$ magnitudes calculated on the basis of their TGAS parallaxes (blue filled circles) and plotting with red lines the $PM_K$ relations obtained using instead Benedict et al.'s  HST parallaxes for the five RR Lyrae stars with (solid red line) and without (dashed red line) 
Lutz-Kelker corrections \citep{Lutz-Kelker}. Green circles represent star VY~Pyx with the $M_K$ magnitude calculated on the basis of Benedict et al. HST parallax (open circle) and TGAS parallax (filled circles),  respectively. The TGAS parallax makes  VY~Pyx  nicely follow the $PM_K$ relation defined by the 5 RR Lyrae stars, both in the formulation based on their TGAS  parallaxes 
(black solid line)  and that based on the  Benedict et al.'s parallaxes (red solid lines).  
 
As anticipated in the discussion of  VY~Pyx, HST parallaxes have been measured by \citet{Benedict2011} for five RR Lyrae stars. The comparison between Hipparcos, TGAS and \citet{Benedict2011} for these five variables is summarised  in the lower portion of Table~\ref{tab:varhst} and graphically shown  in Fig.~\ref{fig:hhtrrls} for  Hipparcos {\it versus} HST (upper panel), TGAS {\it versus} Hipparcos (middle panel) 
 and TGAS {\it versus} HST (lower panel), respectively.
Errors of the Hipparcos parallaxes are much larger than those of the  HST and TGAS measures and, except for RR Lyrae itself, the Hipparcos parallaxes differ significantly from the HST values, whereas  the TGAS and  HST parallaxes agree within 1$\sigma$ for RR~Lyr,  SU~Dra, UV~Oct, and XZ~Cyg. On the other hand, the 1$\sigma$ agreement of Hipparcos, TGAS and HST parallax values for RR Lyrae itself is particularly satisfactory, also in light of the much reduced error bar in the TGAS value: 0.23~mas, to compare with 0.64~mas in Hipparcos.
For the remaining star, RZ~Cep, \citet{Benedict2011} provide two different parallax values, 2.12 and 2.54~mas \citep{Neeley2015}. We show both values in  Fig.~\ref{fig:hhtrrls}. Although 
\citet{Benedict2011} preferred value for this star is 2.12~mas (corresponding to the grey filled circle in Fig.~\ref{fig:hhtrrls}) the alternative  value of 2.54~mas is in much better agreement with the TGAS parallax of RZ~Cep and nicely places the star on the bisector of the HST and TGAS parallaxes. 
To conclude, as already noted for the classical Cepheids  (see Fig.~\ref{fig:hhtccs}),  best agreement is found between the TGAS 
 and the HST parallaxes confirming once again the higher reliability of the TGAS parallaxes and the improvement with respect to Hipparcos. 
 
Fig.~\ref{fig:Ben}  deserves further comments. 
There is a systematic zero-point offset of about 0.14~mag between the $PM_K$ relation inferred from the HST parallaxes of \citet{Benedict2011} for the 5 RR Lyrae stars without applying any Lutz-Kelker correction (red dashed line) and the relation (black solid line) obtained with the  $M_K$ magnitudes inferred from the TGAS parallaxes (blue filled circles). The latter was obtained by 
 linear least squares fit of the $M_K$ magnitudes based on the TGAS parallaxes, adopting the same slope as in \citet{Benedict2011}, that is $-$2.38 from \citet{Sol2008}  and without applying  Lutz-Kelker corrections. Since there is good agreement between the TGAS and HST parallaxes of these 5 RR Lyrae stars, the observed zero point offset between $PM_K$ relations might hint to some systematic effect in the method used to compute these relations. Indeed, as discussed in detail in Section~\ref{sec:meth}, direct transformation of parallaxes to absolute magnitudes and linear least squares fit is 
 not advisable in presence of large errors as those affecting the parallaxes of these stars and this might have induced systematic effects.
 
  We remind the reader that although globally the possible systematic errors in the TGAS parallaxes are well below their formal errors\footnote{We remind that \citet{cas17}   claim that also formal errors of TGAS parallaxes may be overestimated.}, there could still be some systematic effects at a typical level of $\pm0.3$~mas depending on the sky position and the colour of the source (\citealt{lindegren16}). 
However, the question of these additional systematic errors is still under very much debate within DPAC and its value has often been recognized as an overestimate,  
which is why uncertainties smaller than 0.3~mas can be found in the TGAS catalog.     
  In principle, the nominal 
uncertainties quoted in the TGAS catalog already contemplate all sources of 
variance including the systematic uncertainties and a safety margin. 
Therefore, there should be no need to add the 0.3~mas extra-variance. 
Furthermore, the zero point error in the parallaxes is of the order of 
$-$0.04~mas  (\citealt{Arenou2017}), hence, does not seem to support the need for the 
extra-variance.
Additionally, while the analysis of regional/zonal effects (for example in quasars) 
shows differences across various regions of the sky, these systematic effects are spatially correlated and not totally
random over the celestial sphere. Hence,  they become an important issue only if analysing a particular 
region of the sky, like star clusters. 
 However, in all-sky studies like those presented in this paper, and particularly for the RR Lyrae stars, 
 which are not concentrated in any specific part of the sky (see Fig.~\ref{fig:skydistr}) this systematic effect does  not influence the global zero-point of the derived $PL$, $PLZ$ and $M_V-\rm{[Fe/H]}$ relations.
 
  \citet{Arenou2017} report systematic zero-points respectively of $-0.014\pm0.014$~mas  and $-0.07\pm0.02$~mas  in the TGAS parallaxes of 207 classical Cepheids and 130 RR Lyrae stars they have analysed,  and an average shift of $-0.034\pm0.012$~mas  when combining the two samples. We have not found in the literature information about systematic effects affecting the HST parallaxes.   
 Nevetheless, the direct star-by-star comparison of the parallaxes in Table~\ref{tab:varhst} and Fig.~\ref{fig:hhtrrls} does not seem to show evidence for  the presence of a systematic difference between  the TGAS and HST parallaxes of the fairly small sample (3 classical Cepheids, 1 Type~II Cepheid and 5 RR Lyrae stars) for which a direct comparison with the HST is possible.}

 \begin{table*}
\footnotesize
\caption[]{Comparison between Hipparcos,TGAS and HST parallaxes\label{tab:varhst}}
\begin{center}
\begin{tabular}{l c c c c c c c l}
\hline
\hline
\noalign{\smallskip}
Name   & ID$_{\rm Hipparcos}$*&$\varpi_{\rm Hipparcos}$ & $\sigma\varpi_{\rm Hipparcos}$& 
$\varpi_{\rm TGAS}$ & $\sigma\varpi_{\rm TGAS}$ &$\varpi_{\rm HST}$ & $\sigma\varpi_{\rm HST}$& HST Reference\\
   &  &(mas) & (mas) &  (mas) & (mas) & (mas) & (mas) & \\
\hline
\noalign{\smallskip}
\multicolumn{9}{c}{\textbf{Classical Cepheids}}\\
\hline
\noalign{\smallskip}
FF~Aql**  & 93124 &  2.110 &$\pm$ 0.330  & 1.640     &     $\pm$0.89  & 2.810     &
$\pm$0.180  &   \citet{ben07}    \\ 
SS~CMa &36088 &  0.400 &$\pm$1.780 &0.686     &   $\pm$0.234    & 0.348 &   
$\pm$0.038     &    \citet{cas16}   \\
SY~Aur &24281 &  -1.840 & $\pm$1.720 &0.687     &   $\pm$0.255    & 0.428     &
$\pm$0.054   &   \citet{riess14}     \\
\noalign{\smallskip}
\hline 
\noalign{\smallskip}
\multicolumn{9}{c}{\textbf{Type II Cepheids}}\\
\noalign{\smallskip}
\hline
\noalign{\smallskip}
VY~Pyx& 434736 & 5.00 &$\pm$ 0.44  & 3.85 &$\pm$ 0.28 & 6.44&$\pm$ 0.23  & \citet{ben07}    \\
\noalign{\smallskip} 
\hline
\noalign{\smallskip}
\multicolumn{9}{c}{\textbf{RR Lyrae stars}}\\
\hline
\noalign{\smallskip}
RR Lyr & 95497  &3.46 &  $\pm$0.64  &3.64&   $\pm$0.23    &  3.77     &  $\pm$0.13 &
   \citet{Benedict2011}   \\ 
RZ Cep & 111839  & 0.59 &$\pm$1.48& 2.65&  $\pm$0.24  &2.12 (2.54)***  &  $\pm$0.16     & 
\citet{Benedict2011}     \\
SU Dra &  56734    & 0.20 &   $\pm$1.13&1.43& $\pm$0.29 & 1.42      & $\pm$0.16    
&   \citet{Benedict2011}      \\
UV Oct & 80990      &    2.44   &    $\pm$0.81  & 2.02& $\pm$0.22      & 1.71  &  
$\pm$0.10       &  \citet{Benedict2011}     \\
XZ Cyg & 96112   &      2.29 &  $\pm$0.84&1.56&$\pm$0.23&      1.67  &  $\pm$0.17   
&   \citet{Benedict2011}     \\
\noalign{\smallskip}
\hline 
\end{tabular}
\end{center}
* \citet{van07b}\\
**\cite{gallenne2012} have estimated the distance to FF Aql via  interferometric Baade-Wesselink technique, the corresponding parallax is 2.755$\pm$0.554 mas.\\
*** Two different parallax values are provided for this star by  \citet{Benedict2011}, in the table we list both values.\\
\normalsize
\end{table*}

\begin{figure}[t!]
\centering
\includegraphics[trim=0 150 0 50 clip, width=0.95\linewidth]{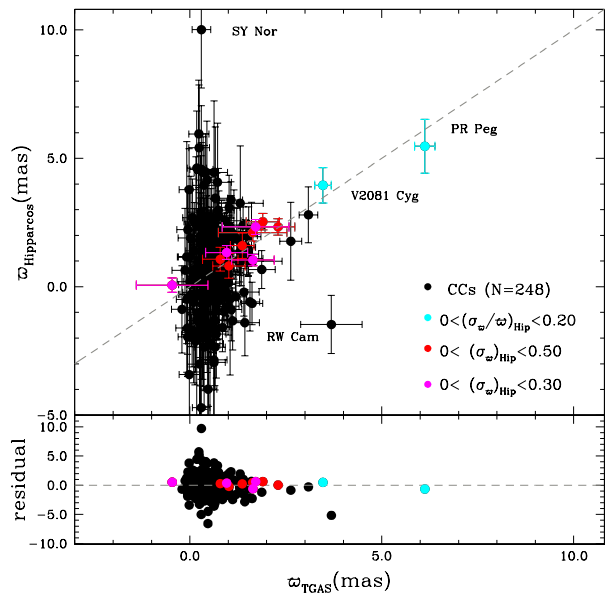}
\caption[]{Comparison between Hipparcos and TGAS parallaxes, obtained from a sample of 248 classical Cepheids which have both measurements. 
Red and magenta filled circles represent stars with $(\sigma_{\varpi})_{\rm Hipparcos} <$ 0.50 and 0.30~mas, respectively, cyan filled circles are two stars 
with $(\sigma_{\varpi}/\varpi)_{\rm Hipparcos} <0.20$, namely, V2081 Cyg and PR Peg.  A dashed line shows the bisector. Residuals are: TGAS $-$ Hipparcos parallax values.
}
\label{fig:compCCs}
\end{figure}

\begin{figure}
\includegraphics[trim=0 150 0 75 clip, width=0.95\linewidth]{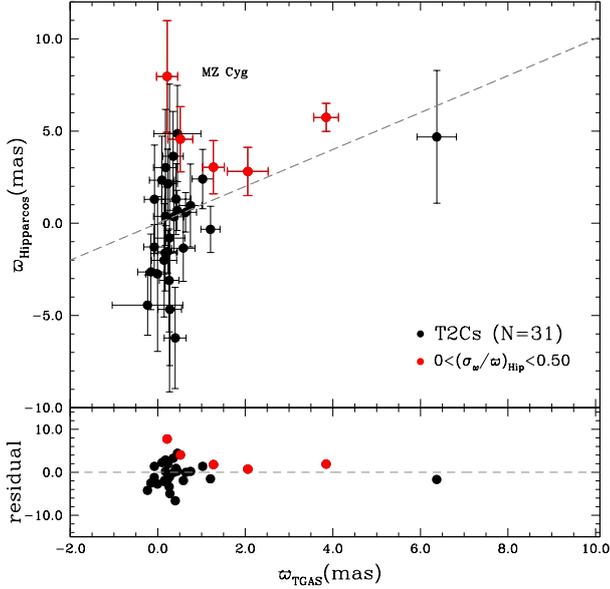}
\caption{Comparison between Hipparcos and TGAS parallaxes, obtained from the sample of 31 Type~II  Cepheids (T2Cs in the label) which have both measurements.  Red filled circles represent stars with 
$(\sigma_{\varpi}/\varpi)_{\rm Hipparcos} <0.50$. A dashed line shows the bisector.  Residuals are: TGAS $-$ Hipparcos parallax values.
}
\label{fig:compT2Cs}
\end{figure}

\begin{figure}[t!]
\centering
\includegraphics[trim=0 150 0 75 clip, width=0.95\linewidth]{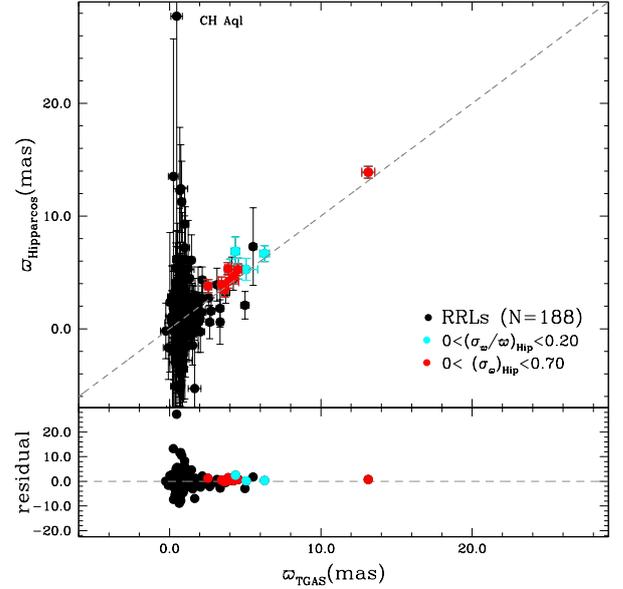}
\caption[]{Comparison between Hipparcos and TGAS parallaxes, obtained from a sample of 188  RR Lyrae stars (black filled circles) which have both measurements. 
Cyan filled circles mark sources with $(\sigma_{\varpi}/\varpi)_{\rm Hipparcos} <0.20$.  
Red filled circles are  RR Lyrae stars 
with $(\sigma_{\varpi})_{\rm Hipparcos} <$ 0.70.  
A dashed line shows the bisector. Residuals are: TGAS $-$ Hipparcos parallax values.
}
\label{fig:compRRLs}
\end{figure}

\begin{figure}[t!]
\centering
\includegraphics[trim=0 160 0 160 clip, width=0.95\linewidth]{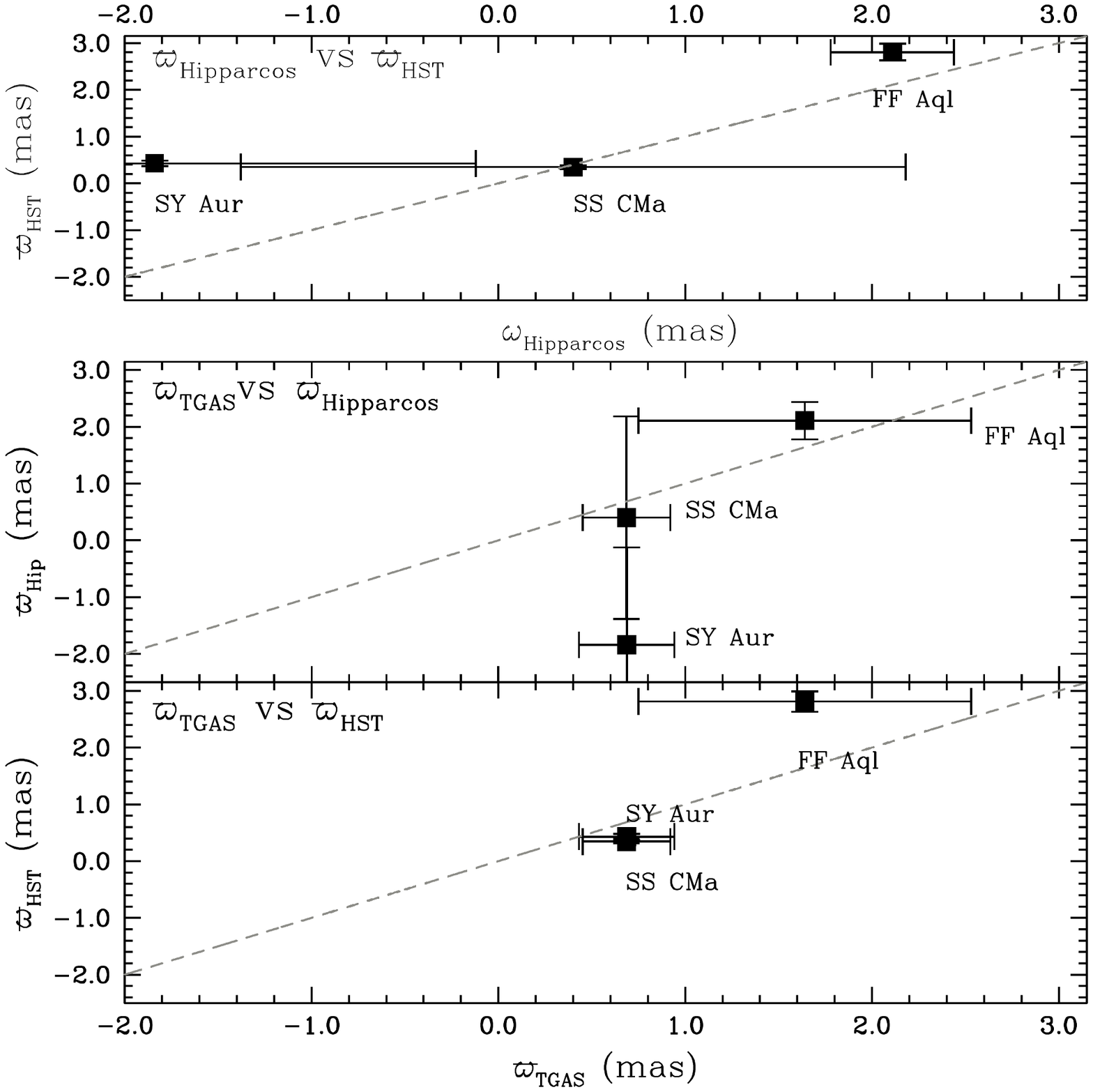}
\caption[]{Comparison between Hipparcos and HST parallax (upper panel), 
TGAS and Hipparcos parallax (middle panel)  and TGAS and HST parallax (lower panel)  for the classical Cepheids FF Aql, SY Aur and SS CMa. 
FF Aql is the brightest star in our sample of 331 classical Cepheids and is known to be component of a binary system. A dashed line shows the bisector.}
\label{fig:hhtccs}
\end{figure}

\begin{figure}[t!]
\centering
\includegraphics[trim=0 160 0 120 clip, width=0.95\linewidth]{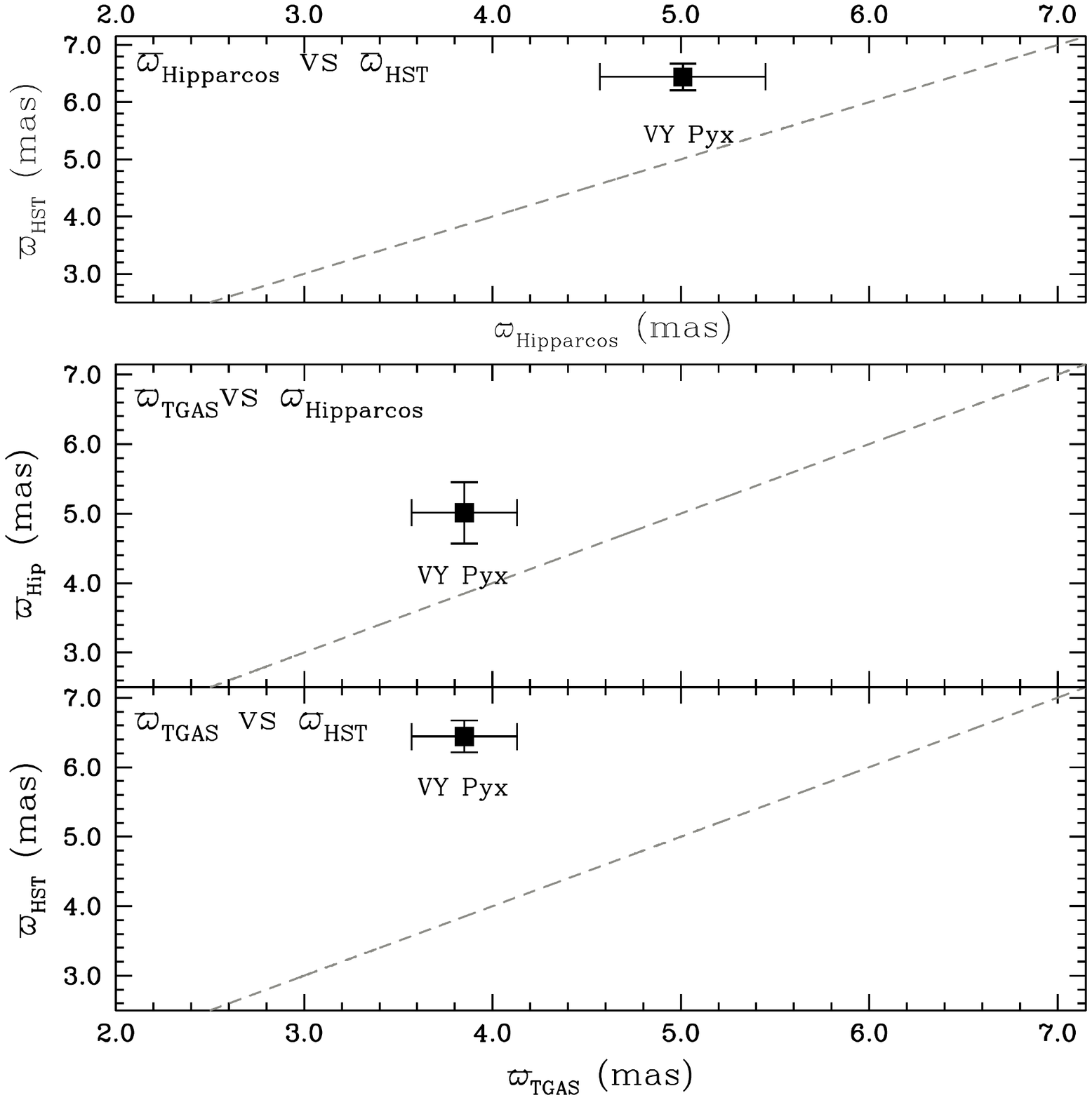}
\caption[]{Comparison between Hipparcos and HST parallax (upper panel), 
TGAS and Hipparcos parallax (middle panel)  and TGAS and HST parallax (lower panel)  for the Type II Cepheid VY Pyx. Dashed lines show the bisectors.}
 \label{fig:hhtt2}
\end{figure}

\begin{figure}
   \centering
   \includegraphics[trim=40 150  20 100 clip, width=\linewidth]{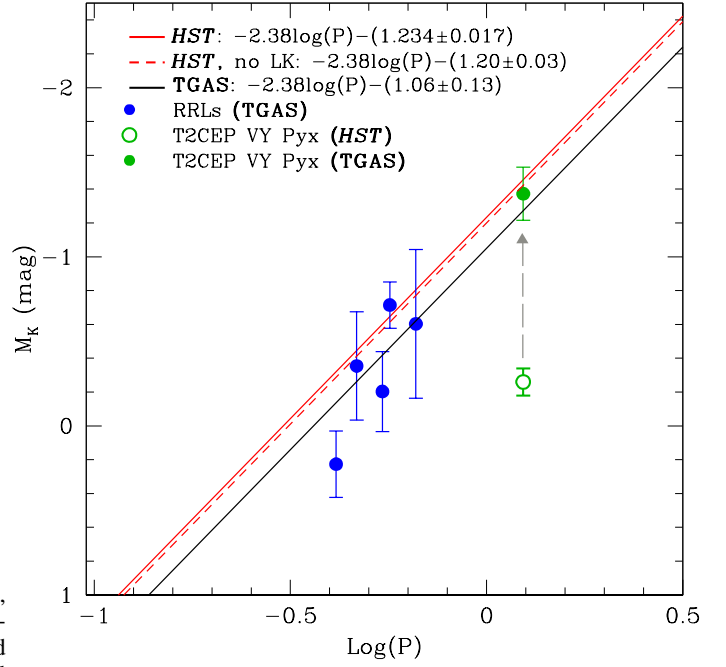}
   \caption{Weighted linear least squares fit performed over the $M_K$ magnitude of the five RR Lyrae stars in \citet{Benedict2011} using the $M_K$ values inferred from the HST parallaxes with (red solid line) and 
   without (red dashed line) Lutz-Kelker correction and the $M_K$ values (blue filled circles) inferred from the TGAS 
   parallaxes (black line).  
   Green filled and open circles show the Type II Cepheid VY Pyx with the $M_K$ magnitude determined from the TGAS and HST parallax, respectively. The star was not used in the fit. } 
\label{fig:Ben}   
    \end{figure}

\begin{figure}[t!]
\centering
\includegraphics[trim=0 160 0 120 clip, width=0.95\linewidth]{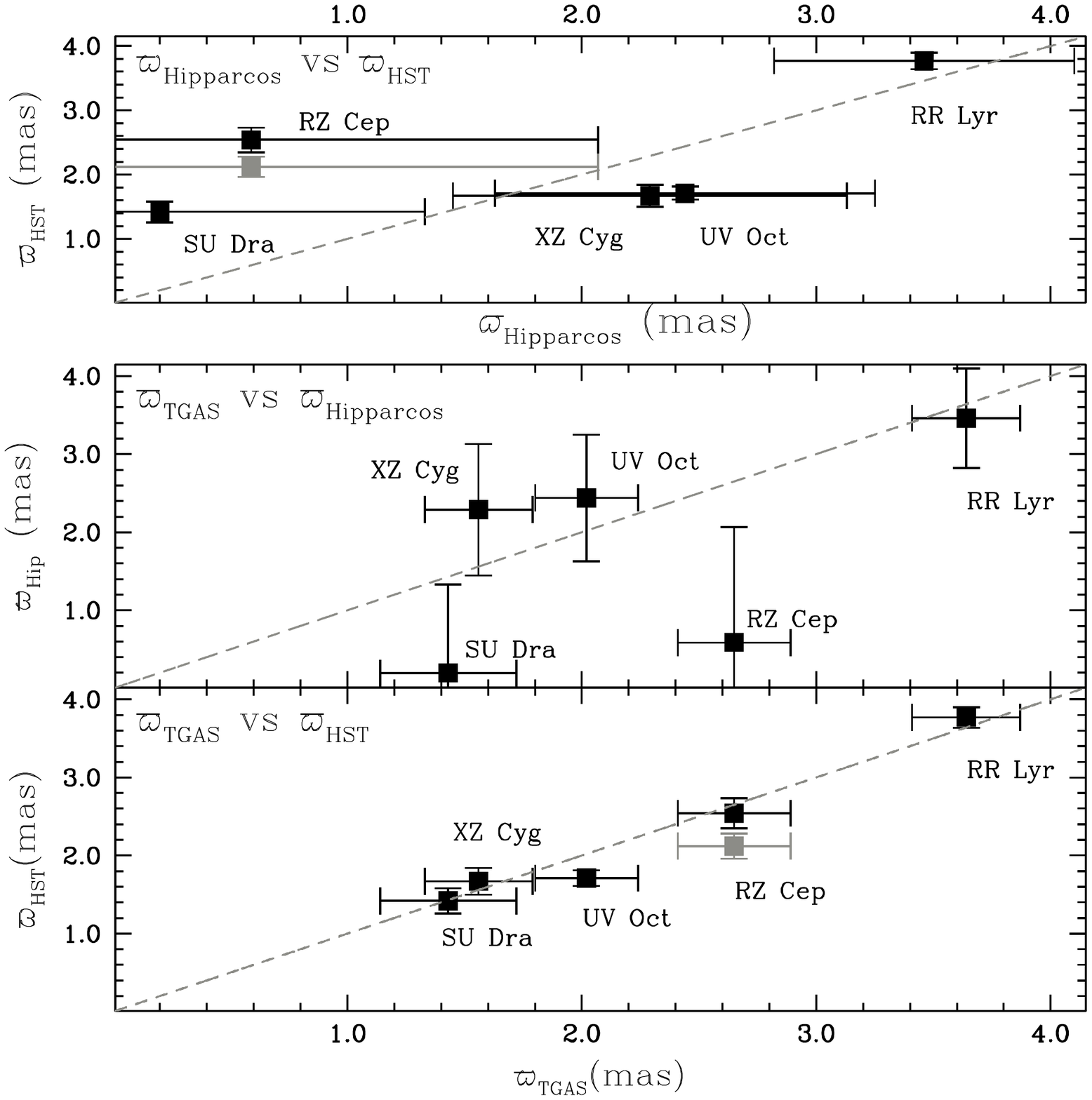}
\caption[]{Comparison between Hipparcos and HST parallaxes (upper panel), 
TGAS and Hipparcos (middle panel),  TGAS and HST (lower panel)  for the RR Lyrae stars: RR Lyr, RZ Cep, SU Dra, XZ Cyg and UV Oct. Two values   
from  \citet{Benedict2011} are shown  for RZ~Cep:  2.12~mas (grey filled square) and 2.54~mas (black filled square). TGAS parallax for RZ~Cep is in good agreement with the larger,  less favourite value in \citet{Benedict2011}. Dashed lines show the bisectors.}
 \label{fig:hhtrrls}
\end{figure}

\subsection{Comparison with parallaxes inferred by theoretical model fitting of the light curves}\label{sec:modelfit}

\begin{figure}
\includegraphics[width=\linewidth]{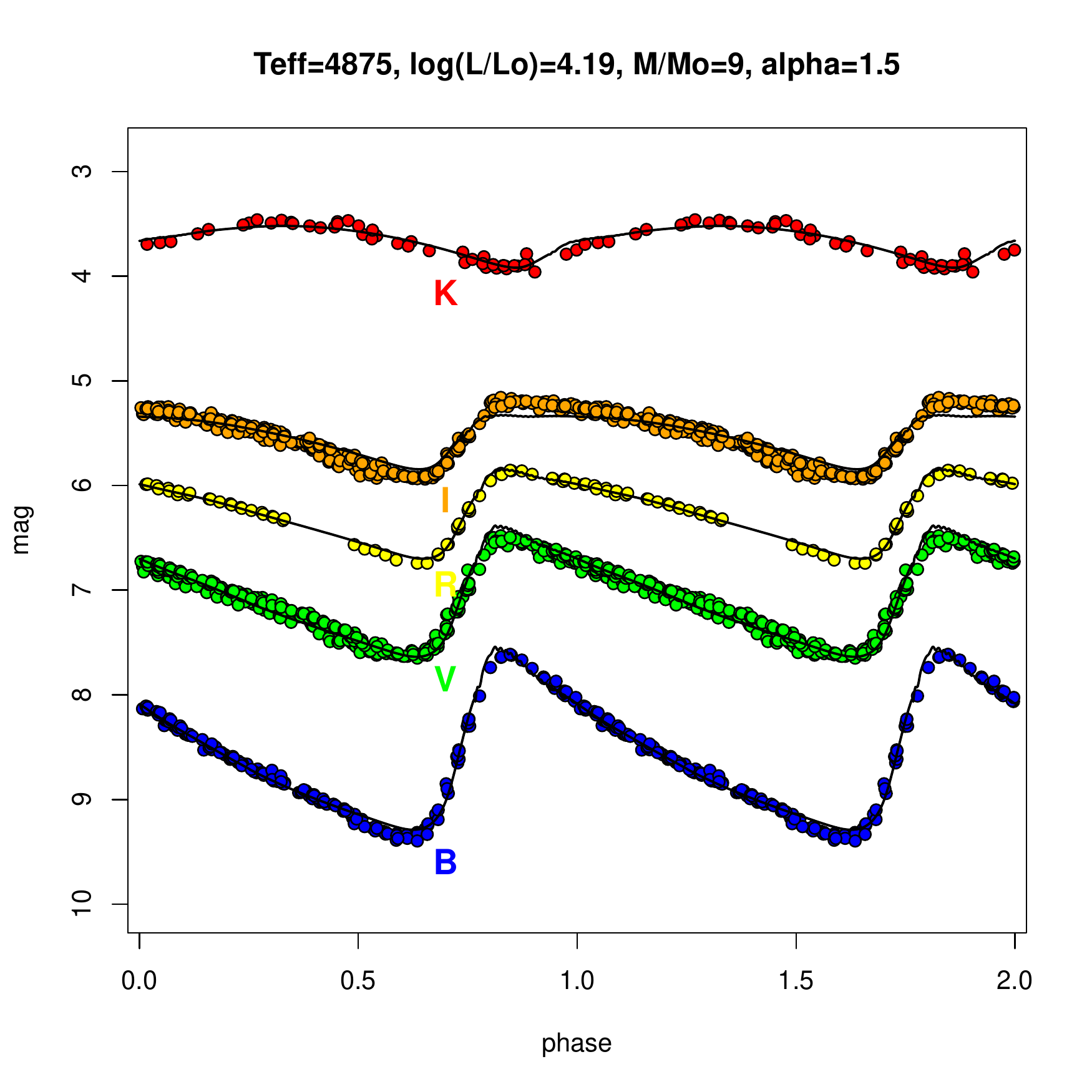}
\caption{Model fitting of the fundamental mode classical Cepheid RS Pup, the second longest period Cepheid in our sample with P=41.528~days. The model fitting provides a parallax: $\varpi_{\rm FIT}$= 0.58 $\pm$ 0.03~mas, in excellent agreement with the TGAS parallax for this star:  $\varpi_{\rm TGAS}$= 0.63 $\pm$ 0.26~mas.}
 \label{fig:rspup}
\end{figure}

An independent method to infer the distance (hence the parallax) of a pulsating star is the ``model fitting'' of the multi-wavelength star light curves  through nonlinear convective pulsation models 
\citep[see e.g.][and references therein]{mc05,kw06,marconi13a,marconi13b}. 
Indeed, one of the advantages of nonlinear hydrodynamical codes that involve a detailed treatment of the coupling between pulsation and convection is that they are able to predict the variation of any relevant quantity along the pulsation cycle. The direct comparison between observed and predicted  light curve as based on an extensive set of models with the period fixed to the observed value but varying the mass, the luminosity, the effective temperature and the chemical composition, allows us to obtain a best fit model and in turn to constrain not only the distance but also the intrinsic stellar properties of the  pulsating star under study. This approach was first applied to a Magellanic Classical Cepheid \citep{was97} and a field RR Lyrae \citep{bcm00} and later extended to cluster members \citep{md07,marconi13b} and variables for which radial velocity curves were also available \citep[see e.g.][and references therein]{difa02,natale08,marconi13a,marconi13b}. Furthermore,  the method was successfully applied to a number of different classes of pulsating stars in the LMC  \citep[see e.g.][and references therein]{bcm02, mc05, mcna07, marconi13a, marconi13b}, also by different teams \citep[see also][]{was97,kw02,kw06}, always obtaining consistent results. 
 Because of the significant amount of time and computing resources required by the model fitting technique, here we applied this method only to three  classical Cepheids, for which multi-band light curves are available in the literature,  that we selected among the sample of classical Cepheids we use to derive the $PL$ relations in Sec.~\ref{sec:plccs}. The first case is RS Pup, pulsating in the fundamental mode with a period of 41.528~days. 
The photometric data for this star and for the other two analysed in this section, are taken from a number of papers \citep{w84,ls92,berdni08,mp11} and well sample the light variations in the different filters.  Fig.~\ref{fig:rspup} shows the results of model fitting the star light curves in the $B,V,R,I,K$ bands. RS Pup is  the second longest period classical Cepheids in our sample  
and is known to be  surrounded by a nebula reflecting the light from the central star \citep[see e.g.][]{kerve08,kervella14}, thus  allowing an independent geometric evaluation of the distance to be obtained from the light echoes propagating in the star circumstellar nebula, corresponding to a parallax  $\varpi_{K14}$= 0.524 $\pm$ 0.022~mas \citep[see][for details]{kervella14}. This value is consistent within the errors with the TGAS value   
 $\varpi_{\rm TGAS}$= 0.63 $\pm$ 0.26~mas.
The pulsation model best reproducing RS Pup multi-filter light curve corresponds to a 9 $M_{\odot}$ star with an intrinsic 
luminosity $\log{L/L_{\odot}}$=4.19. From the apparent distance moduli obtained with the best fit in the various bands we were able to estimate the extinction correction and the intrinsic distance modulus $\mu_0{\rm (FIT)}=11.1\pm0.1$~mag. This provides a model fitting  parallax $\varpi_{\rm FIT}$= 0.58 $\pm$ 0.03~mas, which is in excellent agreement with the TGAS parallax for the star.

\begin{figure}
\includegraphics[width=\linewidth]{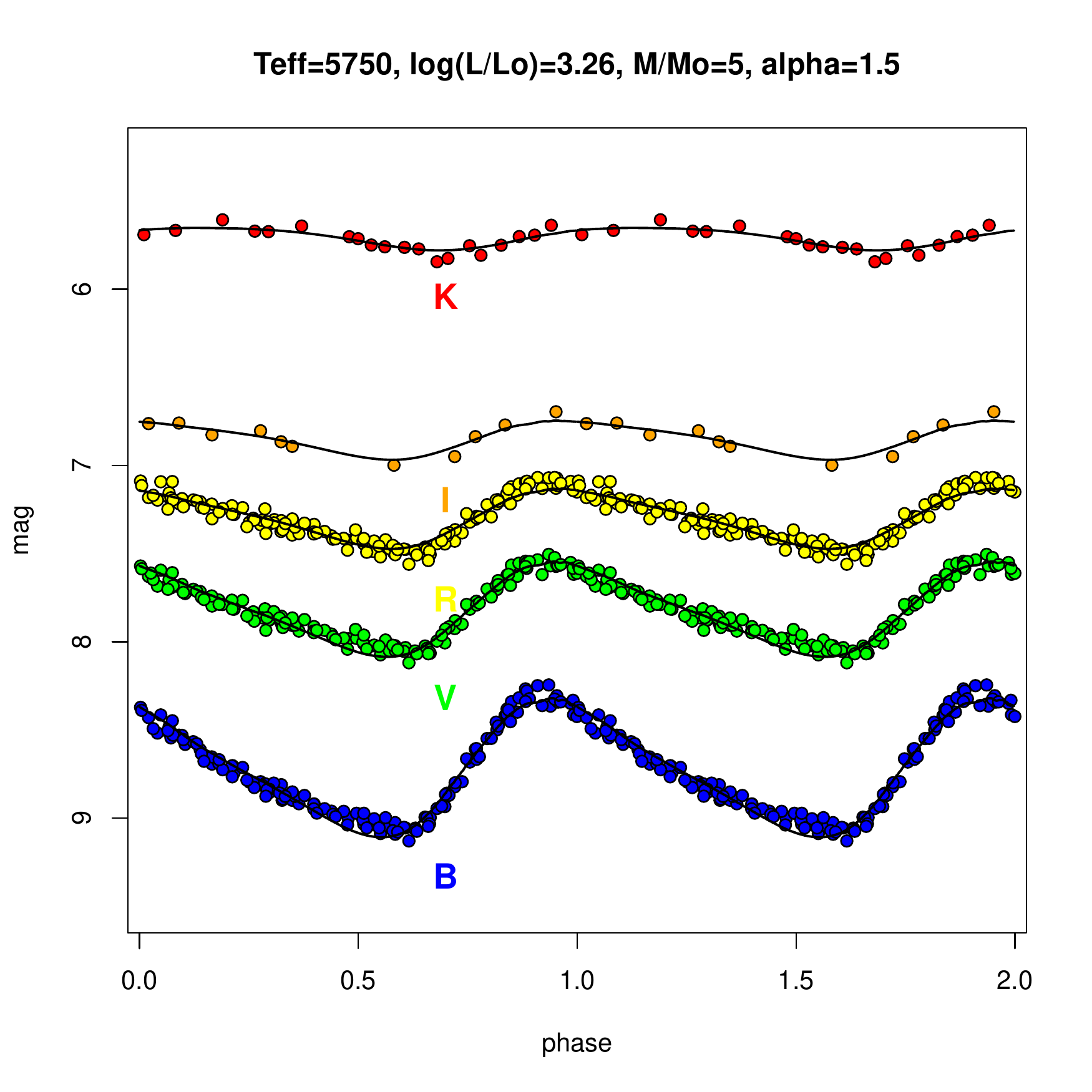}
\caption{Model fitting of the fundamental mode classical Cepheid V1162 Aql, P=5.376~days. The model fitting provides a parallax: $\varpi_{\rm FIT}$= 0.79 $\pm$ 0.04~mas, in agreement, within the errors, with the TGAS parallax for this star:  $\varpi_{\rm TGAS}$= 1.01 $\pm$ 0.29~mas.}
  \label{fig:v1162aql}
\end{figure}

Similarly, we performed the model fitting of V1162 Aql light curves, a Galactic fundamental mode Cepheid of much shorter period (5.376~days), as shown in Fig.~\ref{fig:v1162aql}. 
In this case, the pulsation model best reproducing the multi-filter light curve corresponds to a  5 $M_{\odot}$ star with an intrinsic luminosity $\log{L/L_{\odot}}=3.26$. The inferred model fitting intrinsic distance modulus is  $\mu_0{\rm (FIT)}$=10.5$\pm0.1$, corresponding to a parallax  
$\varpi_{FIT}$= 0.79 $\pm$ 0.04~mas, in agreement with the TGAS value ( $\varpi_{\rm TGAS}$= 1.01 $\pm$ 0.29~mas), within the errors.

\begin{figure}
\includegraphics[width=\linewidth]{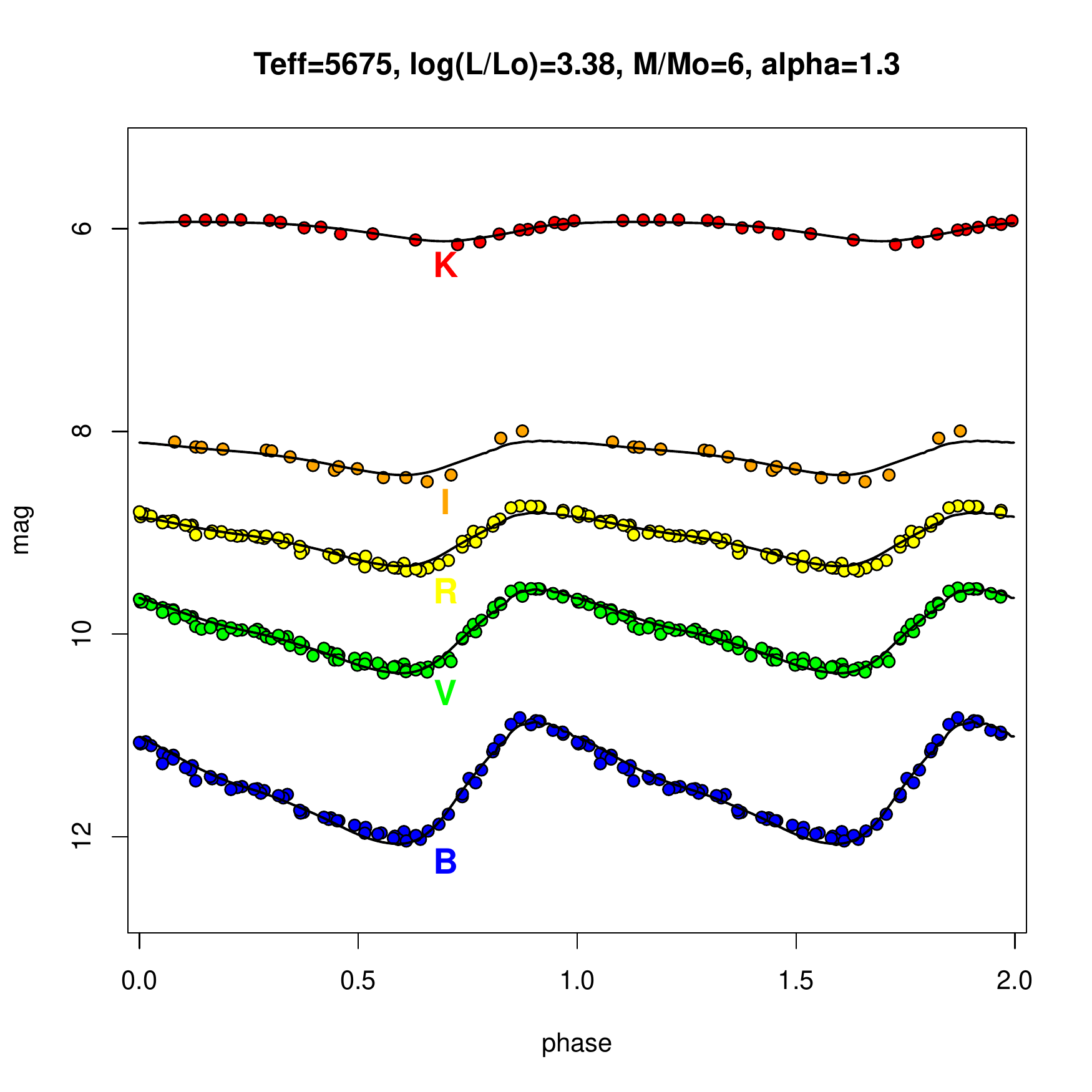}
\caption{Model fitting of the fundamental mode classical Cepheid RS Cas, P=6.296~days. The model fitting provides a parallax: $\varpi_{\rm FIT}$= 0.60 $\pm$ 0.03~mas in significant disagreement 
with  the TGAS parallax for this star:  $\varpi_{\rm TGAS}$= 1.53 $\pm$ 0.32~mas.}
  \label{fig:rscas}
\end{figure}

The third classical Cepheid analysed with the model fitting technique is RS Cas, a Galactic fundamental mode Cepheid with a  period of about 6.296~days. 
When applying our model fitting approach, we obtain the best fit model shown in Fig.~\ref{fig:rscas}, corresponding to a 6 $M_{\odot}$ star with an intrinsic luminosity $\log{L/L_{\odot}}=3.38$. This implies an intrinsic distance modulus  $\mu_0{\rm (FIT)}$=11.1$\pm0.1$ and a pulsation parallax $\varpi_{\rm FIT}$= 0.60 $\pm$ 0.03~mas, much smaller than the TGAS parallax  $\varpi_{\rm TGAS}$= 1.53 $\pm$ 0.32~mas. Consequently, the predicted distance modulus is about 2~mag longer than the TGAS-based value and the absolute magnitude is brighter by the same amount. 
It is interesting to note that an upward shift of approximately 1.5-2~mag would allow  RS Cas  
to match the $PL$ relations in Fig.~\ref{fig:PL_CC_K}. 
This seems to suggest that the TGAS parallax for RS Cas is incorrect. We wonder whether the discrepancy observed for this star may be caused by a companion such as e.g. a white dwarf, which might affect the TGAS measurement. 

Finally, we note that the theoretical model fitting technique has been often applied also to RR Lyrae stars both in and outside the MW (e.g. \citealt{bcm00}, \citealt{difa02}, \citealt{mc05}).
For one of the RR Lyrae star with TGAS parallax, U Com,  \citet{bcm00} measured the parallax by fitting the star multi-band light curves with nonlinear convective pulsation models:
 $\varpi_{\rm FIT}$= 0.63 $\pm$ 0.02~mas. 
 This value agrees within the errors with  TGAS parallax for U Com:  $\varpi_{\rm TGAS}$= 0.46 $\pm$ 0.28~mas.
 
The results presented in this section confirm the predictive capability of the adopted theoretical scenario and the potential of the light curve model fitting technique 
to test and constrain the accuracy of empirical distance determinations. 

\subsection{Comparison with Baade-Wesselink studies.}

\begin{figure}[t!]
\centering
\includegraphics[trim=0 160 0 80 clip, width=0.95\linewidth]{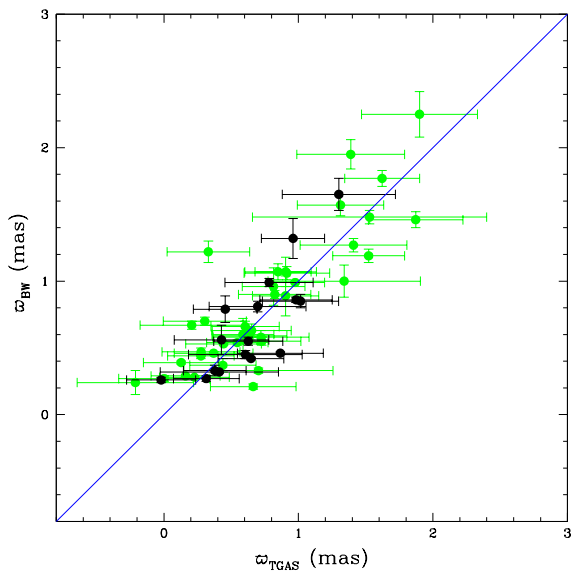}
\caption[]{Comparison between the photometric parallaxes  estimated
via B-W method and the TGAS parallaxes of classical Cepheids. Black and green circles represent single and binary Cepheids, respectively. The blue line is the bisector.}
\label{fig:BW_CC}
\end{figure}

\begin{figure}[t!]
\centering
\includegraphics[trim=0 160 0 80 clip, width=0.95\linewidth]{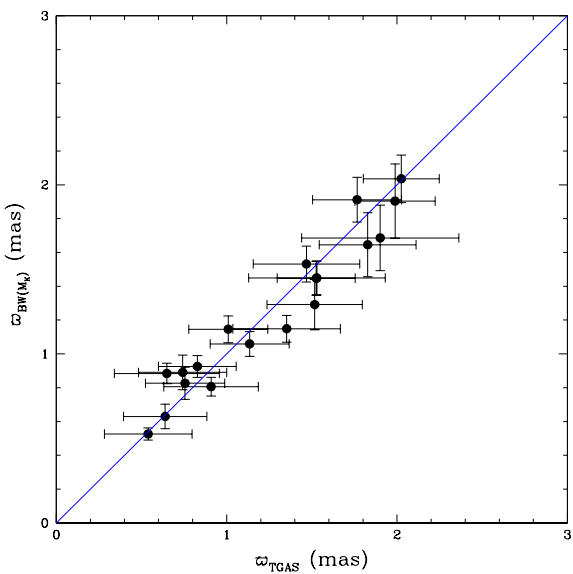}
\caption[]{Comparison between  the photometric parallaxes inferred from the $K$-band absolute magnitudes ($M_{K}$ ) estimated
via  B-W method and the TGAS parallaxes for RR Lyrae stars. The blue line represents the bisector.}
 \label{fig:MkBWgaia}
\end{figure}

Photometric parallaxes of  classical Cepheids and RR Lyrae stars have been often estimated with the Baade-Wesselink (B-W) method in various different implementations
(e.g. the Infrared Surface Brightness technique or the Spectro-Photo-Interferometric modeling approach of \citealt{merand15} and  \citealt{breitfelder16}).

 \citet{fou2007} list in their Table~6,  photometric parallaxes inferred from the application of the Infrared Surface Brightness version of the B-W technique to 
 62 classical Cepheids, among which 54 have a TGAS parallax estimate. The comparison between the TGAS and the B-W parallaxes for these 54 Cepheids is shown in Fig.~\ref{fig:BW_CC}.
 The sample of 54 classical Cepheids contains a large fraction (38) of binary systems. The phenomenon of binarity/multiplicity is rather common among classical Cepheids, as it will be discussed in more detail in Sec.~\ref{sec:binary}.  The presence of a binary companion may prevent an accurate estimate of parallax. Cepheids known to be in binary systems are shown as green circles in Fig.~\ref{fig:BW_CC}, they are more scattered around the bisector line. The r.m.s scatter from the bisector line is
0.28 mag for binary classical Cepheids and reduces to 0.23 mag when the 16
non binary Cepheids are considered. 
 A weighted least squares fit of the relation  $\varpi_{B-W} =\alpha \varpi_{\rm TGAS}$ 
 returns a slope value of $(0.90\pm0.07)$ for the sample of 16 classical Cepheids which are not in binary systems.

 A similar comparison  was also done for the RR Lyrae stars. We considered 19 MW RR Lyrae variables with TGAS parallaxes  and absolute visual  ($M_{V}$)  and $K$-band ($M_{K}$) magnitudes available in the literature from B-W studies (see Table~2 in \citealt{Muraveva2015}). The B-W absolute magnitudes were  
taken from the compilations  in Table~11 of
\citet{cacciari92}  and Table~16 of \citet{Skillen1993} and revised:
(i) assuming for the p factor used to transform  the observed radial velocity to true pulsation velocity the value p = 1.38 proposed by \cite{fernley1994}\footnote{According to table~1 in \cite{fernley1994} this change in the p factor makes the B-W absolute magnitudes to become  systematically brighter by 0.1 mag, on average.}, and (ii)
averaging multiple determinations of individual stars. 
We note that the $K$-band magnitudes of RR Lyrae stars analyzed with the B-W method are in the Johnson photometric system, however, the difference between 2MASS $K_\mathrm{s}$ and Johnson $K$ is small, of about 0.03~mag on average, for the  typical colour of RR Lyrae stars (\citealt{Muraveva2015}) and, anyway,  much smaller than individual errors in the B-W $M_{K}$ magnitudes. 
We have transformed the B-W absolute magnitudes to photometric parallaxes. 
 The direct transformation of parallaxes to absolute magnitudes and vice-versa should be avoided  when parallax uncertainties are large because of the resulting  asymmetric errors (see Sec.~\ref{sec:meth}). However, given the small relative errors in distance 
moduli (less than 3\%) of the 19 RR Lyrae stars analysed with the B-W technique, the uncertainties in the inferred parallaxes are 
symmetric and would not be affected by any reasonable prior 
distribution.
Comparison of the photometric parallaxes inferred from the $M_{K}$ absolute magnitudes of the 19 RR Lyrae stars with the corresponding 
 TGAS
parallaxes  is shown in Fig.~\ref{fig:MkBWgaia}.   
The two independent parallax estimates appear to be in very good agreement within the errors. We performed also the comparison with the parallaxes inferred from the $V$-band absolute magnitudes. A weighted least squares fit of the relations  $\varpi_{M_V (B-W)} =\alpha \varpi_{\rm TGAS}$ 
and $\varpi_{M_K (B-W)} =\alpha \varpi_{\rm TGAS}$ returns slopes of 0.97 and 0.98, respectively, which are both  very close to the bisector slope
$\alpha=1$. 
 To conclude, TGAS parallaxes are in general good agreement with the photometric parallaxes obtained in B-W studies of classical Cepheids and RR Lyrae stars. The agreement is particularly good  for the RR Lyrae stars and seems to support the adoption of the larger p factor proposed by  \cite{fernley1994}. 
 
\section{Selection biases and methods}\label{sec:bias_meth}
\subsection{Biases}\label{sec:bias}

When determining a luminosity calibration (e.g.  a $PL$,  or $PW$, or
$PLZ$ or $M_V - {\rm[Fe/H]}$ relation in our case)
from astrometric data, we have to be very careful in taking into account
possible sources of bias that can affect it, lest the results contain
systematic errors.
A first source of such biases is the way the sample has been obtained, due 
to either censorship (missing data) or to truncation (some selection process done on purpose for the study).
 Selection criteria that directly or indirectly
favour brighter or fainter stars can affect the $PL$, $PW$, $PLZ$ or $M_V - {\rm[Fe/H]}$ relations
derived
from the sample.  Specifically, the so-called Malmquist bias \citep{Malmquist1936} caused by  the limitation in apparent
magnitude of the TGAS subset has to be taken into account.
In this Section we attempt to discuss qualitatively the effects of the
different selection filters that result in the samples used in this
work to infer the luminosity calibration. It is well known that biased
samples may result in biased estimates of the (linear) model
parameters.  Our samples are the result of several processing stages
each with a different impact on the resulting sample. As it turns out,
the very limited size of our samples, and the magnitude of the
uncertainties mask all of these effects.

\subsubsection{Trucations/censorships in the generation of the TGAS catalog}\label{sec:truncations}
\begin{figure}
 \begin{center}
  \includegraphics[width=8cm]{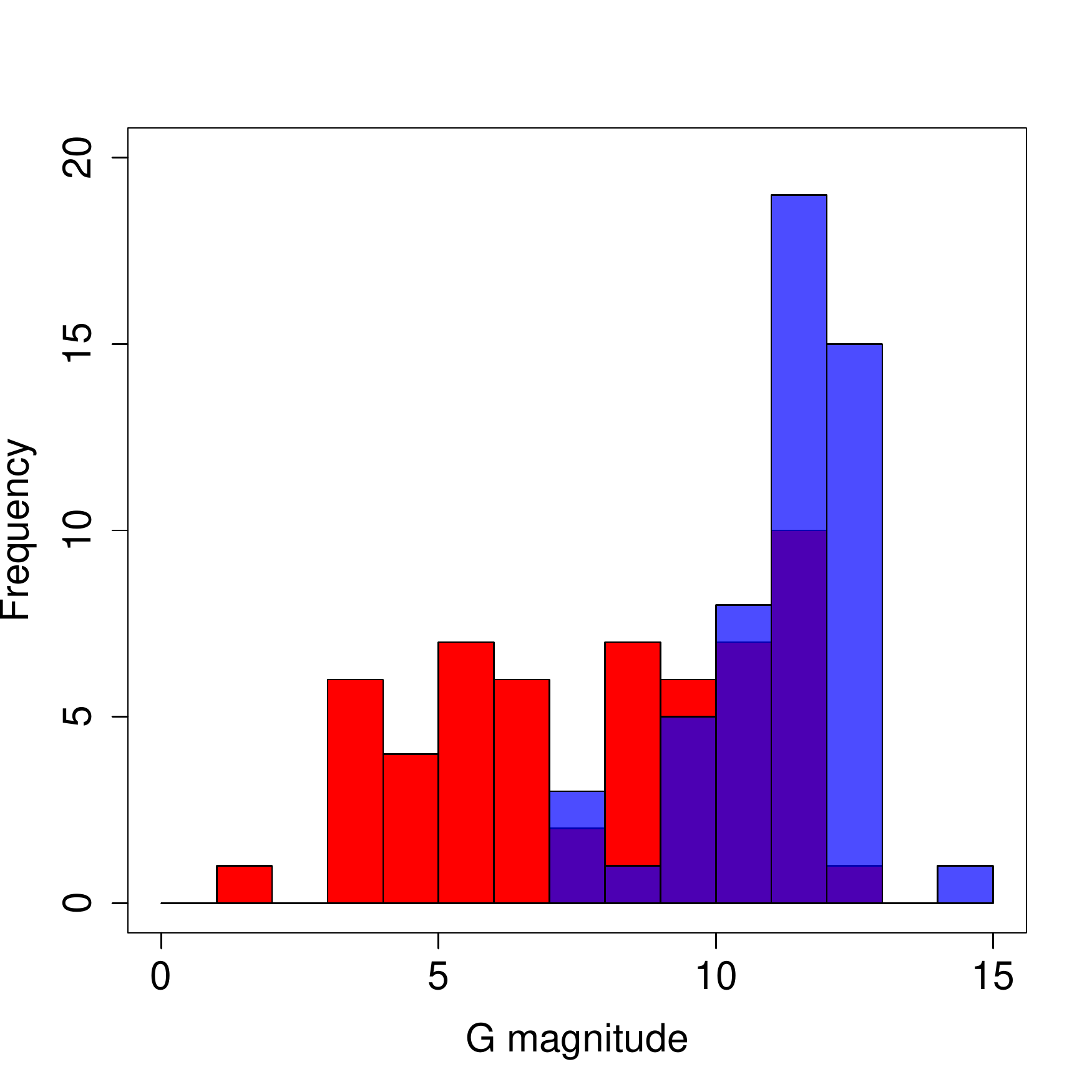}
 \caption{Histogram of the apparent magnitudes of known Tycho-2 RR~Lyrae stars (blue) and classical Cepheids (red) not included in the
   TGAS catalog.}
 \label{TGASMissing}
 \end{center}
\end{figure}

We first discuss the truncation of the samples at the bright
end. Stars brighter than $B_T=2.1$ or $V_T=1.9$~mag were excluded from the
Tycho-2 catalog. In particular, 17588 stars included in the Hipparcos
catalog were not included in Tycho-2. Some of these were actually
included in TGAS, but TGAS is itself affected by removal of many (but not all) sources
brighter than $G\sim$ 7 mag, so the effects of the truncation are subtle and
difficult to assess.  We have checked the existing catalogs to
identify known Tycho-2 classical Cepheids and RR~Lyrae stars missing from
the TGAS catalog. It turns out that 54 known RR~Lyrae stars and 57 classical
Cepheids  in Tycho-2 are missing from the TGAS catalog. Only two Type~II Cepheids are missing in the TGAS catalogue, hence in the following we focus on RR~Lyrae star and classical Cepheids samples and do not discuss Type~II Cepheids any further. The
distributions  in  apparent $G$ magnitude of the missing RR Lyrae stars and Cepheids are shown in Fig.~\ref{TGASMissing}. We see that the truncation  at the bright end of the
Tycho-2 sample does not affect the RR Lyrae sample, while it results in the loss of 
24 classical Cepheids. This 
represents a major loss that can seriously affect the
inferences of the $PL$ and $PW$ relations for classical Cepheids. Fortunately, 21 of these 24 sources
are also in the Hipparcos catalog and we can gauge its impact in the
inference. 
It turns out that the Hipparcos parallaxes and periods are fully
consistent with the distributions from our TGAS sample both in the 2D
plane and in their marginal distributions. We can therefore conclude
that this loss did not bias our sample in any respect.

High proper motion ($\mu>3.5$ arcsec/yr) Tycho-2 stars are also missing
from the TGAS catalog, although this selection has no effect on our
samples, at least to the level that can be checked with Tycho-2 proper
motions (that is, the Tycho-2 proper motions of the known RR~Lyrae and
classical Cepheids missing from TGAS are well below the limit of 3.5
arcsec/yr). So we can discard this trucation as a potential source of
biases.

As discussed in \cite{gaiacol-brown}, the Gaia DR1 has completeness
issues in dense areas of the sky, where the limiting magnitude of the
catalog can be brighter by several magnitudes. This could result in a
potential loss of Cepheids in dense regions of the disk, were it not
for the bright limiting magnitude of the Tycho-2 catalog. RR~Lyrae
stars are not concentrated in the disk (see Fig.~\ref{fig:skydistr}) and therefore
are even less likely to be affected. 

Another truncation/censorship of the Tycho-2 sample comes from the rejection by
the photometric processing of sources with less than 5 transits or
extreme red or blue colour indices. The cut in the number of transits
results in a loss of sources with projected positions near the
ecliptic that should not bias our samples in any respect. Biases will
appear if there is a correlation between the position in the celestial
sphere and the brightness of the sources, such that the sources missed
due to an insufficient number of transits were predominantly bright or
faint (we discard direct correlations with the period). In principle,
the number of transits depends on the scanning law and hence, any
correlation if present should be negligible. 

The selection of sources based on the estimated G$_{BP}$-G$_{RP }$ can
be very simplistically reduced to rejection of sources outside the
range [0.5,3.5] mag \cite[but see][for details]{gaiacol-brown}. This may
result in a loss of stars in the highest extinction regions of the
disk. If periods are not taken into account and only the instability
strip is considered, the brighter stars will be redder and thus, more
prone to be rejected by the photometric reduction pipeline due to
interstellar reddening beyond the 3.5 mag limit. In principle, this should
only affect significantly the sample of Cepheids (because RR~Lyrae stars  are
not concentrated in the disk where most of the extinction occurs). For
these samples of brightest sources and a fixed period however, the
brighter sources are at the blue edge of the instability strip, so at
any given period in the $PL$, $PW$ or $PLZ$ relations, we are more likely to lose the
fainter stars. In summary, we may expect a bias present in the bright
part of the $PL$ relations in the sense of an underrepresentation of the
fainter pulsators. 

\begin{figure}
 \begin{center}
 \includegraphics[width=8cm]{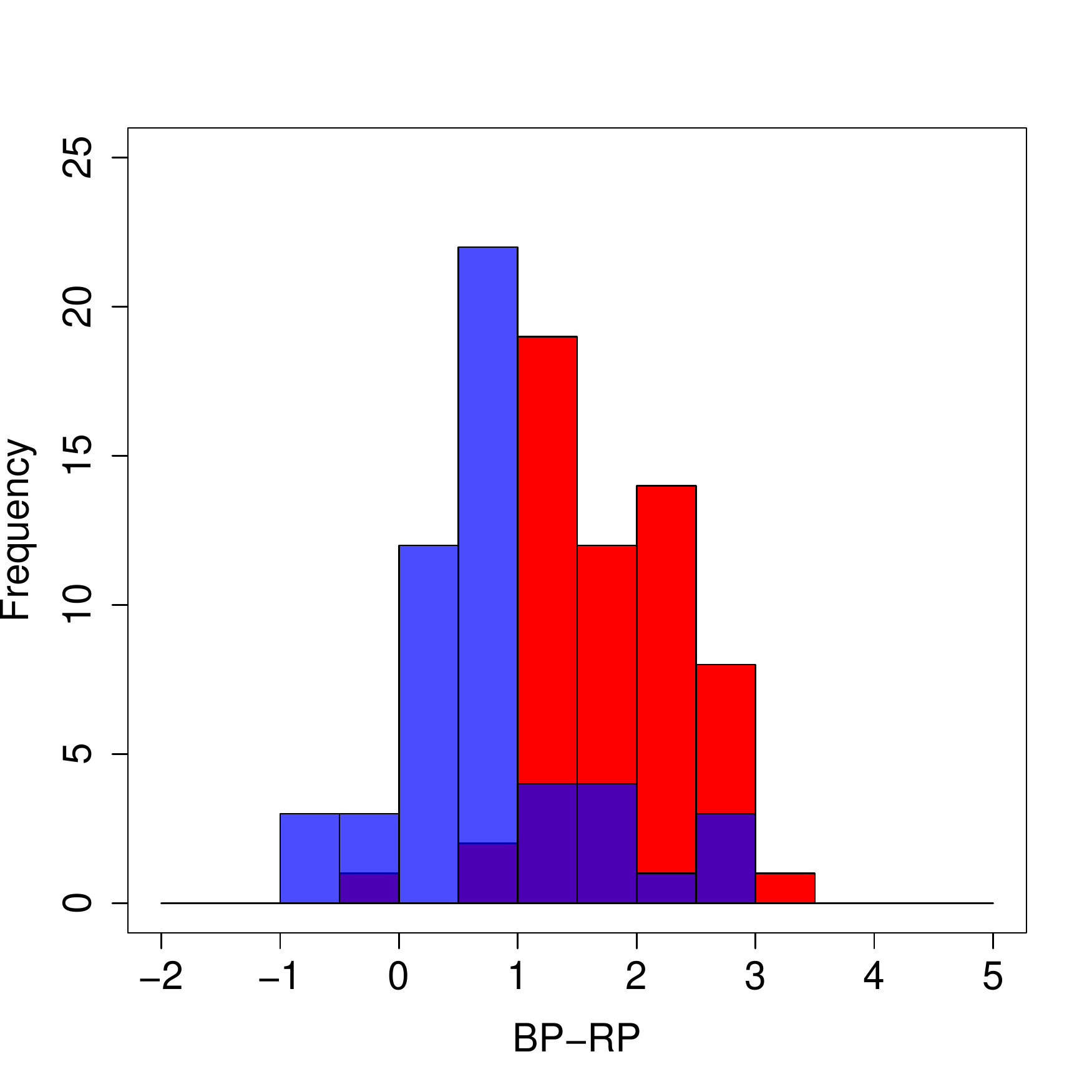}
\caption{Histogram of the estimated G$_{BP}$-G$_{RP}$ colour indices
   of known Tycho-2 RR~Lyrae stars (blue) and classical Cepheids (red)  missing
   from the TGAS catalog.}
\label{bprprrl}
 \end{center}
\end{figure}

We have estimated the $G$, $G_{BP}$ and $G_{RP}$ magnitudes for Tycho-2
sources missing from TGAS using the photometric relations by 
\citet{jordi10}. Figure~\ref{bprprrl}
shows the distribution of estimated $G_{BP}-G_{RP}$ colour indices
for the samples of classical Cepheids (red) and RR~Lyrae stars (blue).
As expected, all values are bluer than the 3.5 mag limit and hence,
reddening is not responsible for the exclusion of this sources from the
catalog. On
the opposite side, however, the TGAS catalog misses at least 16 
RR~Lyrae (but only one Cepheid) type pulsators with colour indices
bluer than 0.5. This should affect more strongly the fainter members
of the instability strip that are, on average, bluer.

Finally, sources with astrometric uncertainties beyond 1 mas/yr in
proper motion, 20 mas in position, or 1 mas in parallax are also
excluded from the TGAS catalog. All these uncertainties depend on the
source colour and apparent magnitude, and the complex TDI-gating
scheme makes a more detailed quantification of the effects very
difficult. In general, the fainter and/or redder the star, the larger
the uncertainty, with the apparent magnitude having a much stronger
effect on the uncertainties \cite[see e.g.][]{Bruijneetal}. We see from
simulations that the net effect is a bias that affects predominantly
the long period end of the $PL$, $PW$ or $PLZ$ distributions in the sense of shifting the
average absolute magnitude for a fixed period towards brighter (more
negative) values. Hence, it has the opposite effect of the Malmquist
bias which tends to flatten the slope of the $PL$ relations. Indeed, we
see from these simulations that in some cases, the two biases cancel
their effects on the slope and only a zero point shift remains as the
net result of the two biases.

\subsubsection{Selections caused by availability of external data}\label{sec:selections}
So far, we have analysed the effects of the Malmquist bias already
present in the Tycho-2 sample, and those emerging from the various
truncations/censorships of the Tycho-2 sample carried out in the generation of the
TGAS catalog. In this work we apply a last filter to retain only
sources in TGAS with $V$, $I$, $J$, $K_{\rm s}$ or $W1$ magnitudes and periods available
(hereafter, external data, see Sections~\ref{sec:dceps}, \ref{sec:t2ceps}, \ref{sec:rrls}). Therefore, the effect of this last filter
remains to be discussed. If the filter left the distribution of TGAS
pulsators unchanged, then no further bias would be introduced in the
analyses. In the following paragraphs we study the differences under
the light of the available data.

We have compared the one-dimensional empirical distributions in
parallax, absolute magnitudes, and periods (with respect
to OGLE distributions of the LMC fundamental mode classical Cepheids and the LMC and Galactic bulge RR Lyrae stars). 
In the following, we will apply the
Anderson-Darling test \citep{anderson1952} which is a frequentist
approach to a problem that would be better approached using Bayesian
methods. Unfortunately, the frequentist approach (with all its
limitations) will have to suffice until we can put forward a
reasonable parametric model of the biases.

For the RR~Lyrae parallaxes, the Anderson-Darling test yields a
p-value of 0.00008, well below any reasonable significance level. The
main difference between the two distributions is a clear excess of
parallaxes greater than 2 mas in the sample of TGAS sources without
external data. When the uncertainties are taken into account using
bootstrapping, the net result is that 42\% (80\%) of the experiments
falls below the common 0.01 (0.05) significance level.

If we remove sources with negative parallaxes, we can apply the
Anderson-Darling test to check if the two samples of absolute
magnitudes could have been drawn from the same distribution. The
result is a p-value equal to $9\cdot 10^{-7}$, with the set of stars
without external data showing a median absolute G magnitude
(1.2$\substack{+0.8\\-0.7}$ with the uncertainties quoting the first
to third quantile range) fainter than the sample used for the $PL$
relations (0.7$\pm{0.5}$). The bootstrapping experiment
performed to include the uncertainties into the analysis yields a 38\%
(68\%) of experiments with values below the 0.01 (0.05) significance
level. 

In both cases (parallaxes and absolute magnitudes), the evidence for a
difference seems inconclusive when the uncertainties are taken into
account, but there are very significant hints that indicate an
underrepresentation of faint sources in the TGAS sample. 
This evidence
is made more robust if we re-estimate the p-values using absolute
magnitudes derived from distance estimates obtained along the lines
suggested by \citet{2015PASP..127..994B}. In this latter case we use the prior for the distance derived from the
Bayesian models of the PL relationships (in the hierarchical model, this
prior is inferred as part of the model as explained in Section~\ref{sec:meth}). The final p-value obtained with these
absolute magnitudes is 0.003, again below the two significance levels
quoted above. 

For the parallaxes of classical Cepheids, the Anderson-Darling test
yields a p-value of 0.01. In this case again, the difference is the
excess of large parallaxes in the sample of sources without external
data with respect to the TGAS sample. The excess is much
smaller than in the RR~Lyrae case. When the uncertainties are taken
into account, the result is that 11\% (45\%) of the experiments fall
below the 0.01 (0.05) significance level.

For absolute magnitudes (and again, removing stars with negative
parallaxes) we obtain a p-value of 0.22 (without bootstrapping the
uncertainties) and 0.02\% (0.002\%) of bootstrap samples below the
0.01 (0.05) significance level. Hence, if the difference is real, we
do not expect it to bias our $PL$ inferences significantly.

Overall, the complexity in the censorships of the samples available 
this paper is not easy to
interpret and makes it difficult to produce a reliable estimation of all
possible biases introduced
by them.  We identified at least three clear sources of biases: (i) the
loss of the bluest RR Lyrae stars, (ii) the absence in  the Gaia DR1 of the sources with largest  
astrometric uncertainties (that is, the exclusion from the TGAS catalog of sources with uncertainties 
larger than 1 mas/yr in proper motion, 20 mas in position 
or 1 mas in parallax), 
and (iii) the selection of sources with external data available. 
Although we have assumed that the local systematic correlations 
described in Section~\ref{comp_par} average out for the two stellar types 
discussed, the lack of a detailed description of these correlations 
prevents us from completely discarding potential biases in the case of 
classical Cepheids which are not distributed uniformly in the celestial 
sphere. The results presented in this paper have then to be considered as
preliminary, and to be
superseded by results from further releases of Gaia data allowing the
use of samples less affected
by uncontrolled censorship effects.

\subsection{Methods}\label{sec:meth}

 An often, commonly used way to obtain the luminosity calibration (e.g.  a $PL$,  $PW$, $PLZ$ or $M_V - {\rm[Fe/H]}$ relation) is the direct transformation 
to distance (and then absolute magnitude) by parallax inversion and then the least squares fit of the derived parameters.
However, in the presence of parallaxes with errors larger than 10\% 
 this method has significant disadvantages that we discuss in the following.

The  absolute magnitude of a star is calculated from its parallax
by using the following relation:

\begin{equation}\label{eq:abs_mag}
   M = m_0 + 5 \log(\varpi) - 10
\end{equation}

\noindent where $M$ is  the absolute magnitude, $m_0$ is the apparent
dereddened
magnitude of the star, and $\varpi$ is the parallax in mas.
Notice that although the errors in the Gaia parallaxes are
well behaved, approximately Gaussian and thus symmetrical,
the errors in these derived magnitudes will not be so. The
logarithm in the expression makes the derived error in $M$
asymmetrical and can thus lead to biases. In particular,
the application of the least squares method to a fit using
these values is generally not advisable, since this method relies
on the errors of the fitted values being Gaussian or at least symmetrical. Furthermore,
negative parallaxes cannot be used in such fitting processes.

In order to include stars with negative parallax measurements and to
avoid non-linear transformations when fitting the  $PL$,  $PW$, $PLZ$ or $M_V - {\rm[Fe/H]}$  relations,
we follow
the prescription of \citet{Arenou1999}, who define an astrometric based
luminosity (ABL) as

\begin{equation}
a = 10^{0.2M} = \varpi 10^{0.2{m_0}-2}
\end{equation}
\par\noindent
where $M$ is the absolute magnitude, $\varpi$ the parallax in mas, and
$m_0$ the
extinction corrected apparent magnitude. Using the ABL instead of the
absolute
magnitude is preferable, as the parallax is used linearly, leading to
symmetrical
error bars as the parallax errors themselves. Additionally, there is no
additional Lutz-Kelker bias due to sample
truncation, as stars
with negative parallax can be included (though biases due to
earlier sample selection will remain present, see Sec.~\ref{sec:bias}).
 For example, assuming that the stars follow a
relation between period and absolute magnitude of the form:

\begin{equation}\label{eq:pl_fit_lsq}
M = \alpha \textrm{logP} + \rho
\end{equation}
where $P$ is the period and $\alpha$ and $\rho$ are the slope and zero
point of the
$PL$ and $PW$ relations, we fit
\begin{equation}\label{eq:pl_fit}
10^{0.2(\alpha \textrm{log}P + \rho)} = \varpi 10^{0.2{m_0}-2}
\end{equation}
using the weighted non-linear least squares.
The same approach could be used to fit the $PLZ$ and  $M_{V}-{\rm
[Fe/H]}$ relations of RR Lyrae stars. We just apply the non-linear least
squares
fit to the equations:
\begin{equation}\label{eq:plz_fit}
10^{0.2(\alpha \textrm{logP} + \beta \textrm{[Fe/H]}+ \rho)} = \varpi
10^{0.2{m_0}-2}
\end{equation}
and
\begin{equation}\label{eq:mz_fit}
10^{0.2(\alpha \textrm{[Fe/H]} + \rho)} = \varpi 10^{0.2{m_0}-2}
\end{equation}
\par\noindent

\begin{figure}
 \begin{center}
 \includegraphics[width=8cm]{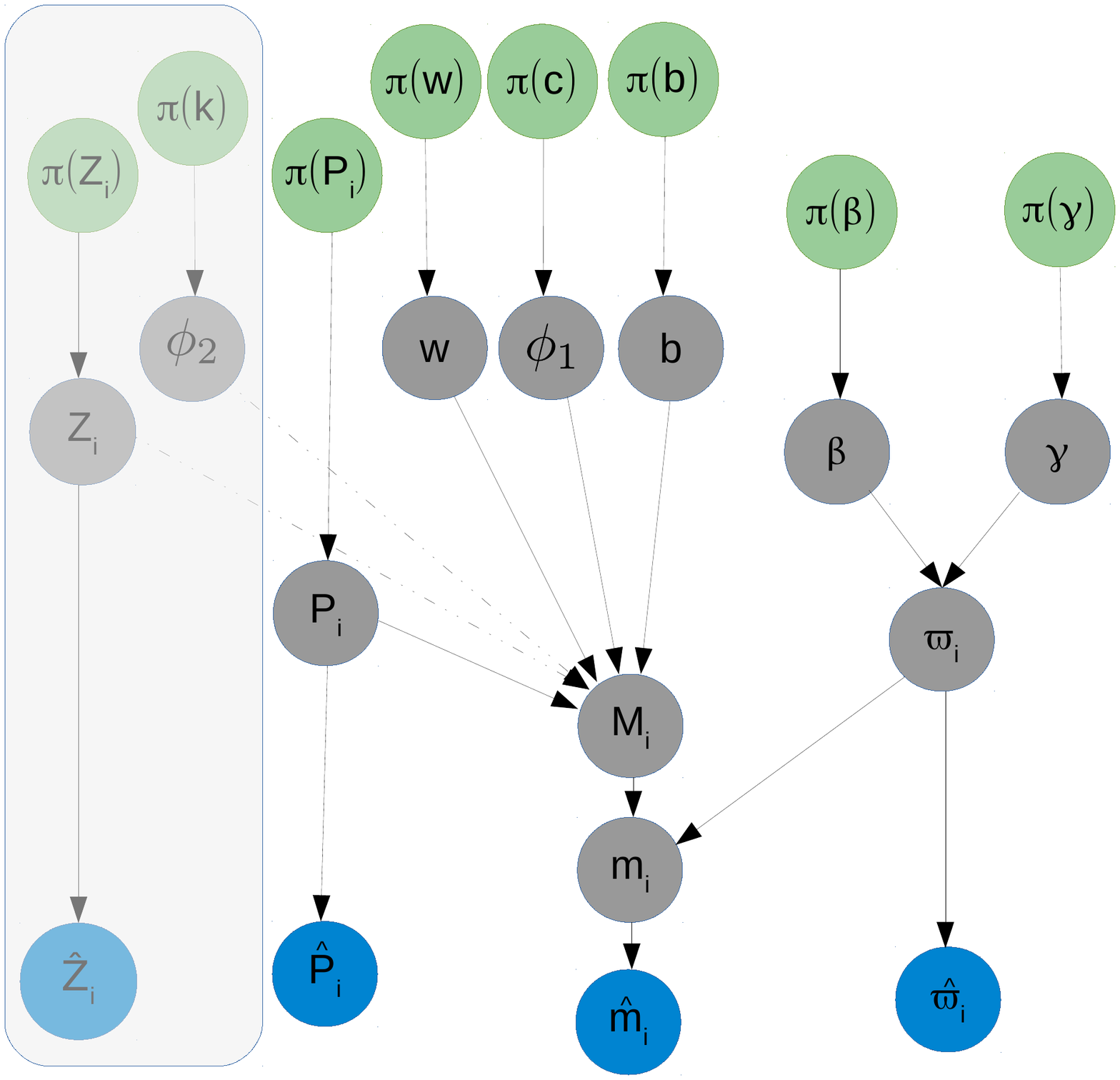}
 \caption{Directed Acyclic Graph that represents the forward model
   used to infer the $PL(Z)$ relation coefficients.}
 \label{DAG}
 \end{center}
\end{figure}

 Finally, the use of a Bayesian approach to fit the $PL$, $PW$, $PLZ$ or $M_{V}-{\rm[Fe/H]}$ relations followed by the variable stars, provides
an excellent alternative option.
The Bayesian estimate of the $PL(Z)$ relationships is accomplished by
means of a hierarchical model \cite[that will be described elsewhere,
but see][for a similar approach]{Sesar2016} encoded in the
directed acyclic graph shown in Fig.~\ref{DAG}. It shows the
measurements at the bottom level: periods ($\hat{P}_i$), apparent
magnitudes ($\hat{m}_i$), metallicities (when appropriate,
$\hat{Z}_i$) and parallaxes ($\hat{\varpi}_i$). The subindex $i$ runs
from 1 to the total number of stars $N$ in each sample. Our model
assumes that the measurements (denoted by
$\{\hat{d}_i=(\hat{m}_i, \hat{P}_i,
\hat{\varpi}_i,\hat{Z}_i); i:1,2,...,N\}$) are realizations from
normal distributions centred at the true (unknown) values (denoted as
$P_i, Z_i, m_i$, and $\varpi_i$) and with standard deviations given by
the measurement uncertainties. We express the most general $PLZ$
relation as

\begin{equation}
M = b + c\cdot\log(P) + k\cdot Z + \epsilon 
\end{equation}

where $\epsilon$ is a Gaussian distributed random variable

\begin{equation}
\epsilon \sim \mathcal{N}(0,w)
\end{equation}

that represents the intrinsic dispersion of the relation (not due to
measurement uncertainties).

The parallaxes are assumed to be drawn from a log-normal prior of
parameters $\beta$ and $\gamma$. The model is hierarchical in the
sense that this prior is inferred as part of the model, and $\beta$
and $\gamma$ are themselves prescribed by their own priors (see
Table \ref{tab:priors} for a full list of priors).

\begin{table*}
\begin{center}
\begin{tabular}{c}
  \hline

  \hline 
  
   $p(m_i|\varpi_i,\phi_1,b,\phi_2,w,P_i,Z_i)\sim\mathcal{N}(m_i|\tan(\phi_1)\cdot\log_{10}(P_i)+\tan(\phi_2)\cdot Z_i+b-5\cdot\log_{10}(\varpi_i)-10,w)$\\
   $p(\varpi_i|\beta,\gamma) \sim  \ln\mathcal{N}(\varpi_i|\beta,\gamma)$\\
   $\pi(\gamma)\sim \exp(1)$\\
   $p(\beta)\sim \mathcal{N}(0,2)$\\
   $\pi(\phi_1)\sim \mathcal{N}(0,3.14/2)$\\
   $\pi(\phi_2)\sim \mathcal{N}(0,3.14/2)$\\
   $\pi(b)\sim \mathcal{N}(0,10)$\\
   $\pi(w)\sim \exp(10)$\\
   $\pi(Z_i) \sim \mathcal{N}(Z_{i}|0,5)$\\
   $\pi(P_i) \sim \mathcal{N}(P_{i}|0,3)$\\
\hline
\end{tabular}

\caption {Prior ($\pi$) definitions for the hierarchical Bayesian model of the $PL(Z)$ relations. The symbol $\sim$ has to be read as "{\it is distributed according to}". The $\exp$ expression has to be interpreted not as the exponential analytical function but as the exponential probability density distribution. We use the $\pi$ symbol to refer to the prior probability and use 3.14 to refer to the half-length of the circle.}

\label{tab:priors}
\end{center}
\end{table*}

We have tried several prior definitions and the results are
insensitive to the prior choice except for unreasonable setups.

In the following analysis we apply all three methods: (i) the direct transformation
of the parallaxes to absolute magnitudes (Eq.~\ref{eq:abs_mag}) and the weighted 
linear least squares fitting (LSQ) in the period-absolute magnitude plane ($PL$, $PW$, $PLZ$ relations) and the 
absolute magnitude-metallicity plane ($M_{V}-{\rm [Fe/H]}$ relation); (ii) the use of the ABL to perform the non-linear weighted
least
squares fit of Eqs.~\ref{eq:pl_fit}-\ref{eq:mz_fit}; (iii) the realization of the Bayesian fitting approach. The Bayesian  solution corresponds to the maximum a posteriori (MAP) values, 
and the r.m.s. represents the dispersion of the unweighted residuals 
with respect to this MAP solution. 
The ABL and Bayesian methods are
preferred to avoid biases and we include the LSQ fitting  for comparison
purposes and to also allow comparison with published results using this method.

\section{Classical Cepheids}\label{sec:dceps}

\begin{figure}[t!]
\centering
\includegraphics[trim=30 150 0 100 clip, width=1.05\linewidth]{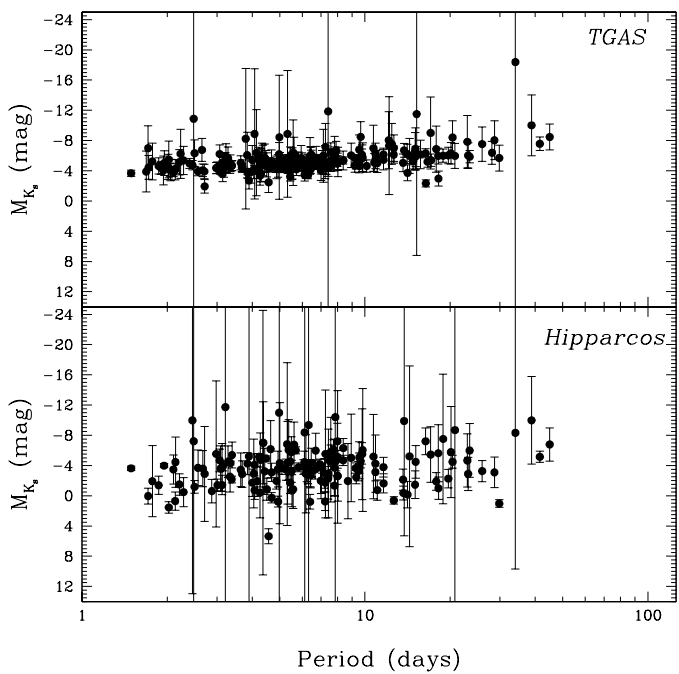}
\caption[]{Classical Cepheid $PM_{K_\mathrm{s}}$ relations using TGAS (upper panel) and  Hipparcos (lower panel)  parallaxes, respectively.
 The upper panel shows the absolute magnitudes, $M_{K_\mathrm{s}}$, of 221 classical Cepheids with 
positive parallaxes from TGAS. The lower panel shows  156 classical Cepheids with positive Hipparcos parallaxes. 
}
\label{fig:PL_Hip_Gaia}
\end{figure}

TGAS parallaxes are available in {\it Gaia} DR1 for 331 Galactic classical Cepheids. We have collected from the literature $V$, $I$ and 2MASS $K_\mathrm{s}$ 
 photometry,  $E(B-V)$ reddening values, period and classification for the whole sample, making an effort to assemble an as much as possible uniform and homogeneous catalogue.  
For the mean magnitudes whenever available we collected intensity-averaged mean $V,I,K_\mathrm{s}$ magnitudes based on a complete sampling of the light curves, in several cases we 
computed ourselves the intensity-averaged mean values from the light curves published in the literature.
Period and $V,I,K_\mathrm{s}$ photometry values have been taken from different sources, among which primarily \citet{groe1999}, \citet{berdni00}, \citet[DDO Database of Galactic Classical Cepheids]{ddo}, \citet{ngeow2012}, the GCVS (\citealt{gcvs})  and the ASAS3 catalogue (\citealt{Pojmanski2002}). $E(B-V)$ reddening values and related errors were taken from \citet{ddo}, \citet{groe1999}, \citet{Turner2001}, \citet{fou2007} and \citet{Pejcha2012}. In a few cases we specifically estimated the reddening for this study from the available photometry. Metal abundances  were mainly taken from \citet{gen14}  and for a few stars from  \citet{ngeow2012}. Information about duplicity is from \citet{Klag2009} and \citet{anderson16}.

We found in the literature period values for 312 classical Cepheids (94\%) of our sample, $E(B-V)$ reddening values for 276 stars (83\%), photometry in the $V$, $I$ and $K_\mathrm{s}$ bands for 297 (90\%), 250 (76\%) and 292 (88\%) classical Cepheids, respectively.  We provide the complete dataset of the 331 Galactic classical Cepheids in Table~\ref{tab:cephall}. Hipparcos parallaxes are 
 available for 248 of them. 
  Among the 248 classical Cepheids with parallax measured by both Hipparcos and {\it Gaia} we selected 228 with $K_\mathrm{s}$ magnitude, reddening and period available in the literature. To correct the $K_\mathrm{s}$ magnitudes for extinction we used the extinction law of  \citet{Cardelli1989}  and adopted for the ratio of total
to selective extinction the value $\rm R_V=3.1$, thus deriving $K_\mathrm{s,0}=K_\mathrm{s}-0.35E(B-V)$. We then transformed the TGAS and Hipparcos parallaxes to absolute $M_{K_\mathrm{s}}$ magnitudes  applying Eq.~\ref{eq:abs_mag}. This transformation was  possible only  for stars with positive parallax values, namely, 221 out of 228 stars with TGAS parallax and only 156 out of 228 for Hipparcos.  The corresponding 
$PM_{K_\mathrm{s}}$ relations are shown in the upper and lower panels of Fig.~\ref{fig:PL_Hip_Gaia} for the TGAS and Hipparcos parallaxes, respectively. 
The improvement  
produced  by the TGAS parallaxes is impressive. The scatter is very much reduced, the sample is about 30\%  larger and, although error bars are still very large, a clear $PM_{K_\mathrm{s}}$ relation can now be seen in the data. 
Since 
 other factors (e.g. multiplicity, see Sec.~\ref{sec:binary}), besides errors in the parallax measurements, may contribute to the large dispersion 
 seen in the upper panel of Fig.~\ref{fig:PL_Hip_Gaia}, 
  we cleaned the sample from  binary systems and retained for analysis only single, fundamental mode, classical Cepheids.

\begin{figure}[t!]
\centering
\includegraphics[trim=30 200 0 100 clip, width=1.05\linewidth]{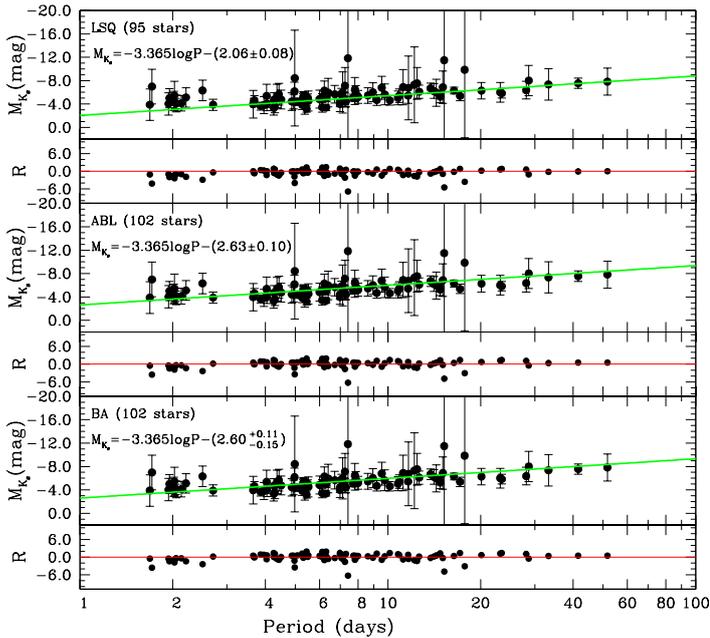}
\caption[]{ Classical Cepheid $PL$ relations in the $K_\mathrm{s}$-band  
obtained i) by linear least squares fitting  the  stars' absolute magnitudes inferred from direct transformation of the TGAS parallaxes (upper panel); ii) via non-linear least squares fit  and the ABL method (middle panel); and iii) using a Bayesian approach (bottom panel). The slope of the fit is adopted from \citet{fou2007}. The bottom part of each panel shows the residuals from the best fit line.}
\label{fig:PL_CC_K}
\end{figure}

\subsection{Binarity-mutiplicity among Cepheids}\label{sec:binary}
Classical Cepheids are Population~I stars, therefore occurrence
of binaries among them is a common phenomenon. However, it is not
easy to detect the presence of a companion because it is the supergiant
Cepheid that dominates the system. For brightest (naked-eye) Cepheids, the frequency of binaries (including 
systems consisting of more than two stars) exceeds 50\%, and
there is a selection effect towards fainter visual magnitudes that
results in a deficit in known binaries (\citealt{szabados03}).
For classical Cepheids in the MW, there exists an on-line data base developed and mantained at the Konkoly Observatory\footnote{http://www.konkoly.hu/CEP/intro.html}
 in which
one can find the actual list of known binaries involving a Cepheid
component. A few further binary Cepheids are known from \citet{anderson16}.
When performing astrometric reduction of the {\it Gaia} data, these stars
have to be treated as binary systems because the orbital motion (if 
not taken into account) falsifies the resulting trigonometric 
parallax (\citealt{szabados97}). Moreover, the brightness of the Cepheid
($G$, $G_{BP}$, and $G_{RP}$ magnitudes) has to be corrected for the 
photometric contribution of the companion star. If the available photometric, spectroscopic, and astrometric data
are insufficient to characterize the binary system or get reliable
apparent and absolute magnitudes of the Cepheid component, that
particular Cepheid cannot be used as a calibrator in 
determining the zero point of the $PL$ relationship. 
 Combining information from the on-line data base of the Konkoly Observatory, \citet{Klag2009} and  \citet{anderson16} we find that 198 (60\%) of the 331 
classical Cepheids in our sample are binaries.
They were not used to compute the $PL$ relations. Once known binaries are discarded our sample reduces to 133 non-binary classical Cepheids. 
The selection for binarity  we have operated may sound very strict, however, the reduced scatter of the comparison with photometric parallaxes 
obtained with the Baade-Wesselink technique (see Fig.~\ref{fig:BW_CC}) when binary Cepheids are discarded indicates that ours is a safe  approach.

\subsection{Derivation of the classical Cepheid $PL$ and $PW$ relations}\label{sec:plccs}

 We found information on the period, $V$, $I_c$, $K$ photometry and reddening only for 125 of the 
133 non-binary classical Cepheids. Furthermore, three of them, namely, BB~Her, EV~Aql and V733~Aql are classified as Type II Cepheids in the McMaster Cepheid Photometry and Radial Velocity Data Archive\footnote{http://crocus.physics.mcmaster.ca/Cepheid/}. We discarded them from the sample of classical Cepheids but did not include them in the list of Type~II Cepheids since their  classification is uncertain. After this additional cleaning we are left with a  sample of 122 bona-fide classical Cepheids.

Fundamental mode (F) and first-overtone (FO) classical Cepheids  follow different $PL$ relations. 
 Information on the pulsation mode of the 122 classical Cepheids in the clean sample is available from the studies by  \citet{Udalski1999}, \citet{szi07}, \citet{Klag2009}, \citet{Molnar2014} and  \citet{Inno15},  by which it turns out that our sample contains 102 fundamental mode classical Cepheids. In order to remove any scatter caused by mixing multiple  $PL$ relations and given the small number  of FO pulsators (20),  in the following analysis  we have considered only  the F classical Cepheids. 
 The error distribution of the TGAS parallaxes for the 102 F classical Cepheids is shown by the magenta histogram in Fig.~\ref{fig:CCcumul}. 
Relevant information ({\it Gaia}/Hipparcos/Tycho-2 Ids, $G,\langle V \rangle, \langle I \rangle,\langle K_\mathrm{s} \rangle$ magnitudes, period, reddening, metallicity, parallax and parallax errors) for them is summarised in the first 102 rows of
Table~\ref{tab:cephall}. Typical errors of the $V,I,K_\mathrm{s}$ apparent mean magnitudes are estimated to be of about 0.02 mag.
 
We note that  $G$-band magnitudes are available for all the classical Cepheids in our sample, however,  we decide to not compute 
 $G$-band $PL$s  
 primarily because {\it Gaia}  $G$-band has a too large throughput (330 - 1050~nm) encompassing roughly from  
the $U$ to the $Y$ spectral ranges.  The intrinsic width of the classical Cepheids instability strip varies significantly going  from $U$ to  $K$ passbands and  
the dispersion of the Cepheids $PL$ varies accordingly, being tightest in the NIR (Sec.~\ref{sec:intro} and, e.g.  fig.~4 of \citealt{madore91}) in this making the $K$-band the best  choice for testing  
TGAS parallaxes with the classical Cepheid $PL$ relations. Furthermore, the  $G$-band magnitudes available for TGAS classical Cepheids in
{\it Gaia} DR1 are  the straight average of, often,  only a few measurements unevenly sampling the cyclic light variation of these  stars, hence further 
enhancing  the  scatter of the $G$-band $PL$. The situation will improve significantly with {\it Gaia} DR2 because both better sampled $G$-band light curves will be released 
for the TGAS Cepheids and, more importantly, because  $G_{RP}$ magnitudes (640 - 1050~nm) will become available, allowing 
the {\it Gaia} $G_{RP}$  $PL$ relation to be built  for classical Cepheids.

 We computed $PL$ relations in the $V$, $I$ and ${K_\mathrm{s}}$ bands (hereinafter, $PM_{V}$, $PM_{I}$, $PM_{K_\mathrm{s}}$ and $(V,V-I)$, $(K_\mathrm{s},V-K_\mathrm{s})$ $PW$ relations [$PW(V,V-I)$, $PW(K_\mathrm{s},V-K_\mathrm{s})$]  relations for the 102 F classical Cepheids in our sample. To correct for extinction the $V$, $I$ and $K_\mathrm{s}$ magnitudes we adopted the extinction relations from \citet{Cardelli1989} and $R_V=3.1$, thus obtaining: $V_0=V-3.1E(B-V)$,  $I_0=I-1.48E(B-V)$ and $K_\mathrm{s,0}=K_\mathrm{s}-0.35E(B-V)$.
The Wesenheit magnitudes used in this paper are defined as follows: $W(V,I)=V-2.55(V-I)$ \citep{fou2007} and $W(V,K_\mathrm{s})$=$K_\mathrm{s}-0.13\,(V-K_\mathrm{s})$ \citep{ripe12}. As pointed out in Sec.~\ref{sec:intro}, the  $PW$ relations have several advantages with respect to normal $PL$ relations, because the effect of errors on the reddening estimates is in principle removed, in practice greatly mitigated, and because the colour term in the $W$ magnitude definition takes into account and partially corrects for the finite colour width of the instability strip, thus reducing the associated uncertainty on the distance determinations. It is also worth noticing that in the case of  the $W(V,K_\mathrm{s})$, the $PW$ is equivalent to the $PLC$ relation in the same filters (see, e.g. \citealt{ripe12}). Since  
errors of the TGAS parallax are large 
 we did not attempt to derive both slope and zero point but  
 fixed the slope of the $PL$ and $PW$ relations and 
used the TGAS parallaxes just to estimate the zero points. We adopted the slopes from 
\citet{fou2007} for the $PM_{V}$, $PM_{I}$, $PM_{K_\mathrm{s}}$ and $PW(V,V-I)$ relations and from \citet{ripe12} for the $PW(K_\mathrm{s},V-K_\mathrm{s})$ relation. 

The $PL$ and $PW$ relations obtained applying the three different approaches described in Sec.~\ref{sec:meth} to the various passbands considered in this paper ($V$, $I$ and ${K_\mathrm{s}}$) are  
 reported in Table~\ref{tab:CCs_relations} and shown in Fig.~\ref{fig:PL_CC_K} for the case of the $PK_\mathrm{s}$ relations. 
Specifically, the upper panel of Fig.~\ref{fig:PL_CC_K} shows the weighted least squares fit of the absolute $M_{K_\mathrm{s}}$ magnitudes obtained by direct transformation of the parallaxes (Eq.~\ref{eq:abs_mag}, hereinafter referred to as LSQ). The fit was possible only for 95 classical Cepheids in the sample for which TGAS parallaxes have positive values. 
The $PM_{V}$, $ PM_{I}$, $PM_{K_\mathrm{s}}$, $PW(V,V-I)$ and $PW(K_\mathrm{s},V-K_\mathrm{s})$ relations  
were then computed using the ABL method  and a weighted non-linear least squares fit in the form of Eq.~\ref{eq:pl_fit}.  Using the ABL  approach the whole sample of 102 classical Cepheids, without discarding stars with negative parallaxes, could be used. The $PM_{K_\mathrm{s}}$ relation derived with the ABL method is shown in the middle panel of Fig.~\ref{fig:PL_CC_K}.
Finally, we used our Bayesian approach to fit the $PL$ and $PW$ relations of the whole sample of 102 classical Cepheids, the resulting $PM_{K_\mathrm{s}}$ relation is 
presented in the bottom panel of Fig.~\ref{fig:PL_CC_K}.  
Comparison of the results in Table~\ref{tab:CCs_relations} shows  that the ABL and Bayesian approaches are generally in good agreement to each other and provide brighter absolute magnitudes (hence longer distances) than  the direct transformation of parallaxes and  the LSQ fit which absolute magnitudes 
are about $\sim$ 0.5 mag systematically fainter. 
However, we also note that the r.m.s. scatter of all relations is very large, due to the large parallax uncertainties.  The r.m.s.  of the Bayesian solution  is particularly large  as it represents the unweighted residuals with respect to the maximum a posteriori solution. 

\begin{table*}
\caption[]{$PL$ and $PW$ relations for classical Cepheids with zero point based on TGAS parallaxes.\label{tab:CCs_relations}}
\begin{center}
\begin{tabular}{l c c}
\hline
\hline
\noalign{\smallskip}
\tiny
 & Relation & RMS \\
 & & (mag)  \\
 \noalign{\smallskip}
 \hline
\noalign{\smallskip}
$PM_{V}$ 95 objects (LSQ) & $ -2.678\log P-(1.00\pm0.08$) & 0.74 \\
$PM_{V}$ 102 objects (ABL)  &  $-2.678\log P-(1.54\pm0.10)$ & 0.85  \\
$PM_{V}$ 102 objects (BA)  & $ -2.678 \log P - (1.49\substack{+0.12\\-0.11} ) $& 1.31 \\
\noalign{\smallskip}
\hline 
\noalign{\smallskip}
$PM_{I}$ 95 objects (LSQ) & $-2.98\log P-(1.28\pm0.08)$ & 0.78 \\
$PM_{I}$ 102 objects  (ABL) & $-2.98\log P-(1.84\pm0.10)$ & 0.87  \\
$PM_{I}$ 102 stars (BA) & $-2.98 \log P - (1.80\substack{+0.13\\-0.12})$ & 1.33  \\
\noalign{\smallskip}
\hline
\noalign{\smallskip}
$PM_{\rm K_\mathrm{s}}$ 95 objects (LSQ) & $-3.365\log P-(2.06\pm0.08)$ & 0.74  \\
$PM_{\rm K_\mathrm{s}}$ 102 objects  (ABL) & $-3.365\log P-(2.63\pm0.10)$ & 0.88 \\
$PM_{K_S}$ 102 stars (BA) & $ -3.365 \log P - (2.60\substack{+0.11\\-0.15} ) $& 1.33  \\ 
\noalign{\smallskip}
\hline 
\noalign{\smallskip}
$PW(V,V-I)$ 95 objects (LSQ) & $-3.477\log P-(2.21\pm0.08)$ & 0.77  \\
$PW(V,V-I)$ 102 objects (ABL) & $-3.477\log P-(2.82\pm0.11)$ & 0.90   \\
$PW(V,V-I)$ 102 stars (BA) & $-3.477 \log P-(2.63\pm 0.13)$ &  1.36 \\
\noalign{\smallskip}
\hline 
\noalign{\smallskip}
$PW(K_\mathrm{s},V-K_\mathrm{s})$ 95 objects  (LSQ) & $-3.32\log P-(2.32\pm0.08)$ & 0.73    \\
$PW(K_\mathrm{s},V-K_\mathrm{s})$ 102 objects  (ABL) &  $-3.32\log P-(2.87\pm0.10)$ & 0.87  \\
$PW(K_\mathrm{s},V-K_\mathrm{s})$ 102 stars (BA) & $ -3.32 \log P  -(2.81\substack{+0.14\\-0.12}) $& 1.33  \\
\noalign{\smallskip}
\hline 
\noalign{\smallskip}
\end{tabular}
\end{center}
\normalsize
\end{table*}

\section{Type II Cepheids}\label{sec:t2ceps}

Type II Cepheids are pulsating variables that belong to the Population
II star family. They have been studied by several authors
(\citealt{Wallerstein1984}, \citealt{Gingold1985}, \citealt{har85},
\citealt{Bono1997}, \citealt{Wallerstein2002}, \citealt{Feast2008},
\citealt{Matsunaga2011}, \citealt{Sos2008},
\citealt{Ripepi2015}) and are usually divided in three classes: BL
Herculis (BL Her) with periods between 1 and 4~days, W Virginis (W Vir)
with period from 4 to 20~days and RV Tauri (RV Tau) with periods from
20 to 150~days. The light curves of BL Her and W Vir stars can be almost
sinusoidal or highly non-sinusoidal, those of  RV
Tauri have alternating minima.  It is believed that BL Her variables
are low-mass stars (0.5-0.6 $M_{\odot}$) that start the
central He burning on the blue side of the RR Lyrae star gap and crossing the instability strip at  
about 0.5-1.5~mag brighter than the RR Lyrae stars (hence the longer
periods). Hydrodynamic models indicate that they pulsate in the
fundamental radial mode \citep[see e.g.][and references therein]{Marconi2007}.
The W Vir variables, instead, cross the instability strip during their blue-loop excursions
(``blue-nose'') from the asymptotic giant branch (AGB) during helium-shell flashes. They are
$\sim$2-4~mag brighter than the RR Lyrae stars. As the BL Her, also the W Vir stars are
thought to pulsate in the radial fundamental mode (see, e.g. \citealt{lem15}, and references therein).

The RV Tau  are post-AGB stars  that cross the 
instability strip at high luminosity during their path to the white
dwarf cooling sequence. It is still unclear whether it is appropriate to include them in the 
same class as the BL Her and W Vir stars, since they
have different evolutionary histories  (\citealt{Wallerstein2002}).
In addition to the three above types, \citet{Sos2008} suggested the
existence of the so-called peculiar W Vir (pW Vir) stars. They 
exhibit peculiar light curves and, at constant period, are
generally brighter than normal Type~II Cepheids. Although their 
true nature remains uncertain, it is likely that the pW Vir are
part of binary systems.

It has been known since  a long time that the Type~II Cepheids follow a $PL$ relation 
\citep{Nemec1994} in the optical. \citet{Kubiak2003} later 
found that all  Type~II Cepheids with periods in the range  of 0.7 to about 10~days in the OGLE~II \citep{Udalski1992} sample follow the same $PL$
relation as then also confirmed by \citet{Pritzl2003} and
\citet{Matsunaga2006} for Type~II Cepheids in Galactic globular clusters, by
\citet{Groenewegen2008} for Type~II Cepheids  in the Galactic bulge,  and by \citet{Sos2008}
for the LMC sample, on the basis of OGLE~III data.
\citet{Matsunaga2011} using single-epoch
data showed that the Type~II Cepheids  follow 
tight $PL$ relations in the $J,H,K_\mathrm{s}$ passbands. 
This result has been  confirmed by \citet{Ripepi2015} based on 
multi-epoch $J,K_{\mathrm{s}}$ photometry for 130 Type~II Cepheids observed  in the LMC  by 
the {\it VISTA survey of the Magellanic Clouds system,
  VMC} \citep{Cioni2011}.  These authors found that  unique $PL$ and 
$PW$ relations holds both for BL Her and W Vir variables, whereas  RV Tau stars
follow different and more dispersed relationships.  
In light of this and the different evolutionary history we have not 
considered the RV Tau stars in our analysis.
\citet{Ripepi2015} also found that the metallicity dependence of the 
Type~II Cepheid $PL$ and $PW$
relations is small, if any. Therefore, the metallicity was neglected in our analysis. 

\begin{figure}[t!]
\centering
\includegraphics[trim=30 200 0 100 clip, width=1.05\linewidth]{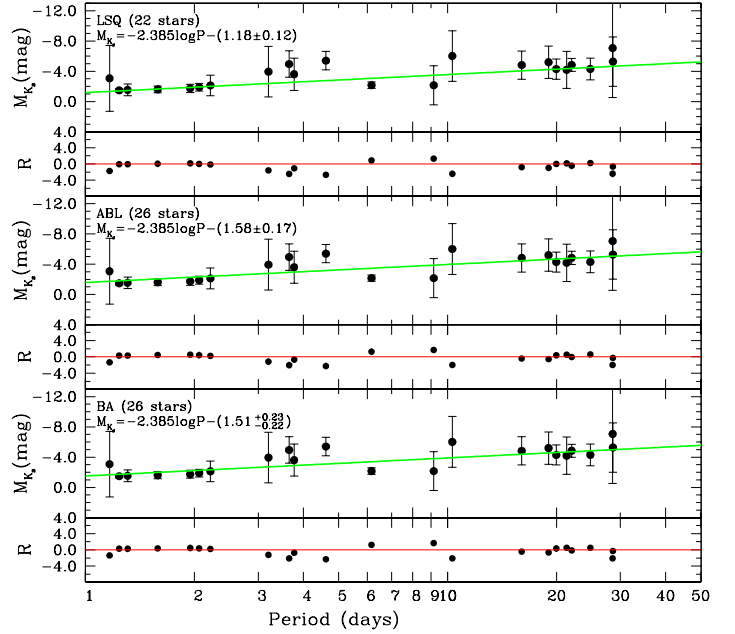}
\caption[]{$PL$ relation in the $K_\mathrm{s}$ band of the Type~II~Cepheids obtained i) by linear least squares fit of the stars' absolute magnitudes inferred by direct transformation of the TGAS parallaxes (upper panel); ii) via non-linear least squares fit  and the ABL method (middle panel), and iii) using the Bayesian approach (bottom panel).  The slope of the fit is adopted from \citet{Ripepi2015}. The bottom part of each panel shows the residuals from the best fit line.}
\label{fig:PL_T2CEP_K}
\end{figure}

The sample of 31 Type~II Cepheids that have TGAS parallax  (see
Section~\ref{sec:samples}) contains 12 BL Her, 14 W Vir and 5 RV Tau
stars. Excluding the RV Tau variables we are left with a sample of 26 stars spanning  the period range from 1.16 to 30.0 d. 
 The error distribution of the TGAS parallaxes for these 26 Type~II~Cepheids is shown by the grey histogram in Fig.~\ref{fig:T2Ccumul}.
$J$ and $K_\mathrm{s}$ photometry for them was taken from
the 2MASS catalogue \citep{Cutri2003}, 
pulsation periods from the GCVS
\citep{gcvs},  $E(B-V)$ reddening values and related errors were taken from the NASA/IPAC Infrared Science Archive\footnote{http://irsa.ipac.caltech.edu/applications/DUST/}, which is based on the reddening maps of \citet{schl2011}. 
To de-redden the $K_\mathrm{s}$ and $J$ magnitudes we applied the extinction laws $A_J=0.87E(B-V)$ and $A_K=0.35E(B-V)$ from \citet{Cardelli1989}, adopting $R_V=3.1$.
We also calculated the Wesenheit magnitude
$W(K_\mathrm{s},J)=K_\mathrm{s}-0.69(J-K_\mathrm{s})$ and fitted the
$PM_{K_\mathrm{s}}$, $PM_J$ and $PW(K_\mathrm{s}, J-K_\mathrm{s})$ relations
adopting the slopes derived by \citet{Ripepi2015} for BL Her and W
Wir stars in the LMC. 
 For the Type~II~ Cepheids we only have single-phase 2MASS $J$ and $K_\mathrm{s}$ magnitudes instead of magnitudes averaged over the whole pulsation cycle, which introduced  additional uncertainty. The typical amplitude of the Type~II Cepheids in these bands is $\sim0.3$ mag, hence, we 
adopted mean  errors of the apparent $J$ and $K_\mathrm{s}$ magnitudes of 0.15~mag.
The sample of 26 Type~II Cepheids contains four stars with
negative parallaxes. 
 Hence, we performed the linear least squares fitting of the 
absolute magnitudes only for the 22 Type~II Cepheids with a positive parallax, 
while the ABL  and Bayesian methods were applied to 
the whole sample of 26 Type~II Cepheids. As with the classical Cepheids, we show graphically in Fig.~\ref{fig:PL_T2CEP_K} only the  $PM_{K_\mathrm{s}}$ relation and 
summarise in   
Table~\ref{tab:T2CEP_relations} the relations obtained in all various bands considered in the paper.  Similarly to what we found for classical Cepheids, the r.m.s. of all relations is fairly large, 
 the ABL and Bayesian approaches are generally in good agreement to each other and, on average, about 0.4 mag brighter than found with the LSQ fit.

\begin{table*}
\caption[]{$PL$ and $PW$ relations for Type~II Cepheids based on TGAS parallaxes.\label{tab:T2CEP_relations}}
\begin{center}
\begin{tabular}{l c c l}
\hline
\hline
\noalign{\smallskip}
\tiny
 & Relation & RMS \\
 & & (mag) & \\
 \noalign{\smallskip}
 \hline
\noalign{\smallskip}
$PM_{J}$ 22 objects (LSQ) & $ -2.19\log P-(0.97\pm0.13$) & 0.88 \\
$PM_{J}$ 26 objects (ABL)  &  $-2.19\log P-(1.50\pm0.20)$ & 1.25  \\
$PM_{J}$ 26 objects (BA) & $ -2.19 \log P - (1.36 \substack{+0.26\\-0.25} ) $& 1.15  \\

\noalign{\smallskip}
\hline 
\noalign{\smallskip}
$PM_{K_\mathrm{s}}$ 22 objects (LSQ) & $-2.385\log P-(1.18\pm0.12)$ & 0.81  \\
$PM_{K_\mathrm{s}}$ 26 objects  (ABL) & $-2.385\log P-(1.58\pm0.17)$ & 1.10  \\
$PM_{K_\mathrm{s}}$ 26 objects (BA) & $-2.385 \log P - (1.51\substack{+0.23\\-0.22} ) $& 1.14  \\

\noalign{\smallskip}
\hline
\noalign{\smallskip}
$PW(K_\mathrm{s},J-K_\mathrm{s})$ 22 objects  (LSQ) & $-2.52\log P-(1.34\pm0.10)$ & 0.80  \\
$PW(K_\mathrm{s},J-K_\mathrm{s})$ 26 objects  (ABL) &  $-2.52\log P-(1.59\pm0.13)$ & 1.04  \\
$PW(K_\mathrm{s},J-K_\mathrm{s})$ 26 objects  (BA) & $ -2.52 \log P - (1.66\substack{+0.21\\-0.22} ) $& 1.13  \\

\noalign{\smallskip}
\hline 
\end{tabular}
\end{center}
\normalsize
\end{table*}

\section{RR Lyrae stars}\label{sec:rrls}

\begin{figure}[t!]
\centering
\includegraphics[trim=30 150 0 110 clip, width=1.05\linewidth]{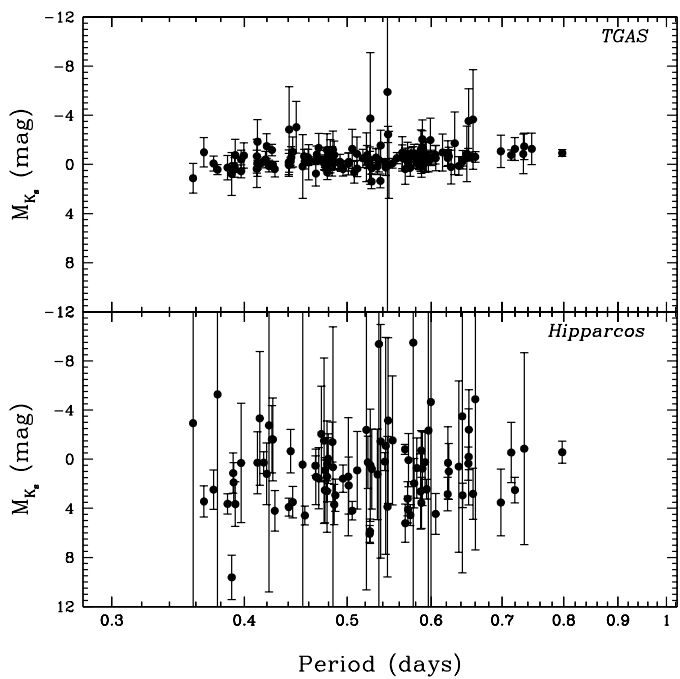}
\caption[]{RR Lyrae star $PM_{K_\mathrm{s}}$ relations using TGAS (upper panel) and  Hipparcos (lower panel)  parallaxes, respectively.
 The upper panel shows the absolute magnitudes, $M_{K_\mathrm{s}}$, of 143 RR Lyrae stars with 
positive TGAS parallaxes. The lower panel shows 91 RR Lyrae stars with positive Hipparcos parallaxes.}
\label{fig:PL_RR_Hip_Gaia}
\end{figure}

TGAS parallaxes are published in {\it Gaia} DR1 for 364 MW RR Lyrae stars.  Photometry and metallicity for most of these stars is available in the literature, although sparse through many different papers and catalogues.  
\citet{Dambis2013} have collected and homogenised the literature data of  403 MW RR Lyrae stars for which they  publish period, pulsation mode, interstellar visual absorption 
($A_V$),  iron abundance ([Fe/H]) on the \citet{ZW1984} metallicity scale (for 402 stars), and  intensity-averaged mean magnitudes in the Johnson {\it V}, 2MASS $K_\mathrm{s}$ and Wide-field Infrared Survey Explorer ({\it WISE}) W$_1$ (3.4~$\mu$m) passbands for 384, 403 and 398 stars, respectively. The pulsation periods are taken from the ASAS3 (\citealt{Pojmanski2002}, \citealt{Maintz2005}) and GCVS \citep{gcvs} catalogues. The interstellar extinction values are estimated from the three-dimensional model by \citet{Drimmel2003}. 
 These authors do not provide individual errors for the extinction values but  compared their extinction estimates with those derived from near-infrared colour-magnitude diagrams of different MW fields based on 2MASS  data, finding differences smaller than 0.05~mag. We adopt this value as the mean uncertainty of the extinction.  The mean $V$ magnitudes were calculated from nine overlapping sets of observations (see details in \citealt{Dambis2013} and references therein); the $K_\mathrm{s}$-band mean magnitudes from  2MASS  single-epoch observations of \citet{Cutri2003} and applying the phase-correction procedure described in \citet{Feast2008}\footnote{For 32 RR Lyrae stars in \citet{Dambis2013} sample the 2MASS magnitudes do not have phase correction, however, only one of them falls in our sample of 364.}. Mean magnitudes in the mid-infrared $W1$ passband are calculated from the {\it WISE} single-exposure database. 
We cross-matched our sample of 364 RR Lyrae stars with TGAS parallaxes against the catalogue of \citet{Dambis2013} and found 200 sources in common. They span the period range from 
0.27 to 0.80 d. The error distribution of the TGAS parallaxes for these 200 RR Lyrae stars is shown by the blue histogram in Fig.~\ref{fig:RRcumul}. 
The complete dataset ({\it Gaia}/Hipparcos/Tycho-2 Ids, period, pulsation mode, $G$, $\langle V \rangle$,  $\langle K_{\mathrm{s}} \rangle$,  $\langle W_1 \rangle$ magnitudes, $A_V$ and [Fe/H] values) along with 
TGAS parallaxes and errors 
for the total sample of 364 RR Lyrae stars is presented in Table~\ref{tab:rrlsall}. For the first 200 entries in the table, the literature values are  from \citet{Dambis2013}, for the remaining 164 sources the literature information is mainly taken from the GCVS \citep{gcvs}. 

Both Hipparcos and TGAS parallaxes are available for 145 RR Lyrae stars out of 200. As for the classical Cepheids we transformed the TGAS and Hipparcos parallaxes to absolute
$M_{\rm K_\mathrm{s}}$ magnitudes applying Eq. 1. This transformation was
possible only for stars with positive parallax values, namely, 143
out of 145 stars with TGAS parallax and only 
91 out of 145 for Hipparcos. We ``fundamentalized''  the periods of the RRc stars by adding 0.127 to the $\log P$.
The corresponding $PM_{\rm K_\mathrm{s}}$ relations are shown in the upper and lower panels
of Fig.~\ref{fig:PL_RR_Hip_Gaia} for TGAS and Hipparcos 
parallaxes, respectively.  
The former shows the significant improvement of the TGAS parallaxes for RR Lyrae stars. This is much more impressive  than for classical Cepheids and reveals a $PM_{\rm K_\mathrm{s}}$ 
relation  which becomes markedly visible if compared to Hipparcos'. 

Since \citet{Dambis2013}  provide a homogeneous dataset  for a fairly large sample of RR Lyrae stars we used their sample to study the $PL$ relations in the $K_{\mathrm{s}}$ and $W1$ passbands, the $PLZ$ relation in the $K_{\mathrm{s}}$ band and the optical $M_{V}-{\rm [Fe/H]}$ relation adopting the three different approaches described in Sec.~\ref{sec:meth}. 
The occurence of RR Lyrae stars in binary systems is an extremely rare event. Only  one RR Lyrae star, TU UMa, is confirmed to be a member of a binary star system with an orbital period of approximately 23 yr (\citealt{wade99}). Hence, we do not expect extra-scatter in the RR Lyrae relations due to this effect. Indeed, TU UMa is in the sample of  200 RR Lyrae stars that we used to derive the relations for the RR Lyrae stars described in the following sections  and it is found to fall very nicely on the best fit line of the various relations.

We note that we did not use the $G$-band magnitudes to compute the $PL$, $PLZ$ or the $M_{V}-{\rm [Fe/H]}$ relations of the RR Lyrae stars, the reason being the same as discussed for  classical Cepheids in  Sec.~\ref{sec:plccs} and even more so for the RR Lyrae stars. In fact, as clearly shown by fig.~2 in \citet{Cat2004}, the slope of the RR Lyrae $PL$ relation changes from positive to negative values moving from the blue to the red edges of {\it Gaia} $G$ passband, being roughly zero at its center.  Hence, such a large passband should not be used 
to derive the RR Lyrae $PL$ and  $M_{V}-{\rm [Fe/H]}$ relations. 
\subsection{Derivation of the RR Lyrae stars $PL$ and $PLZ$ relations in the $K_{\mathrm{s}}$ passband}\label{sec:RR_PLZ}

\begin{figure}
\includegraphics[trim=30 200 0 100 clip, width=1.05\linewidth]{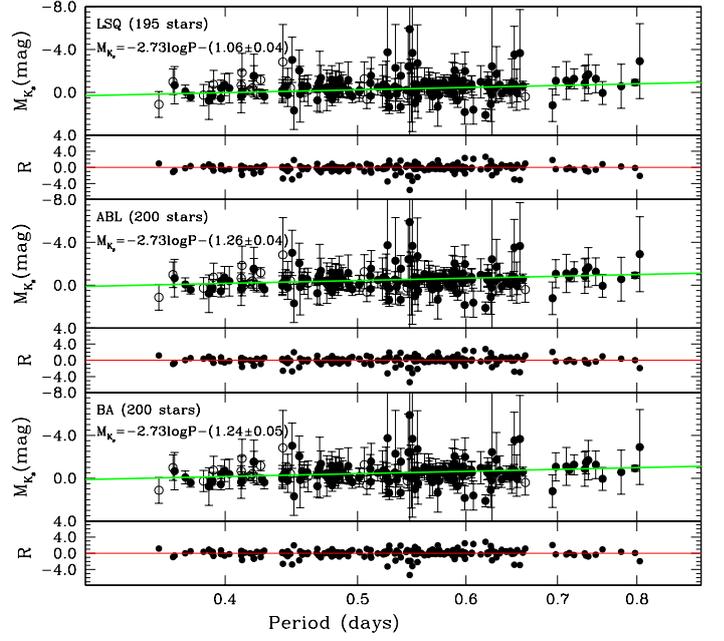}
\caption{RR Lyrae $PL$ relation in the $K_\mathrm{s}$ obtained i) by linear least squares fitting  the  stars' absolute magnitudes inferred from direct transformation of the TGAS parallaxes (upper panel); ii) via non-linear least squares fit  and the ABL method (middle panel),  and iii)  using the Bayesian approach (bottom panel). The slope of the fit is taken from \citet{Muraveva2015}.
Filled and empty circles represent fundamental-mode (RRab)  and first overtone (RRc) stars, respectively. The bottom part of each panel shows the residuals from the best fit line.}
  \label{fig:PL_RRL_K}
\end{figure}

  The near-infrared $PM_{K}Z$ relation of RR Lyrae stars has been studied by many different authors 
   and,  
  as summarised in Section~\ref{sec:intro}, coefficients and zero point of the relation differ significantly from one study to the other,   with the literature values for the dependence of $M_{\rm K_\mathrm{s}}$ on period 
  ranging from 
  $-2.101$ \citep{Bono2003} to  $-2.73$ \citep{Muraveva2015}  and the dependence on metallicity ranging from 0.03 \citep{Muraveva2015} to 0.231 \citep{Bono2003}. We  used the sample of 200 RR Lyrae stars with TGAS parallaxes  along with $K_\mathrm{s}$ magnitudes, period and metallicity values from \citet{Dambis2013} to fit the $PM_{K_\mathrm{s}}$ and $PM_{K_\mathrm{s}}Z$ relations. The $K_\mathrm{s}$ magnitudes were dereddened  using the  $V$-band  absorption values ($A_V$)  in  \citet{Dambis2013} and \citet{Cardelli1989}'s  $A_K/A_V$ = 0.114 extinction law for the $K$ band. 
The periods of the first overtone RR Lyrae stars (RRc) were ``fundamentalized''  as described in the previous section.  
As with the Cepheids we did not attempt to derive both slope and zero point of the RR Lyrae relations 
but used the TGAS parallaxes only to estimate the zero points.
Specifically, we adopted the slope of the $PM_{K_\mathrm{s}}$ relation from \citet{Muraveva2015},  who studied RR Lyrae stars in the LMC.  
The upper panel of Fig.~\ref{fig:PL_RRL_K} shows the linear least squares fit of  the  absolute magnitudes inferred from the direct transformation of the TGAS parallaxes. Five RR Lyrae stars  in the sample have a negative parallax value, hence, 
the least squares fit could be applied  only to 195 RR Lyrae stars. 
Middle and bottom panels of  Fig.~\ref{fig:PL_RRL_K} show the $PM_{K_\mathrm{s}}$ relations obtained with the ABL and the Bayesian approaches, respectively. The whole sample of 200
RR Lyrae stars was used with these two methods. 
The $PM_{K_\mathrm{s}}$ relations obtained with the three different approaches are summarised in the first three rows of Table~\ref{tab:RR_relations}. Once again, the r.m.s of all relations is significantly large,  the zero-points obtained with the ABL and Bayesian approaches are  in good agreement to each other and, on average,  about 0.2 mag brighter than found with the LSQ fit. 
 We also note that  the 
zero-points obtained with the ABL and Bayesian approaches are in perfect agreement with the zero-point of the $PM_{K_\mathrm{s}}Z$  relation obtained in  \citet{Muraveva2015} using the HST parallaxes of four RR Lyrae stars from \citet{Benedict2011}, (eq.~ 6 in \citealt{Muraveva2015}). 

To fit the RR Lyrae near-infrared $PM_{K_\mathrm{s}}Z$ relation we used the metallicities in  \citet{Dambis2013}, which are on the Zinn  \& West metallicity scale, and took the slope of the period term from \citet{Muraveva2015}, who used  the metallicity scale  defined in \citet{Grat2004}. This is  0.06~dex systematically higher than the Zinn \& West scale, hence, we transformed  \citet{Dambis2013}'s metallicities accordingly. Furthermore, \citet{Dambis2013} do not provide errors for the metallicities. We assumed them to be of 0.2~dex for all stars, as an average between 
spectroscopic determinations, which uncertainties generally are of the order of 0.1 dex and photometric metallicities whose typical errors can be as large as 0.3 dex. 
The $PM_{K_\mathrm{s}}Z$ relations obtained with the three different approaches are provided in rows 4 to 6 of Table~\ref{tab:RR_relations}. The zero-points obtained with the ABL and Bayesian approaches are  in good agreement to each other and, on average,  about 0.1 mag brighter than found with the LSQ fit. 
The dependence of the $M_{K_\mathrm{s}}$ magnitude on metallicity is always found to be negligible considering the current uncertainties.   

\subsection{Derivation of the RR Lyrae stars $PL$ relation in the $W_1$ passband.}

The mid-infrared $PW_1$ relation of RR Lyrae variables has been  studied by \citet{Madore2013}, \citet{Dambis2014}, \citet{Klein2014} and \citet{Neeley2015},  who found the slope of the $PM_{W_1}$ dependence on period to range from   $-$2.332 (\citealt{Neeley2015} for the {\it Spitzer} 3.6~$\mu$m passband) to $-$2.44 (\citealt{Madore2013}, for the WISE passbands). 
The metallicity dependence of the RR Lyrae $PL$ relations is known to decrease with increasing the wavelength from near to mid infrared (see Section~\ref{sec:intro}). Since in this paper the dependence on metallicity of
the  $M_{K_\mathrm{s}}$ magnitudes was found to be consistent with zero (see Section~\ref{sec:RR_PLZ}) we do not expect any dependence on metallicity of the  $PM_{W_1}$  relation
to be detectable, given the current, large uncertainties. Hence to compute the $PL$ relation in the $W_1$ passband we  adopt the slope of the $PM_{W_1}$ relation from \citet{Madore2013}
who neglect the metallicity term. 
Apparent $W_1$ magnitudes are available, from \citet{Dambis2013}, for 198 of the 200 RR Lyrae stars in our sample. Five of them have negative parallaxes, hence,  
when deriving the $PM_{W_1}$ relation with the LSQ fit  we could use only 193 stars. ABL and Bayesian approaches were applied instead to the whole sample of 198 stars. To correct for the extinction we used the relation $A_{W_1}/A_V$ = 0.065 from \citet{Madore2013}.  
The $PM_{W_1}$ relations obtained with the three different approaches are summarised in rows 7 to 9 of Table~\ref{tab:RR_relations}. 
 The zero-point of the $PM_{W_1}$ derived from the LSQ fit is 0.24~mag fainter than the zero-point in \citet{Madore2013} ($-1.26\pm0.25$)  which is based on the HST parallaxes of four RR Lyrae stars from \citet{Benedict2011}, while the zero-points obtained with the ABL and Bayesian methods are about 0.2 mag  brighter and well in agreement, within the errors, with \citet{Madore2013}.

\begin{table*}
\caption[]{$PL$, and $M_{V}-{\rm [Fe/H]}$  relations for RR Lyrae stars based on TGAS parallaxes.}\label{tab:RR_relations}
\begin{center}
\begin{tabular}{l c c l}
\hline
\hline
\noalign{\smallskip}
\tiny
 & Relation & RMS \\
 & & (mag) \\
 \noalign{\smallskip}
 \hline
\noalign{\smallskip}
$PM_{K_\mathrm{s}}$ 195 stars (LSQ) & $-2.73\log P - (1.06 \pm 0.04)$ & 0.84  \\
$PM_{K_\mathrm{s}}$ 200  stars (ABL) & $-2.73\log P - (1.26 \pm 0.04)$ & 0.90  \\
$PM_{K_\mathrm{s}}$ 200  stars (BA) & $-2.73\log P - (1.24 \pm 0.05)$ & 1.02  \\

\noalign{\smallskip}
\hline 
\noalign{\smallskip}
$PM_{K_\mathrm{s}}Z$ 195 stars (LSQ) & $-2.73\log P - (0.01 \pm 0.07){\rm [Fe/H]} - (1.08 \pm 0.09)$ & 0.84 \\
$PM_{K_\mathrm{s}}Z$ 200  stars (ABL) & $-2.73\log P + (0.07 \pm 0.07){\rm [Fe/H]} - (1.17 \pm 0.10)$ & 0.89 \\
$PM_{K_\mathrm{s}}Z$ 200 stars (BA)  & $-2.73 \log P - (0.01\substack{+0.11\\-0.07}){\rm [Fe/H]} - (1.22\substack{+0.40\\-0.09}) $& 1.02 & \\

\noalign{\smallskip}
\hline
\noalign{\smallskip}
$PM_{W_1}$ 193 stars (LSQ) & $-2.44\log P - (1.02 \pm 0.04)$ & 0.82  \\
$PM_{W_1}$ 198 stars (ABL) &  $-2.44\log P - (1.21 \pm 0.04)$ & 0.87  \\
$PM_{W_1}$ 198 stars (BA) & $-2.44 \log P - (1.20\pm 0.05) $ & 1.02 \\

\noalign{\smallskip}
\hline
\noalign{\smallskip}
$M_{V}-{\rm [Fe/H]}$ 195 stars (LSQ) & $0.214{\rm [Fe/H]} + (1.01\pm0.04)$ & 0.82 \\
$M_{V}-{\rm [Fe/H]}$ 200 stars (ABL) & $0.214{\rm [Fe/H]} + (0.82\pm0.04)$ & 0.87 \\
$M_{V}-{\rm [Fe/H]}$ 200 stars (BA) & $0.214{\rm [Fe/H]} + (0.88\substack{+0.04\\-0.06})$ & 1.21 \\

\noalign{\smallskip}
\hline 
\end{tabular}

\end{center}
\normalsize
\end{table*}

\subsection{Derivation of the optical $M_{V}-{\rm [Fe/H]}$ relation of RR Lyrae stars}\label{sec:MvFe}

Finally, we used our sample of 200 RR Lyrae stars to compute the luminosity-metallicity relation, $M_{V}-{\rm [Fe/H]}$ that RR Lyrae stars conform to in the optical. Results are summarised in the bottom three rows of 
Table~\ref{tab:RR_relations} for the LSQ fit, ABL method and Bayesian approach, separately. The direct transformation of the parallaxes to absolute magnitudes was possible only for 195 stars with positive parallaxes, while,  the ABL and Bayesian approaches were applied to the whole sample of 200 RR Lyrae stars. We corrected the $V$ apparent  magnitudes for extinction using $A_V=3.1E(B-V)$ \citep{Cardelli1989}. 
We adopted the slope of \citet{Clementini2003} and \citet{Grat2004} who studied the luminosity-metallicity relation of RR Lyrae stars in the LMC.  
As in Sec.~\ref{sec:RR_PLZ}, we transformed \citet{Dambis2013} metallicities to the scale adopted in \citet{Clementini2003} and \citet{Grat2004}.  
The $M_{V}-{\rm [Fe/H]}$ relation derived by the direct transformation of the TGAS parallaxes provides an absolute magnitude of $M_V=0.69 \pm 0.04$~mag for RR Lyrae stars with metallicity [Fe/H]=$-$1.5~dex. This is significantly fainter than $M_V=0.45\pm0.05$~mag obtained for the same metallicity by \citet{Benedict2011} using HST parallaxes, while agrees, within the errors, with the value  of $M_V=0.66\pm0.14$~mag derived by \citet{Cat2008} for RR Lyrae itself ([Fe/H]=$-$1.48~dex).

The  $M_{V}-{\rm [Fe/H]}$ relations obtained by applying the ABL and Bayesian approaches lead to $M_V=0.50 \pm 0.04$~mag and $M_V=0.56^{+0.04}_{-0.06}$~mag at  [Fe/H]=$-$1.5~dex, respectively, which is marginally consistent within the relative errors  with the absolute magnitude derived by \citet{Benedict2011}. 
 
\section{Comparison of results from different relations and conclusions}\label{sec:lmc}

In  this section we use 
the TGAS-based $PL, PW, PLZ$ and  $M_V-{\rm [Fe/H]}$ relations of classical Cepheids, Type~II Cepheids and RR Lyrae stars derived with the three alternative fitting approaches 
presented in the previous sections to infer the distance to the LMC.  We considered only the $PM_{K_\mathrm{s}}$, $PW(V,I)$, $PW(V,K_\mathrm{s})$ relations for  the classical Cepheids and  for  the RR Lyrae stars only the $PM_{K_\mathrm{s}}$, $PM_{K_{\rm s}}Z$ and $M_{V}-{\rm [Fe/H]}$ relations, because they are the most relevant in distance scale studies. 
This comparison  may allow the user to apprehend potential and expected level of systematics of the TGAS  parallaxes for these primary standard candles of the cosmological distance ladder. It also gives some perspective on the effects of handling parallaxes in parallax or distance (absolute magnitude) space.

The LMC is a fundamental anchor of the extragalactic distance ladder and its first step. Over the last couple of decades the distance modulus of the LMC has been measured countless times using different Population~I and  II distance  indicators and many  independent techniques (see e.g. \citealt{Gib2000}, \citealt{Ben2002}, \citealt{Clementini2003}, \citealt{Sch2008} and 
 a recent compilation of literature values by \citealt{deGr2014}), most of which are now 
 converging on a  median value  of $\mu_{\rm LMC}=18.49\pm0.09$~mag, that is well in agreement  with the value of  $(m-M)_0$= 18.493 $\pm$ 0.008 (stat) $\pm$ 0.047 (syst)~mag derived by \citet{pietrz13} using  8 eclipsing binaries in the  LMC bar. 

\begin{figure*}
\includegraphics[trim= 0 140 0 100  clip,width=12cm]{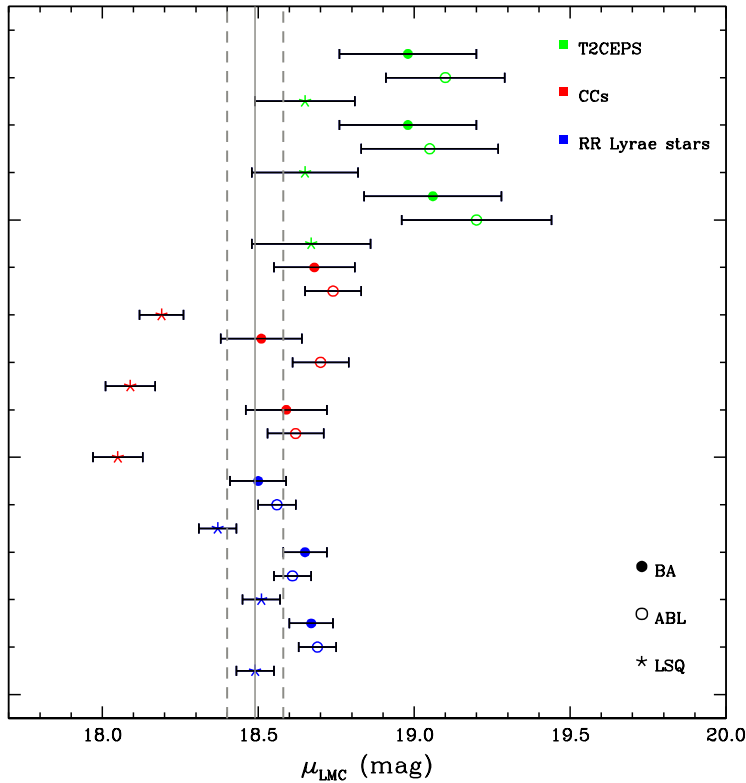}
\caption{Values of the LMC distance modulus obtained from the $PL$, $PLZ$ and $M_{V}-{\rm [Fe/H]}$  relations for RR Lyrae stars (blue symbols), and some of the different TGAS-based $PL$ and $PW$ relations for classical (CCs; red symbols) and Type~II Cepheids (T2CEPs; green symbols) derived in this paper. 
Asterisks, open circles and filled circles indicate results obtained using the linear least square fit (LSQ),  the ABL  method and the Bayesian approach, respectively.
From  bottom to top, RR Lyrae stars (blue symbols): $PM_{K_{\rm s}}$ relations for 195/200/200 stars (LSQ, ABL, BA methods)   and 
slope from 
\citet{Muraveva2015}; $PM_{K_{\rm s}}Z$ relations for 195/200/200 stars (LSQ, ABL, BA methods)   and slope of the dependence on period from
\citet{Muraveva2015}; $M_{V}-{\rm [Fe/H]}$ relations for 195/200/200 stars (LSQ, ABL, BA methods) with slope from \citet{Clementini2003}. 
For classical Cepheids (red symbols): 
$PM_{K_{\rm s}}$ and $PW(V,V-I)$ relations for  95/102/102 stars (LSQ, ABL, BA methods) with slopes from \citet{fou2007}; 
 and $PW(K_{\rm s},V-K_{\rm s})$ relation for 95/102/102 stars (LSQ, ABL, BA methods) with slope from  \citet{ripe12}.
For Type~II Cepheids (green symbols): $PM_{J}$, $PM_{K_{\rm s}}$ and $PW(K_{\rm s},J-K_{\rm s})$ relations 
for 22 /26/26  stars (LSQ, ABL, BA methods) with slopes from \citet{Ripepi2015}.}
\label{fig:DMlmc}
\end{figure*}

The distance moduli derived for the LMC from some of the various relations obtained in this paper are summarised 
in Table~\ref{tab:rr_lmc},  with results for classical Cepheids, Type~II Cepheids and RR Lyrae stars shown in the upper, middle and lower  portions of the table, respectively. They were calculated according to the following procedure:  for the classical Cepheids we adopted the $PM_{K_\mathrm{s}}$ and $PW(V,I)$ relations derived for the LMC variables by \citet{fou2007} (raws 4 and 2 from the bottom of their table~8) and the $PW(V,K_\mathrm{s})$ relation 
 by \citet{ripe12} (their table~4). For the Type~II Cepheids we used the LMC relations in table~5 of \citet{Ripepi2015}. Finally, for the RR Lyrae stars we applied the relations from \citet{Muraveva2015} and \citet{Clementini2003}. We then subtracted from the zero points of these relations, which are in apparent magnitude,  our corresponding TGAS-based zero points. 
 Finally, errors in  the distance moduli were calculated as the r.m.s of our relations divided by the square root of  the number of sources used in the fit.  Different values are graphically compared in Fig.~\ref{fig:DMlmc} 
 where we adopt as a reference value the LMC distance modulus from \citet{pietrz13}.
\begin{table*}
\caption[]{Distance moduli of the LMC obtained from some of the TGAS-based $PL$, $PW$, $PLZ$ and $M_{V}-{\rm [Fe/H]}$ relations derived in this study.}
\begin{center}
\label{tab:rr_lmc}
\begin{tabular}{l c c c}
\hline
\hline
\noalign{\smallskip}
Relation & $\mu$ (LSQ) & $\mu$ (ABL method) & $\mu$ (Bayesian method) \\
    & (mag) &  (mag)  & (mag) \\
\noalign{\smallskip}
 \hline
\multicolumn{3}{c}{\textbf{Classical Cepheids}}\\
 \hline
\noalign{\smallskip}
$PM_{\rm K_s}$ for 95 (102) objects   & $18.05\pm0.08$ & $18.62\pm0.09$ & $18.59\pm0.13$ \\
$PW(V,V-I)$ for 95 (102) objects & $18.09\pm0.08$ & $18.70\pm0.09$ & $18.51\pm0.13$ \\
$PW(K_s,V-K_s)$ for 95 (102) objects & $18.19\pm0.07$ & $18.74\pm0.09$ & $18.68\pm0.13$ \\
\noalign{\smallskip}
\hline
\noalign{\smallskip}
\multicolumn{3}{c}{\textbf{Type II Cepheids}}\\
\noalign{\smallskip}
\hline
\noalign{\smallskip}
$PM_{J}$ for 22 (26) objects  & $18.67\pm0.19$ & $19.20\pm0.24$ & $19.06\pm0.22$ \\
$PM_{K_s}$ for 22 (26) objects  &$18.65\pm0.17$ & $19.05\pm0.22$ & $18.98\pm0.22$ \\
$PW(K_s,J-K_s)$  for 22 (26) objects & $18.65\pm0.16$ & $19.10\pm0.19$ & $18.98\pm0.22$ \\
\noalign{\smallskip}
\hline
\noalign{\smallskip}
\multicolumn{3}{c}{\textbf{RR Lyrae stars}}\\
\hline
\noalign{\smallskip}
$PM_{K_s}$ for 195 (200) objects & $18.49 \pm 0.06$  & $18.69\pm 0.06$ & $18.67\pm0.07$ \\
$PM_{K_s}Z$ for 195 (200) objects & $18.51 \pm 0.06$  & $18.61 \pm0.06$  & $18.65\pm0.07$\\
$M_{V}-{\rm [Fe/H]}$ for 195 (200) objects & $18.37 \pm 0.06$  & $18.56 \pm 0.06$ & $18.50\pm0.09$\\
\noalign{\smallskip}
\hline
\noalign{\smallskip}
\end{tabular}
\end{center}
\normalsize
\end{table*}

We find that there is a good consistency, within the errors, between results obtained with the ABL and Bayesian approaches for all three types of variables. On the other hand, the LSQ fit  provides systematically shorter moduli.  This discrepancy is larger (about 0.5-0.6 mag, on average) for the classical Cepheids, reduces to 0.4-0.5 mag for the Type~II Cepheids,  and is the smallest one, 0.2 mag,  for the RR Lyrae stars. 
When compared with the  LMC distance modulus of \citet{pietrz13}, results from the Bayesian approach applied to classical Cepheids are always well consistent within the errors with the canonical value, the ABL results infer slightly longer distances than \citet{pietrz13}, whereas results from the LSQ fit are 0.2-0.4 mag   shorter than currently accepted 
in the literature. 
For the Type~II Cepheids, all three methods provide longer moduli than the canonical value, by  0.3-0.5 mag the ABL and Bayesian approaches and by 0.2 mag  the LSQ fit,  hence being still consistent,  within the errors, with \citet{pietrz13} estimate. 
The RR Lyrae stars used in this exercise  are twice in number the classical Cepheids and more than 7 times the Type~II Cepheids. The 
results obtained for the RR Lyrae stars show a much better agreement among the three methods and also a reasonably good agreement with the literature, once again confirming the impressive improvement in quality and statistics of the TGAS parallaxes for RR Lyrae stars  compared to Hipparcos.

However, taken at face value the results summarised in Table~\ref{tab:rr_lmc} and Fig.~\ref{fig:DMlmc}  span an uncomfortable, large  range of over one magnitude around the commonly accepted value of 18.5~mag for the distance modulus of the LMC.
Because errors are  still fairly  large, it is not clear, at this stage,  whether this hints to some systematics in  
the parallax derivation  that may affect in different way the different types of variables used in this paper. For instance,  since no chromatic corrections were applied to derive the TGAS parallaxes,
a colour effect could affect more classical Cepheids, which are both intrinsically redder and more reddened, than RR Lyrae stars or Type~II Cepheids.
Nevertheless, it is not easy to interpret these results unless in light of the still very large uncertainties affecting the TGAS parallaxes and, perhaps,  the relatively small sample of variable stars that could be used, for instance  in the analysis of the Type~II Cepheids.  
We also remind the reader that the complexity in the censorships of the samples available prevented us from producing a more reliable estimation of the possible biases introduced by them.
 As such the results presented in this paper have to be considered as preliminary, and to be superseded by results from further releases of {\it Gaia} data allowing the use of samples with more accurate parallaxes and less 
  affected by uncontrolled censorship effects.   

\include{ack}

\clearpage

\pagestyle{empty}
\begin{landscape}
 \begin{table}
\caption[]{Dataset for the classical Cepheids}
\tiny
\label{tab:cephall}
\sisetup{round-mode=places}

\begin{tabular}{l c c l S[round-precision=2] c S[round-precision=2]
S[round-precision=2] S[round-precision=4] S[round-precision=3] S[round-precision=3]
S[round-precision=3] S[round-precision=3] S[round-precision=3]  c c c}
\hline
\hline
\noalign{\smallskip}
Name & ID$_{{\it Gaia}}$ & ID$_{\rm Hipparcos}$& ID$_{Tycho2}$& {$\varpi_{Hip}$}&
$\sigma\varpi_{Hip}$ &{$\varpi_{\rm TGAS}$} & {$\sigma\varpi_{\rm TGAS}$} & {P} & {$G_{\it Gaia}$}&
{$\sigma_{G}$}& 
{$\langle K_\mathrm{s} \rangle$}& {$\langle V \rangle$}&{$\langle I \rangle$}&
{E($B$-$V$)}& ${\rm
[Fe/H]}$& Ref\\
& & & & {(mas)}& (mas) &{(mas)} & {(mas)} & {(days)} & {(mag)}& {(mag)}&  {(mag)}& 
{(mag)}& {(mag)}&  (mag)&(dex)&\\
\noalign{\smallskip}
\hline
\noalign{\smallskip}
  AY Cen & 5334506130758436352 & 55726 &   & 0.26 & 1.29 & 0.54025868302707 &
0.22113039033006 & 5.30975 & 8.43081820789701 & 0.01349812530365 & 6.249 & 8.813 &
 7.693 & 0.38 & 0.05 & 1,2,3,8,9\\
  BE Mon & 3133819107955689984 & 31905 &   & -0.10 & 2.44& 0.53351566308143 &
0.23536980400087 & 2.70551 & 10.0774204267615 & 0.01484972757816 & 7.676 &
10.56802484 & 9.24160012 & 0.565 & 0.05 & 1,2,3,9,10,11\\
  BR Vul & 1827869808377481216 & 97309 & & -2.35 & 1.67 & 0.65621633389887 &
0.26481326889195 & 2.04623193333087 & 10.0359929990972 & 0.01740087151505 & 7.136
& 10.686 & 9.021 & 0.911 &  {-}  & 1,3\\
  CR Cep & 2008504450538203776 & 112430 &   & 1.85 & 1.04 & 0.96362273526233 &
0.23791620806700 & 6.23315 & 8.90631026748576 & 0.01080406869406 & 5.873 &
9.64678155 & 7.97366171 & 0.709 & 0.0 & 1,2,3,7,9,10,11\\
  CR Ser & 4147381362033178624 & 89013 &  & -2.95 & 2.27 & 0.62673064582949 &
0.29504787825877 & 5.30141 & 9.81983724714790 & 0.01463077560295 & 6.555 &
10.85583614 & 8.89952126 & 0.961 &   {-} & 1,3,9,10,11\\
  \hline
  \noalign{\smallskip}
\end{tabular}
\footnotesize
\\
(1) \citet{van07b}; (2) \citet{gen14}; (3) \citet{gcvs}; (4) \citet{szi07}; (5)
\citet{roma08}; (6)  \citet{lem07} and \citet{lem08}; (7) \citet{luc11}; (8)
\citet{lul11}; (9) \citet{berdni00}; (10) \citet{groe1999}; (11) \citet{ngeow2012}.
The first 102 entries are fundamental mode classical Cepheids that we used in the
analysis described in Sec.~\ref{sec:plccs}.\\
This table is published in its entirety in the electronic version of the journal, a
portion is shown here for guidance regarding its form and content. 
\normalsize
\end{table}
 \begin{table}
\caption[]{Dataset for the Type II Cepheids}
\tiny
\label{tab:type2all}
\sisetup{round-mode=places}
\begin{tabular}{l c c l S[round-precision=2] c S[round-precision=2]
S[round-precision=2] S[round-precision=4] S[round-precision=3] S[round-precision=3]
S[round-precision=3] c  c}
\hline
\hline
\noalign{\smallskip}
Name & ID$_{{\it Gaia}}$ & {ID$_{\rm Hipparcos}$}& ID$_{Tycho2}$& {$\varpi_{Hip}$}&
$\sigma\varpi_{Hip}$ &{$\varpi_{\rm TGAS}$} & {$\sigma\varpi_{\rm TGAS}$} & {P} & {$G_{\it Gaia}$}&
{$\sigma_{G}$}& {$K_\mathrm{s}$}& {E($B$-$V$)}\\

& & & & {(mas)}& (mas) &{(mas)} & {(mas)} & {(days)} & {(mag)}& {(mag)}&  {(mag)}& 
(mag)\\
\noalign{\smallskip}
\hline
\noalign{\smallskip}
  BL Her & 4527596846604488448 & 88242 & 1562-1255-1 & 1.27 & 2.23 &
0.74833133169189 & 0.25727358797745 & 1.3074502 & 10.1439253914554 &
0.01608697333622 & 9.116 & 0.07\\
  RT Tra & 5828480455598115584 & 81157 & 9042-226-1 & 1.26 & 1.5 & 1.02576784096990
& 0.23777768343277 & 1.9461124 & 9.58694915283852 & 0.01305802408204 & 8.268 &
0.43\\
    RU Cam & 1108841774211669504 & 35681 & 4364-97-1 & 0.71 & 0.8 & 0.63697685896294
& 0.25018354340524 & 22.0 & 8.18558929672985 & 0.00410422887402 & 6.154 & 0.08\\
   SW Tau & 3283721025728185728 & 20587 & 78-1341-1 & 2.8 & 1.44 & 1.27165921569942
& 0.24780404887744 & 1.583584 & 9.38366728495032 & 0.01833985359968 & 7.958 &
0.282\\ 
   TX Del & 1734124244402708608 & 102853 & 516-1779-1 & 0.04 & 1.69 &
1.20368259629773 & 0.21873303383923 & 6.165907 & 8.94896652981389 &
0.01549419196302 & 7.465 & 0.1\\  
\hline 
\noalign{\smallskip}
\end{tabular}
\footnotesize
\\Note: $K_\mathrm{s}$ magnitudes from 2MASS \citep{Cutri2003}. Periods from
\citet{gcvs}. Colour excess from \citet{har85}.\\
This table is published in its entirety in the electronic version of the journal, a
portion is shown here for guidance regarding its form and content. 
\normalsize
\end{table}

 \begin{table}
\caption[]{Dataset for the RR Lyrae stars}
\tiny
\label{tab:rrlsall}
\sisetup{round-mode=places}

\begin{tabular}{l c c l S[round-precision=2] c S[round-precision=2]
S[round-precision=2] S[round-precision=4] c S[round-precision=3]
S[round-precision=3] S[round-precision=3] S[round-precision=3]  c c }
\hline
\hline
\noalign{\smallskip}
Name & ID$_{Gaia}$ & ID$_{\rm Hipparcos}$& ID$_{Tycho2}$& {$\varpi_{Hip}$}&
$\sigma\varpi_{Hip}$ &{$\varpi_{\rm TGAS}$} & {$\sigma\varpi_{\rm TGAS}$} & {P} & Mode &
{$G_{Gaia}$}& {$\sigma_{G}$}& {$\langle V \rangle$}& {$\langle K_\mathrm{s} \rangle
$}& {A$_{V}$}& ${\rm
[Fe/H]}$\\

& & & & {(mas)}& (mas) &{(mas)} & {(mas)} & {(days)} & & {(mag)}& {(mag)}&  {(mag)}&
 {(mag)}& (mag)&(dex)\\
\noalign{\smallskip}
\hline
\noalign{\smallskip}
 AA    CMi & 3111925220109675136 & 35281 & 164-182-1 & 2.84 & 3.47 &
0.81946982714046 & 0.23064935247928 & 0.4764 & AB & 11.5132153179761 &
0.02221651735744 & 11.552 & 10.287 & 0.257 & $-$0.55 \\
  AB    UMa & 1546016668386865792 & 59411 & 3455-362-1 & 0.14 & 1.94 &
0.93481019007481 & 0.26620098431843 & 0.5996 & AB & 10.7326885948591 &
0.01012328278142 & 10.899 & 9.623 & 0.068 & $-$0.72\\
    AE    Boo & 1234729395962169216 & 72342 & 1478-225-1 & 0.32 & 2.00 &
1.21132181778807 & 0.26417315206282 & 0.315 & C & 10.5261616807417 &
0.01490147743620 & 10.664 & 9.73 & 0.07 & $-$1.47\\
     AF    Vel & 5360400626025377536 & 53213 & 8207-1400-1 & 1.45 & 2.18 &
1.15670427795945 & 0.24731727156623 & 0.5275 & AB & 11.3935808199721 &
0.00985536630578 & 11.389 & 10.042 & 0.407 & -1.64\\
   AL    CMi & 3143813565573130880 &  & 192-43-1 & {-}  & {-}  & 1.12286547363676 &
0.40436107961229 & 0.5506 & AB & 11.8401243327053 & 0.01217484109889 & 11.931 &
10.767 & 0.024 & -0.85\\    
\hline 
\end{tabular}
\footnotesize
\\
This table is published in its entirety in the electronic version of the journal, a
portion is shown here for guidance regarding its form and content. 
\normalsize
\end{table}

\normalsize
\end{landscape}
\pagestyle{plain}

\end{document}

%% file: authors-c2.tex
\author{
{\it Gaia} Collaboration
\and G.        ~Clementini                    \inst{\ref{inst:0001}}
\and L.        ~Eyer                          \inst{\ref{inst:0002}}
\and V.        ~Ripepi                        \inst{\ref{inst:0003}}
\and M.        ~Marconi                       \inst{\ref{inst:0003}}
\and T.        ~Muraveva                      \inst{\ref{inst:0001}}
\and A.        ~Garofalo                      \inst{\ref{inst:0006},\ref{inst:0001}}
\and L.M.      ~Sarro                         \inst{\ref{inst:0008}}
\and M.        ~Palmer                        \inst{\ref{inst:0009}}
\and X.        ~Luri                          \inst{\ref{inst:0009}}
\and R.        ~Molinaro                      \inst{\ref{inst:0003}}
\and L.        ~Rimoldini                     \inst{\ref{inst:0012}}
\and L.        ~Szabados                      \inst{\ref{inst:0013}}
\and I.        ~Musella                       \inst{\ref{inst:0003}}
\and R.I.      ~Anderson                      \inst{\ref{inst:0015},\ref{inst:0002}}
\and T.        ~Prusti                        \inst{\ref{inst:0017}}
\and J.H.J.    ~de Bruijne                    \inst{\ref{inst:0017}}
\and A.G.A.    ~Brown                         \inst{\ref{inst:0019}}
\and A.        ~Vallenari                     \inst{\ref{inst:0020}}
\and C.        ~Babusiaux                     \inst{\ref{inst:0021}}
\and C.A.L.    ~Bailer-Jones                  \inst{\ref{inst:0022}}
\and U.        ~Bastian                       \inst{\ref{inst:0023}}
\and M.        ~Biermann                      \inst{\ref{inst:0023}}
\and D.W.      ~Evans                         \inst{\ref{inst:0025}}
\and F.        ~Jansen                        \inst{\ref{inst:0026}}
\and C.        ~Jordi                         \inst{\ref{inst:0009}}
\and S.A.      ~Klioner                       \inst{\ref{inst:0028}}
\and U.        ~Lammers                       \inst{\ref{inst:0029}}
\and L.        ~Lindegren                     \inst{\ref{inst:0030}}
\and F.        ~Mignard                       \inst{\ref{inst:0031}}
\and C.        ~Panem                         \inst{\ref{inst:0032}}
\and D.        ~Pourbaix                      \inst{\ref{inst:0033},\ref{inst:0034}}
\and S.        ~Randich                       \inst{\ref{inst:0035}}
\and P.        ~Sartoretti                    \inst{\ref{inst:0021}}
\and H.I.      ~Siddiqui                      \inst{\ref{inst:0037}}
\and C.        ~Soubiran                      \inst{\ref{inst:0038}}
\and V.        ~Valette                       \inst{\ref{inst:0032}}
\and F.        ~van Leeuwen                   \inst{\ref{inst:0025}}
\and N.A.      ~Walton                        \inst{\ref{inst:0025}}
\and C.        ~Aerts                         \inst{\ref{inst:0042},\ref{inst:0043}}
\and F.        ~Arenou                        \inst{\ref{inst:0021}}
\and M.        ~Cropper                       \inst{\ref{inst:0045}}
\and R.        ~Drimmel                       \inst{\ref{inst:0046}}
\and E.        ~H{\o}g                        \inst{\ref{inst:0047}}
\and D.        ~Katz                          \inst{\ref{inst:0021}}
\and M.G.      ~Lattanzi                      \inst{\ref{inst:0046}}
\and W.        ~O'Mullane                     \inst{\ref{inst:0029}}
\and E.K.      ~Grebel                        \inst{\ref{inst:0023}}
\and A.D.      ~Holland                       \inst{\ref{inst:0052}}
\and C.        ~Huc                           \inst{\ref{inst:0032}}
\and X.        ~Passot                        \inst{\ref{inst:0032}}
\and M.        ~Perryman                      \inst{\ref{inst:0017}}
\and L.        ~Bramante                      \inst{\ref{inst:0056}}
\and C.        ~Cacciari                      \inst{\ref{inst:0001}}
\and J.        ~Casta\~{n}eda                 \inst{\ref{inst:0009}}
\and L.        ~Chaoul                        \inst{\ref{inst:0032}}
\and N.        ~Cheek                         \inst{\ref{inst:0060}}
\and F.        ~De Angeli                     \inst{\ref{inst:0025}}
\and C.        ~Fabricius                     \inst{\ref{inst:0009}}
\and R.        ~Guerra                        \inst{\ref{inst:0029}}
\and J.        ~Hern\'{a}ndez                 \inst{\ref{inst:0029}}
\and A.        ~Jean-Antoine-Piccolo          \inst{\ref{inst:0032}}
\and E.        ~Masana                        \inst{\ref{inst:0009}}
\and R.        ~Messineo                      \inst{\ref{inst:0056}}
\and N.        ~Mowlavi                       \inst{\ref{inst:0002}}
\and K.        ~Nienartowicz                  \inst{\ref{inst:0012}}
\and D.        ~Ord\'{o}\~{n}ez-Blanco        \inst{\ref{inst:0012}}
\and P.        ~Panuzzo                       \inst{\ref{inst:0021}}
\and J.        ~Portell                       \inst{\ref{inst:0009}}
\and P.J.      ~Richards                      \inst{\ref{inst:0073}}
\and M.        ~Riello                        \inst{\ref{inst:0025}}
\and G.M.      ~Seabroke                      \inst{\ref{inst:0045}}
\and P.        ~Tanga                         \inst{\ref{inst:0031}}
\and F.        ~Th\'{e}venin                  \inst{\ref{inst:0031}}
\and J.        ~Torra                         \inst{\ref{inst:0009}}
\and S.G.      ~Els                           \inst{\ref{inst:0079},\ref{inst:0023}}
\and G.        ~Gracia-Abril                  \inst{\ref{inst:0079},\ref{inst:0009}}
\and G.        ~Comoretto                     \inst{\ref{inst:0037}}
\and M.        ~Garcia-Reinaldos              \inst{\ref{inst:0029}}
\and T.        ~Lock                          \inst{\ref{inst:0029}}
\and E.        ~Mercier                       \inst{\ref{inst:0079},\ref{inst:0023}}
\and M.        ~Altmann                       \inst{\ref{inst:0023},\ref{inst:0089}}
\and R.        ~Andrae                        \inst{\ref{inst:0022}}
\and T.L.      ~Astraatmadja                  \inst{\ref{inst:0022}}
\and I.        ~Bellas-Velidis                \inst{\ref{inst:0092}}
\and K.        ~Benson                        \inst{\ref{inst:0045}}
\and J.        ~Berthier                      \inst{\ref{inst:0094}}
\and R.        ~Blomme                        \inst{\ref{inst:0095}}
\and G.        ~Busso                         \inst{\ref{inst:0025}}
\and B.        ~Carry                         \inst{\ref{inst:0031},\ref{inst:0094}}
\and A.        ~Cellino                       \inst{\ref{inst:0046}}
\and S.        ~Cowell                        \inst{\ref{inst:0025}}
\and O.        ~Creevey                       \inst{\ref{inst:0031},\ref{inst:0102}}
\and J.        ~Cuypers$^\dagger$             \inst{\ref{inst:0095}}
%\and J.        ~Cuypers                      \inst{\ref{inst:0095}}
\and M.        ~Davidson                      \inst{\ref{inst:0104}}
\and J.        ~De Ridder                     \inst{\ref{inst:0042}}
\and A.        ~de Torres                     \inst{\ref{inst:0106}}
\and L.        ~Delchambre                    \inst{\ref{inst:0107}}
\and A.        ~Dell'Oro                      \inst{\ref{inst:0035}}
\and C.        ~Ducourant                     \inst{\ref{inst:0038}}
\and Y.        ~Fr\'{e}mat                    \inst{\ref{inst:0095}}
\and M.        ~Garc\'{i}a-Torres             \inst{\ref{inst:0111}}
\and E.        ~Gosset                        \inst{\ref{inst:0107},\ref{inst:0034}}
\and J.-L.     ~Halbwachs                     \inst{\ref{inst:0114}}
\and N.C.      ~Hambly                        \inst{\ref{inst:0104}}
\and D.L.      ~Harrison                      \inst{\ref{inst:0025},\ref{inst:0117}}
\and M.        ~Hauser                        \inst{\ref{inst:0023}}
\and D.        ~Hestroffer                    \inst{\ref{inst:0094}}
\and S.T.      ~Hodgkin                       \inst{\ref{inst:0025}}
\and H.E.      ~Huckle                        \inst{\ref{inst:0045}}
\and A.        ~Hutton                        \inst{\ref{inst:0122}}
\and G.        ~Jasniewicz                    \inst{\ref{inst:0123}}
\and S.        ~Jordan                        \inst{\ref{inst:0023}}
\and M.        ~Kontizas                      \inst{\ref{inst:0125}}
\and A.J.      ~Korn                          \inst{\ref{inst:0126}}
\and A.C.      ~Lanzafame                     \inst{\ref{inst:0127},\ref{inst:0128}}
\and M.        ~Manteiga                      \inst{\ref{inst:0129}}
\and A.        ~Moitinho                      \inst{\ref{inst:0130}}
\and K.        ~Muinonen                      \inst{\ref{inst:0131},\ref{inst:0132}}
\and J.        ~Osinde                        \inst{\ref{inst:0133}}
\and E.        ~Pancino                       \inst{\ref{inst:0035},\ref{inst:0135}}
\and T.        ~Pauwels                       \inst{\ref{inst:0095}}
\and J.-M.     ~Petit                         \inst{\ref{inst:0137}}
\and A.        ~Recio-Blanco                  \inst{\ref{inst:0031}}
\and A.C.      ~Robin                         \inst{\ref{inst:0137}}
\and C.        ~Siopis                        \inst{\ref{inst:0033}}
\and M.        ~Smith                         \inst{\ref{inst:0045}}
\and K.W.      ~Smith                         \inst{\ref{inst:0022}}
\and A.        ~Sozzetti                      \inst{\ref{inst:0046}}
\and W.        ~Thuillot                      \inst{\ref{inst:0094}}
\and W.        ~van Reeven                    \inst{\ref{inst:0122}}
\and Y.        ~Viala                         \inst{\ref{inst:0021}}
\and U.        ~Abbas                         \inst{\ref{inst:0046}}
\and A.        ~Abreu Aramburu                \inst{\ref{inst:0148}}
\and S.        ~Accart                        \inst{\ref{inst:0149}}
\and J.J.      ~Aguado                        \inst{\ref{inst:0008}}
\and P.M.      ~Allan                         \inst{\ref{inst:0073}}
\and W.        ~Allasia                       \inst{\ref{inst:0152}}
\and G.        ~Altavilla                     \inst{\ref{inst:0001}}
\and M.A.      ~\'{A}lvarez                   \inst{\ref{inst:0129}}
\and J.        ~Alves                         \inst{\ref{inst:0155}}
\and A.H.      ~Andrei                        \inst{\ref{inst:0156},\ref{inst:0157},\ref{inst:0089}}
\and E.        ~Anglada Varela                \inst{\ref{inst:0133},\ref{inst:0060}}
\and E.        ~Antiche                       \inst{\ref{inst:0009}}
\and T.        ~Antoja                        \inst{\ref{inst:0017}}
\and S.        ~Ant\'{o}n                     \inst{\ref{inst:0163},\ref{inst:0164}}
\and B.        ~Arcay                         \inst{\ref{inst:0129}}
\and N.        ~Bach                          \inst{\ref{inst:0122}}
\and S.G.      ~Baker                         \inst{\ref{inst:0045}}
\and L.        ~Balaguer-N\'{u}\~{n}ez        \inst{\ref{inst:0009}}
\and C.        ~Barache                       \inst{\ref{inst:0089}}
\and C.        ~Barata                        \inst{\ref{inst:0130}}
\and A.        ~Barbier                       \inst{\ref{inst:0149}}
\and F.        ~Barblan                       \inst{\ref{inst:0002}}
\and D.        ~Barrado y Navascu\'{e}s       \inst{\ref{inst:0173}}
\and M.        ~Barros                        \inst{\ref{inst:0130}}
\and M.A.      ~Barstow                       \inst{\ref{inst:0175}}
\and U.        ~Becciani                      \inst{\ref{inst:0128}}
\and M.        ~Bellazzini                    \inst{\ref{inst:0001}}
\and A.        ~Bello Garc\'{i}a              \inst{\ref{inst:0178}}
\and V.        ~Belokurov                     \inst{\ref{inst:0025}}
\and P.        ~Bendjoya                      \inst{\ref{inst:0031}}
\and A.        ~Berihuete                     \inst{\ref{inst:0181}}
\and L.        ~Bianchi                       \inst{\ref{inst:0152}}
\and O.        ~Bienaym\'{e}                  \inst{\ref{inst:0114}}
\and F.        ~Billebaud                     \inst{\ref{inst:0038}}
\and N.        ~Blagorodnova                  \inst{\ref{inst:0025}}
\and S.        ~Blanco-Cuaresma               \inst{\ref{inst:0002},\ref{inst:0038}}
\and T.        ~Boch                          \inst{\ref{inst:0114}}
\and A.        ~Bombrun                       \inst{\ref{inst:0106}}
\and R.        ~Borrachero                    \inst{\ref{inst:0009}}
\and S.        ~Bouquillon                    \inst{\ref{inst:0089}}
\and G.        ~Bourda                        \inst{\ref{inst:0038}}
%\and H.        ~Bouy                          \inst{\ref{inst:0173}}
\and A.        ~Bragaglia                     \inst{\ref{inst:0001}}
\and M.A.      ~Breddels                      \inst{\ref{inst:0195}}
\and N.        ~Brouillet                     \inst{\ref{inst:0038}}
\and T.        ~Br\"{ u}semeister             \inst{\ref{inst:0023}}
\and B.        ~Bucciarelli                   \inst{\ref{inst:0046}}
\and P.        ~Burgess                       \inst{\ref{inst:0025}}
\and R.        ~Burgon                        \inst{\ref{inst:0052}}
\and A.        ~Burlacu                       \inst{\ref{inst:0032}}
\and D.        ~Busonero                      \inst{\ref{inst:0046}}
\and R.        ~Buzzi                         \inst{\ref{inst:0046}}
\and E.        ~Caffau                        \inst{\ref{inst:0021}}
\and J.        ~Cambras                       \inst{\ref{inst:0205}}
\and H.        ~Campbell                      \inst{\ref{inst:0025}}
\and R.        ~Cancelliere                   \inst{\ref{inst:0207}}
\and T.        ~Cantat-Gaudin                 \inst{\ref{inst:0020}}
\and T.        ~Carlucci                      \inst{\ref{inst:0089}}
\and J.M.      ~Carrasco                      \inst{\ref{inst:0009}}
\and M.        ~Castellani                    \inst{\ref{inst:0211}}
\and P.        ~Charlot                       \inst{\ref{inst:0038}}
\and J.        ~Charnas                       \inst{\ref{inst:0012}}
\and A.        ~Chiavassa                     \inst{\ref{inst:0031}}
\and M.        ~Clotet                        \inst{\ref{inst:0009}}
\and G.        ~Cocozza                       \inst{\ref{inst:0001}}
\and R.S.      ~Collins                       \inst{\ref{inst:0104}}
\and G.        ~Costigan                      \inst{\ref{inst:0019}}
\and F.        ~Crifo                         \inst{\ref{inst:0021}}
\and N.J.G.    ~Cross                         \inst{\ref{inst:0104}}
\and M.        ~Crosta                        \inst{\ref{inst:0046}}
\and C.        ~Crowley                       \inst{\ref{inst:0106}}
\and C.        ~Dafonte                       \inst{\ref{inst:0129}}
\and Y.        ~Damerdji                      \inst{\ref{inst:0107},\ref{inst:0225}}
\and A.        ~Dapergolas                    \inst{\ref{inst:0092}}
\and P.        ~David                         \inst{\ref{inst:0094}}
\and M.        ~David                         \inst{\ref{inst:0228}}
\and P.        ~De Cat                        \inst{\ref{inst:0095}}
\and F.        ~de Felice                     \inst{\ref{inst:0230}}
\and P.        ~de Laverny                    \inst{\ref{inst:0031}}
\and F.        ~De Luise                      \inst{\ref{inst:0232}}
\and R.        ~De March                      \inst{\ref{inst:0056}}
\and R.        ~de Souza                      \inst{\ref{inst:0234}}
\and J.        ~Debosscher                    \inst{\ref{inst:0042}}
\and E.        ~del Pozo                      \inst{\ref{inst:0122}}
\and M.        ~Delbo                         \inst{\ref{inst:0031}}
\and A.        ~Delgado                       \inst{\ref{inst:0025}}
\and H.E.      ~Delgado                       \inst{\ref{inst:0008}}
\and P.        ~Di Matteo                     \inst{\ref{inst:0021}}
\and S.        ~Diakite                       \inst{\ref{inst:0137}}
\and E.        ~Distefano                     \inst{\ref{inst:0128}}
\and C.        ~Dolding                       \inst{\ref{inst:0045}}
\and S.        ~Dos Anjos                     \inst{\ref{inst:0234}}
\and P.        ~Drazinos                      \inst{\ref{inst:0125}}
\and J.        ~Dur\'{a}n                     \inst{\ref{inst:0133}}
\and Y.        ~Dzigan                        \inst{\ref{inst:0247},\ref{inst:0248}}
\and B.        ~Edvardsson                    \inst{\ref{inst:0126}}
\and H.        ~Enke                          \inst{\ref{inst:0250}}
\and N.W.      ~Evans                         \inst{\ref{inst:0025}}
\and G.        ~Eynard Bontemps               \inst{\ref{inst:0149}}
\and C.        ~Fabre                         \inst{\ref{inst:0253}}
\and M.        ~Fabrizio                      \inst{\ref{inst:0135},\ref{inst:0232}}
%\and S.        ~Faigler                       \inst{\ref{inst:0256}}
\and A.J.      ~Falc\~{a}o                    \inst{\ref{inst:0257}}
\and M.        ~Farr\`{a}s Casas              \inst{\ref{inst:0009}}
\and L.        ~Federici                      \inst{\ref{inst:0001}}
\and G.        ~Fedorets                      \inst{\ref{inst:0131}}
\and J.        ~Fern\'{a}ndez-Hern\'{a}ndez   \inst{\ref{inst:0060}}
\and P.        ~Fernique                      \inst{\ref{inst:0114}}
\and A.        ~Fienga                        \inst{\ref{inst:0263}}
\and F.        ~Figueras                      \inst{\ref{inst:0009}}
\and F.        ~Filippi                       \inst{\ref{inst:0056}}
\and K.        ~Findeisen                     \inst{\ref{inst:0021}}
\and A.        ~Fonti                         \inst{\ref{inst:0056}}
\and M.        ~Fouesneau                     \inst{\ref{inst:0022}}
\and E.        ~Fraile                        \inst{\ref{inst:0269}}
\and M.        ~Fraser                        \inst{\ref{inst:0025}}
\and J.        ~Fuchs                         \inst{\ref{inst:0271}}
\and M.        ~Gai                           \inst{\ref{inst:0046}}
\and S.        ~Galleti                       \inst{\ref{inst:0001}}
\and L.        ~Galluccio                     \inst{\ref{inst:0031}}
\and D.        ~Garabato                      \inst{\ref{inst:0129}}
\and F.        ~Garc\'{i}a-Sedano             \inst{\ref{inst:0008}}
\and N.        ~Garralda                      \inst{\ref{inst:0009}}
\and P.        ~Gavras                        \inst{\ref{inst:0021},\ref{inst:0092},\ref{inst:0125}}
\and J.        ~Gerssen                       \inst{\ref{inst:0250}}
\and R.        ~Geyer                         \inst{\ref{inst:0028}}
\and G.        ~Gilmore                       \inst{\ref{inst:0025}}
\and S.        ~Girona                        \inst{\ref{inst:0284}}
\and G.        ~Giuffrida                     \inst{\ref{inst:0135}}
\and M.        ~Gomes                         \inst{\ref{inst:0130}}
\and A.        ~Gonz\'{a}lez-Marcos           \inst{\ref{inst:0287}}
\and J.        ~Gonz\'{a}lez-N\'{u}\~{n}ez    \inst{\ref{inst:0060},\ref{inst:0289}}
\and J.J.      ~Gonz\'{a}lez-Vidal            \inst{\ref{inst:0009}}
\and M.        ~Granvik                       \inst{\ref{inst:0131}}
\and A.        ~Guerrier                      \inst{\ref{inst:0149}}
\and P.        ~Guillout                      \inst{\ref{inst:0114}}
\and J.        ~Guiraud                       \inst{\ref{inst:0032}}
\and A.        ~G\'{u}rpide                   \inst{\ref{inst:0009}}
\and R.        ~Guti\'{e}rrez-S\'{a}nchez     \inst{\ref{inst:0037}}
\and L.P.      ~Guy                           \inst{\ref{inst:0012}}
\and R.        ~Haigron                       \inst{\ref{inst:0021}}
\and D.        ~Hatzidimitriou                \inst{\ref{inst:0125},\ref{inst:0092}}
\and M.        ~Haywood                       \inst{\ref{inst:0021}}
\and U.        ~Heiter                        \inst{\ref{inst:0126}}
\and A.        ~Helmi                         \inst{\ref{inst:0195}}
\and D.        ~Hobbs                         \inst{\ref{inst:0030}}
\and W.        ~Hofmann                       \inst{\ref{inst:0023}}
\and B.        ~Holl                          \inst{\ref{inst:0002}}
\and G.        ~Holland                       \inst{\ref{inst:0025}}
\and J.A.S.    ~Hunt                          \inst{\ref{inst:0045}}
\and A.        ~Hypki                         \inst{\ref{inst:0019}}
\and V.        ~Icardi                        \inst{\ref{inst:0056}}
\and M.        ~Irwin                         \inst{\ref{inst:0025}}
\and G.        ~Jevardat de Fombelle          \inst{\ref{inst:0012}}
\and P.        ~Jofr\'{e}                     \inst{\ref{inst:0025},\ref{inst:0038}}
\and P.G.      ~Jonker                        \inst{\ref{inst:0315},\ref{inst:0043}}
\and A.        ~Jorissen                      \inst{\ref{inst:0033}}
\and F.        ~Julbe                         \inst{\ref{inst:0009}}
\and A.        ~Karampelas                    \inst{\ref{inst:0125},\ref{inst:0092}}
\and A.        ~Kochoska                      \inst{\ref{inst:0321}}
\and R.        ~Kohley                        \inst{\ref{inst:0029}}
\and K.        ~Kolenberg                     \inst{\ref{inst:0323},\ref{inst:0042},\ref{inst:0325}}
\and E.        ~Kontizas                      \inst{\ref{inst:0092}}
\and S.E.      ~Koposov                       \inst{\ref{inst:0025}}
\and G.        ~Kordopatis                    \inst{\ref{inst:0250},\ref{inst:0031}}
\and P.        ~Koubsky                       \inst{\ref{inst:0271}}
\and A.        ~Krone-Martins                 \inst{\ref{inst:0130}}
\and M.        ~Kudryashova                   \inst{\ref{inst:0094}}
%\and I.        ~Kull                          \inst{\ref{inst:0256}}
\and R.K.      ~Bachchan                      \inst{\ref{inst:0030}}
\and F.        ~Lacoste-Seris                 \inst{\ref{inst:0149}}
\and A.F.      ~Lanza                         \inst{\ref{inst:0128}}
\and J.-B.     ~Lavigne                       \inst{\ref{inst:0149}}
\and C.        ~Le Poncin-Lafitte             \inst{\ref{inst:0089}}
\and Y.        ~Lebreton                      \inst{\ref{inst:0021},\ref{inst:0340}}
\and T.        ~Lebzelter                     \inst{\ref{inst:0155}}
\and S.        ~Leccia                        \inst{\ref{inst:0003}}
\and N.        ~Leclerc                       \inst{\ref{inst:0021}}
\and I.        ~Lecoeur-Taibi                 \inst{\ref{inst:0012}}
\and V.        ~Lemaitre                      \inst{\ref{inst:0149}}
\and H.        ~Lenhardt                      \inst{\ref{inst:0023}}
\and F.        ~Leroux                        \inst{\ref{inst:0149}}
\and S.        ~Liao                          \inst{\ref{inst:0046},\ref{inst:0349}}
\and E.        ~Licata                        \inst{\ref{inst:0152}}
\and H.E.P.    ~Lindstr{\o}m                  \inst{\ref{inst:0047},\ref{inst:0352}}
\and T.A.      ~Lister                        \inst{\ref{inst:0353}}
\and E.        ~Livanou                       \inst{\ref{inst:0125}}
\and A.        ~Lobel                         \inst{\ref{inst:0095}}
\and W.        ~L\"{ o}ffler                  \inst{\ref{inst:0023}}
\and M.        ~L\'{o}pez                     \inst{\ref{inst:0173}}
\and D.        ~Lorenz                        \inst{\ref{inst:0155}}
\and I.        ~MacDonald                     \inst{\ref{inst:0104}}
\and T.        ~Magalh\~{a}es Fernandes       \inst{\ref{inst:0257}}
\and S.        ~Managau                       \inst{\ref{inst:0149}}
\and R.G.      ~Mann                          \inst{\ref{inst:0104}}
\and G.        ~Mantelet                      \inst{\ref{inst:0023}}
\and O.        ~Marchal                       \inst{\ref{inst:0021}}
\and J.M.      ~Marchant                      \inst{\ref{inst:0365}}
\and S.        ~Marinoni                      \inst{\ref{inst:0211},\ref{inst:0135}}
\and P.M.      ~Marrese                       \inst{\ref{inst:0211},\ref{inst:0135}}
\and G.        ~Marschalk\'{o}                \inst{\ref{inst:0013},\ref{inst:0371}}
\and D.J.      ~Marshall                      \inst{\ref{inst:0372}}
\and J.M.      ~Mart\'{i}n-Fleitas            \inst{\ref{inst:0122}}
\and M.        ~Martino                       \inst{\ref{inst:0056}}
\and N.        ~Mary                          \inst{\ref{inst:0149}}
\and G.        ~Matijevi\v{c}                 \inst{\ref{inst:0250}}
%\and T.        ~Mazeh                         \inst{\ref{inst:0256}}
\and P.J.      ~McMillan                      \inst{\ref{inst:0030}}
\and S.        ~Messina                       \inst{\ref{inst:0128}}
\and D.        ~Michalik                      \inst{\ref{inst:0030}}
\and N.R.      ~Millar                        \inst{\ref{inst:0025}}
\and B.M.H.    ~Miranda                       \inst{\ref{inst:0130}}
\and D.        ~Molina                        \inst{\ref{inst:0009}}
\and M.        ~Molinaro                      \inst{\ref{inst:0384}}
\and L.        ~Moln\'{a}r                    \inst{\ref{inst:0013}}
\and M.        ~Moniez                        \inst{\ref{inst:0386}}
\and P.        ~Montegriffo                   \inst{\ref{inst:0001}}
\and R.        ~Mor                           \inst{\ref{inst:0009}}
\and A.        ~Mora                          \inst{\ref{inst:0122}}
\and R.        ~Morbidelli                    \inst{\ref{inst:0046}}
\and T.        ~Morel                         \inst{\ref{inst:0107}}
\and S.        ~Morgenthaler                  \inst{\ref{inst:0392}}
\and D.        ~Morris                        \inst{\ref{inst:0104}}
\and A.F.      ~Mulone                        \inst{\ref{inst:0056}}
\and J.        ~Narbonne                      \inst{\ref{inst:0149}}
\and G.        ~Nelemans                      \inst{\ref{inst:0043},\ref{inst:0042}}
\and L.        ~Nicastro                      \inst{\ref{inst:0398}}
\and L.        ~Noval                         \inst{\ref{inst:0149}}
\and C.        ~Ord\'{e}novic                 \inst{\ref{inst:0031}}
\and J.        ~Ordieres-Mer\'{e}             \inst{\ref{inst:0401}}
\and P.        ~Osborne                       \inst{\ref{inst:0025}}
\and C.        ~Pagani                        \inst{\ref{inst:0175}}
\and I.        ~Pagano                        \inst{\ref{inst:0128}}
\and F.        ~Pailler                       \inst{\ref{inst:0032}}
\and H.        ~Palacin                       \inst{\ref{inst:0149}}
\and L.        ~Palaversa                     \inst{\ref{inst:0002}}
\and P.        ~Parsons                       \inst{\ref{inst:0037}}
\and M.        ~Pecoraro                      \inst{\ref{inst:0152}}
\and R.        ~Pedrosa                       \inst{\ref{inst:0410}}
\and H.        ~Pentik\"{ a}inen              \inst{\ref{inst:0131}}
\and B.        ~Pichon                        \inst{\ref{inst:0031}}
\and A.M.      ~Piersimoni                    \inst{\ref{inst:0232}}
\and F.-X.     ~Pineau                        \inst{\ref{inst:0114}}
\and E.        ~Plachy                        \inst{\ref{inst:0013}}
\and G.        ~Plum                          \inst{\ref{inst:0021}}
\and E.        ~Poujoulet                     \inst{\ref{inst:0417}}
\and A.        ~Pr\v{s}a                      \inst{\ref{inst:0418}}
\and L.        ~Pulone                        \inst{\ref{inst:0211}}
\and S.        ~Ragaini                       \inst{\ref{inst:0001}}
\and S.        ~Rago                          \inst{\ref{inst:0046}}
\and N.        ~Rambaux                       \inst{\ref{inst:0094}}
\and M.        ~Ramos-Lerate                  \inst{\ref{inst:0423}}
\and P.        ~Ranalli                       \inst{\ref{inst:0030}}
\and G.        ~Rauw                          \inst{\ref{inst:0107}}
\and A.        ~Read                          \inst{\ref{inst:0175}}
\and S.        ~Regibo                        \inst{\ref{inst:0042}}
\and C.        ~Reyl\'{e}                     \inst{\ref{inst:0137}}
\and R.A.      ~Ribeiro                       \inst{\ref{inst:0257}}
\and A.        ~Riva                          \inst{\ref{inst:0046}}
\and G.        ~Rixon                         \inst{\ref{inst:0025}}
\and M.        ~Roelens                       \inst{\ref{inst:0002}}
\and M.        ~Romero-G\'{o}mez              \inst{\ref{inst:0009}}
\and N.        ~Rowell                        \inst{\ref{inst:0104}}
\and F.        ~Royer                         \inst{\ref{inst:0021}}
\and L.        ~Ruiz-Dern                     \inst{\ref{inst:0021}}
\and G.        ~Sadowski                      \inst{\ref{inst:0033}}
\and T.        ~Sagrist\`{a} Sell\'{e}s       \inst{\ref{inst:0023}}
\and J.        ~Sahlmann                      \inst{\ref{inst:0029}}
\and J.        ~Salgado                       \inst{\ref{inst:0133}}
\and E.        ~Salguero                      \inst{\ref{inst:0133}}
\and M.        ~Sarasso                       \inst{\ref{inst:0046}}
\and H.        ~Savietto                      \inst{\ref{inst:0443}}
\and M.        ~Schultheis                    \inst{\ref{inst:0031}}
\and E.        ~Sciacca                       \inst{\ref{inst:0128}}
\and M.        ~Segol                         \inst{\ref{inst:0446}}
\and J.C.      ~Segovia                       \inst{\ref{inst:0060}}
\and D.        ~Segransan                     \inst{\ref{inst:0002}}
\and I-C.      ~Shih                          \inst{\ref{inst:0021}}
\and R.        ~Smareglia                     \inst{\ref{inst:0384}}
\and R.L.      ~Smart                         \inst{\ref{inst:0046}}
\and E.        ~Solano                        \inst{\ref{inst:0173},\ref{inst:0453}}
\and F.        ~Solitro                       \inst{\ref{inst:0056}}
\and R.        ~Sordo                         \inst{\ref{inst:0020}}
\and S.        ~Soria Nieto                   \inst{\ref{inst:0009}}
\and J.        ~Souchay                       \inst{\ref{inst:0089}}
\and A.        ~Spagna                        \inst{\ref{inst:0046}}
\and F.        ~Spoto                         \inst{\ref{inst:0031}}
\and U.        ~Stampa                        \inst{\ref{inst:0023}}
\and I.A.      ~Steele                        \inst{\ref{inst:0365}}
\and H.        ~Steidelm\"{ u}ller            \inst{\ref{inst:0028}}
\and C.A.      ~Stephenson                    \inst{\ref{inst:0037}}
\and H.        ~Stoev                         \inst{\ref{inst:0464}}
\and F.F.      ~Suess                         \inst{\ref{inst:0025}}
\and M.        ~S\"{ u}veges                  \inst{\ref{inst:0012}}
\and J.        ~Surdej                        \inst{\ref{inst:0107}}
\and E.        ~Szegedi-Elek                  \inst{\ref{inst:0013}}
\and D.        ~Tapiador                      \inst{\ref{inst:0469},\ref{inst:0470}}
\and F.        ~Taris                         \inst{\ref{inst:0089}}
\and G.        ~Tauran                        \inst{\ref{inst:0149}}
\and M.B.      ~Taylor                        \inst{\ref{inst:0473}}
\and R.        ~Teixeira                      \inst{\ref{inst:0234}}
\and D.        ~Terrett                       \inst{\ref{inst:0073}}
\and B.        ~Tingley                       \inst{\ref{inst:0476}}
\and S.C.      ~Trager                        \inst{\ref{inst:0195}}
\and C.        ~Turon                         \inst{\ref{inst:0021}}
\and A.        ~Ulla                          \inst{\ref{inst:0479}}
\and E.        ~Utrilla                       \inst{\ref{inst:0122}}
\and G.        ~Valentini                     \inst{\ref{inst:0232}}
\and A.        ~van Elteren                   \inst{\ref{inst:0019}}
\and E.        ~Van Hemelryck                 \inst{\ref{inst:0095}}
\and M.        ~van Leeuwen                   \inst{\ref{inst:0025}}
\and M.        ~Varadi                        \inst{\ref{inst:0002},\ref{inst:0013}}
\and A.        ~Vecchiato                     \inst{\ref{inst:0046}}
\and J.        ~Veljanoski                    \inst{\ref{inst:0195}}
\and T.        ~Via                           \inst{\ref{inst:0205}}
\and D.        ~Vicente                       \inst{\ref{inst:0284}}
\and S.        ~Vogt                          \inst{\ref{inst:0491}}
\and H.        ~Voss                          \inst{\ref{inst:0009}}
\and V.        ~Votruba                       \inst{\ref{inst:0271}}
\and S.        ~Voutsinas                     \inst{\ref{inst:0104}}
\and G.        ~Walmsley                      \inst{\ref{inst:0032}}
\and M.        ~Weiler                        \inst{\ref{inst:0009}}
\and K.        ~Weingrill                     \inst{\ref{inst:0250}}
\and T.        ~Wevers                        \inst{\ref{inst:0043}}
\and \L{}.     ~Wyrzykowski                   \inst{\ref{inst:0025},\ref{inst:0500}}
\and A.        ~Yoldas                        \inst{\ref{inst:0025}}
\and M.        ~\v{Z}erjal                    \inst{\ref{inst:0321}}
\and S.        ~Zucker                        \inst{\ref{inst:0247}}
\and C.        ~Zurbach                       \inst{\ref{inst:0123}}
\and T.        ~Zwitter                       \inst{\ref{inst:0321}}
\and A.        ~Alecu                         \inst{\ref{inst:0025}}
\and M.        ~Allen                         \inst{\ref{inst:0017}}
\and C.        ~Allende Prieto                \inst{\ref{inst:0045},\ref{inst:0509},\ref{inst:0510}}
\and A.        ~Amorim                        \inst{\ref{inst:0130}}
\and G.        ~Anglada-Escud\'{e}            \inst{\ref{inst:0009}}
\and V.        ~Arsenijevic                   \inst{\ref{inst:0130}}
\and S.        ~Azaz                          \inst{\ref{inst:0017}}
\and P.        ~Balm                          \inst{\ref{inst:0037}}
\and M.        ~Beck                          \inst{\ref{inst:0012}}
\and H.-H.     ~Bernstein$^\dagger$           \inst{\ref{inst:0023}}
\and L.        ~Bigot                         \inst{\ref{inst:0031}}
\and A.        ~Bijaoui                       \inst{\ref{inst:0031}}
\and C.        ~Blasco                        \inst{\ref{inst:0520}}
\and M.        ~Bonfigli                      \inst{\ref{inst:0232}}
\and G.        ~Bono                          \inst{\ref{inst:0211}}
\and S.        ~Boudreault                    \inst{\ref{inst:0045},\ref{inst:0524}}
\and A.        ~Bressan                       \inst{\ref{inst:0525}}
\and S.        ~Brown                         \inst{\ref{inst:0025}}
\and P.-M.     ~Brunet                        \inst{\ref{inst:0032}}
\and P.        ~Bunclark$^\dagger$            \inst{\ref{inst:0025}}
\and R.        ~Buonanno                      \inst{\ref{inst:0211}}
\and A.G.      ~Butkevich                     \inst{\ref{inst:0028}}
\and C.        ~Carret                        \inst{\ref{inst:0410}}
\and C.        ~Carrion                       \inst{\ref{inst:0008}}
\and L.        ~Chemin                        \inst{\ref{inst:0038},\ref{inst:0534}}
\and F.        ~Ch\'{e}reau                   \inst{\ref{inst:0021}}
\and L.        ~Corcione                      \inst{\ref{inst:0046}}
\and E.        ~Darmigny                      \inst{\ref{inst:0032}}
\and K.S.      ~de Boer                       \inst{\ref{inst:0538}}
\and P.        ~de Teodoro                    \inst{\ref{inst:0060}}
\and P.T.      ~de Zeeuw                      \inst{\ref{inst:0019},\ref{inst:0541}}
\and C.        ~Delle Luche                   \inst{\ref{inst:0021},\ref{inst:0149}}
\and C.D.      ~Domingues                     \inst{\ref{inst:0544}}
\and P.        ~Dubath                        \inst{\ref{inst:0012}}
\and F.        ~Fodor                         \inst{\ref{inst:0032}}
\and B.        ~Fr\'{e}zouls                  \inst{\ref{inst:0032}}
\and A.        ~Fries                         \inst{\ref{inst:0009}}
\and D.        ~Fustes                        \inst{\ref{inst:0129}}
\and D.        ~Fyfe                          \inst{\ref{inst:0175}}
\and E.        ~Gallardo                      \inst{\ref{inst:0009}}
\and J.        ~Gallegos                      \inst{\ref{inst:0060}}
\and D.        ~Gardiol                       \inst{\ref{inst:0046}}
\and M.        ~Gebran                        \inst{\ref{inst:0009},\ref{inst:0555}}
\and A.        ~Gomboc                        \inst{\ref{inst:0321},\ref{inst:0557}}
\and A.        ~G\'{o}mez                     \inst{\ref{inst:0021}}
\and E.        ~Grux                          \inst{\ref{inst:0137}}
\and A.        ~Gueguen                       \inst{\ref{inst:0021},\ref{inst:0561}}
\and A.        ~Heyrovsky                     \inst{\ref{inst:0104}}
\and J.        ~Hoar                          \inst{\ref{inst:0029}}
\and G.        ~Iannicola                     \inst{\ref{inst:0211}}
\and Y.        ~Isasi Parache                 \inst{\ref{inst:0009}}
\and A.-M.     ~Janotto                       \inst{\ref{inst:0032}}
\and E.        ~Joliet                        \inst{\ref{inst:0106},\ref{inst:0568}}
\and A.        ~Jonckheere                    \inst{\ref{inst:0095}}
\and R.        ~Keil                          \inst{\ref{inst:0570},\ref{inst:0571}}
\and D.-W.     ~Kim                           \inst{\ref{inst:0022}}
\and P.        ~Klagyivik                     \inst{\ref{inst:0013}}
\and J.        ~Klar                          \inst{\ref{inst:0250}}
\and J.        ~Knude                         \inst{\ref{inst:0047}}
\and O.        ~Kochukhov                     \inst{\ref{inst:0126}}
\and I.        ~Kolka                         \inst{\ref{inst:0577}}
\and J.        ~Kos                           \inst{\ref{inst:0321},\ref{inst:0579}}
\and A.        ~Kutka                         \inst{\ref{inst:0271},\ref{inst:0581}}
\and V.        ~Lainey                        \inst{\ref{inst:0094}}
\and D.        ~LeBouquin                     \inst{\ref{inst:0149}}
\and C.        ~Liu                           \inst{\ref{inst:0022},\ref{inst:0585}}
\and D.        ~Loreggia                      \inst{\ref{inst:0046}}
\and V.V.      ~Makarov                       \inst{\ref{inst:0587}}
\and M.G.      ~Marseille                     \inst{\ref{inst:0149}}
\and C.        ~Martayan                      \inst{\ref{inst:0095},\ref{inst:0590}}
\and O.        ~Martinez-Rubi                 \inst{\ref{inst:0009}}
\and B.        ~Massart                       \inst{\ref{inst:0031},\ref{inst:0149},\ref{inst:0594}}
\and F.        ~Meynadier                     \inst{\ref{inst:0021},\ref{inst:0089}}
\and S.        ~Mignot                        \inst{\ref{inst:0021}}
\and U.        ~Munari                        \inst{\ref{inst:0020}}
\and A.-T.     ~Nguyen                        \inst{\ref{inst:0032}}
\and T.        ~Nordlander                    \inst{\ref{inst:0126}}
\and K.S.      ~O'Flaherty                    \inst{\ref{inst:0601}}
\and P.        ~Ocvirk                        \inst{\ref{inst:0250},\ref{inst:0114}}
\and A.        ~Olias Sanz                    \inst{\ref{inst:0604}}
\and P.        ~Ortiz                         \inst{\ref{inst:0175}}
\and J.        ~Osorio                        \inst{\ref{inst:0163}}
\and D.        ~Oszkiewicz                    \inst{\ref{inst:0131},\ref{inst:0608}}
\and A.        ~Ouzounis                      \inst{\ref{inst:0104}}
\and P.        ~Park                          \inst{\ref{inst:0002}}
\and E.        ~Pasquato                      \inst{\ref{inst:0033}}
\and C.        ~Peltzer                       \inst{\ref{inst:0025}}
\and J.        ~Peralta                       \inst{\ref{inst:0009}}
\and F.        ~P\'{e}turaud                  \inst{\ref{inst:0021}}
\and T.        ~Pieniluoma                    \inst{\ref{inst:0131}}
\and E.        ~Pigozzi                       \inst{\ref{inst:0056}}
\and J.        ~Poels$^\dagger$               \inst{\ref{inst:0107}}
\and G.        ~Prat                          \inst{\ref{inst:0618}}
\and T.        ~Prod'homme                    \inst{\ref{inst:0019},\ref{inst:0620}}
\and F.        ~Raison                        \inst{\ref{inst:0621},\ref{inst:0561}}
\and J.M.      ~Rebordao                      \inst{\ref{inst:0544}}
\and D.        ~Risquez                       \inst{\ref{inst:0019}}
\and B.        ~Rocca-Volmerange              \inst{\ref{inst:0625}}
\and S.        ~Rosen                         \inst{\ref{inst:0045},\ref{inst:0175}}
\and M.I.      ~Ruiz-Fuertes                  \inst{\ref{inst:0012}}
\and F.        ~Russo                         \inst{\ref{inst:0046}}
%\and S.        ~Sembay                        \inst{\ref{inst:0175}}
\and I.        ~Serraller Vizcaino            \inst{\ref{inst:0631}}
\and A.        ~Short                         \inst{\ref{inst:0017}}
\and A.        ~Siebert                       \inst{\ref{inst:0114},\ref{inst:0250}}
\and H.        ~Silva                         \inst{\ref{inst:0257}}
\and D.        ~Sinachopoulos                 \inst{\ref{inst:0092}}
\and E.        ~Slezak                        \inst{\ref{inst:0031}}
\and M.        ~Soffel                        \inst{\ref{inst:0028}}
\and D.        ~Sosnowska                     \inst{\ref{inst:0002}}
\and V.        ~Strai\v{z}ys                  \inst{\ref{inst:0640}}
\and M.        ~ter Linden                    \inst{\ref{inst:0106},\ref{inst:0642}}
\and D.        ~Terrell                       \inst{\ref{inst:0643}}
\and S.        ~Theil                         \inst{\ref{inst:0644}}
\and C.        ~Tiede                         \inst{\ref{inst:0022},\ref{inst:0646}}
\and L.        ~Troisi                        \inst{\ref{inst:0135},\ref{inst:0648}}
\and P.        ~Tsalmantza                    \inst{\ref{inst:0022}}
\and D.        ~Tur                           \inst{\ref{inst:0205}}
\and M.        ~Vaccari                       \inst{\ref{inst:0651},\ref{inst:0652}}
\and F.        ~Vachier                       \inst{\ref{inst:0094}}
\and P.        ~Valles                        \inst{\ref{inst:0009}}
\and W.        ~Van Hamme                     \inst{\ref{inst:0655}}
\and L.        ~Veltz                         \inst{\ref{inst:0250},\ref{inst:0102}}
\and J.        ~Virtanen                      \inst{\ref{inst:0131},\ref{inst:0132}}
\and J.-M.     ~Wallut                        \inst{\ref{inst:0032}}
\and R.        ~Wichmann                      \inst{\ref{inst:0661}}
\and M.I.      ~Wilkinson                     \inst{\ref{inst:0025},\ref{inst:0175}}
\and H.        ~Ziaeepour                     \inst{\ref{inst:0137}}
\and S.        ~Zschocke                      \inst{\ref{inst:0028}}
}
\institute{
     INAF - Osservatorio Astronomico di Bologna, via Piero Gobetti 93/3, 40129 Bologna, Italy\relax                                                                                                          \label{inst:0001}
\and Department of Astronomy, University of Geneva, Chemin des Maillettes 51, CH-1290 Versoix, Switzerland\relax                                                                                             \label{inst:0002}
\and INAF - Osservatorio Astronomico di Capodimonte, Via Moiariello 16, 80131, Napoli, Italy\relax                                                                                                           \label{inst:0003}
\and Dipartimento di Fisica e Astronomia, Universit\`{a} di Bologna, Via Piero Gobetti 93/2, 40129 Bologna, Italy\relax                                                                                      \label{inst:0006}
\and Dpto. de Inteligencia Artificial, UNED, c/ Juan del Rosal 16, 28040 Madrid, Spain\relax                                                                                                                 \label{inst:0008}
\and Institut de Ci\`{e}ncies del Cosmos, Universitat  de  Barcelona  (IEEC-UB), Mart\'{i}  Franqu\`{e}s  1, E-08028 Barcelona, Spain\relax                                                                  \label{inst:0009}
\and Department of Astronomy, University of Geneva, Chemin d'Ecogia 16, CH-1290 Versoix, Switzerland\relax                                                                                                   \label{inst:0012}
\and Konkoly Observatory, Research Centre for Astronomy and Earth Sciences, Hungarian Academy of Sciences, Konkoly Thege Mikl\'{o}s \'{u}t 15-17, 1121 Budapest, Hungary\relax                               \label{inst:0013}
\and Department of Physics and Astronomy, The Johns Hopkins University, 3400 N Charles St, Baltimore, MD 21218, USA\relax                                                                                    \label{inst:0015}
\and Scientific Support Office, Directorate of Science, European Space Research and Technology Centre (ESA/ESTEC), Keplerlaan 1, 2201AZ, Noordwijk, The Netherlands\relax                                    \label{inst:0017}
\and Leiden Observatory, Leiden University, Niels Bohrweg 2, 2333 CA Leiden, The Netherlands\relax                                                                                                           \label{inst:0019}
\and INAF - Osservatorio astronomico di Padova, Vicolo Osservatorio 5, 35122 Padova, Italy\relax                                                                                                             \label{inst:0020}
\and GEPI, Observatoire de Paris, PSL Research University, CNRS, 5 Place Jules Janssen, 92190 Meudon, France\relax                                             \label{inst:0021}
\and Max Planck Institute for Astronomy, K\"{ o}nigstuhl 17, 69117 Heidelberg, Germany\relax                                                                                                                 \label{inst:0022}
\and Astronomisches Rechen-Institut, Zentrum f\"{ u}r Astronomie der Universit\"{ a}t Heidelberg, M\"{ o}nchhofstr. 12-14, D-69120 Heidelberg, Germany\relax                                                 \label{inst:0023}
\and Institute of Astronomy, University of Cambridge, Madingley Road, Cambridge CB3 0HA, United Kingdom\relax                                                                                                \label{inst:0025}
\and Mission Operations Division, Operations Department, Directorate of Science, European Space Research and Technology Centre (ESA/ESTEC), Keplerlaan 1, 2201 AZ, Noordwijk, The Netherlands\relax          \label{inst:0026}
\and Lohrmann Observatory, Technische Universit\"{ a}t Dresden, Mommsenstra{\ss}e 13, 01062 Dresden, Germany\relax                                                                                           \label{inst:0028}
\and European Space Astronomy Centre (ESA/ESAC), Camino bajo del Castillo, s/n, Urbanizacion Villafranca del Castillo, Villanueva de la Ca\~{n}ada, E-28692 Madrid, Spain\relax                              \label{inst:0029}
\and Lund Observatory, Department of Astronomy and Theoretical Physics, Lund University, Box 43, SE-22100 Lund, Sweden\relax                                                                                 \label{inst:0030}
\and Laboratoire Lagrange, Universit\'{e} Nice Sophia-Antipolis, Observatoire de la C\^{o}te d'Azur, CNRS, CS 34229, F-06304 Nice Cedex, France\relax                                                        \label{inst:0031}
\and CNES Centre Spatial de Toulouse, 18 avenue Edouard Belin, 31401 Toulouse Cedex 9, France\relax                                                                                                          \label{inst:0032}
\and Institut d'Astronomie et d'Astrophysique, Universit\'{e} Libre de Bruxelles CP 226, Boulevard du Triomphe, 1050 Brussels, Belgium\relax                                                                 \label{inst:0033}
\and F.R.S.-FNRS, Rue d'Egmont 5, 1000 Brussels, Belgium\relax                                                                                                                                               \label{inst:0034}
\and INAF - Osservatorio Astrofisico di Arcetri, Largo Enrico Fermi 5, I-50125 Firenze, Italy\relax                                                                                                          \label{inst:0035}
\and Telespazio Vega UK Ltd for ESA/ESAC, Camino bajo del Castillo, s/n, Urbanizacion Villafranca del Castillo, Villanueva de la Ca\~{n}ada, E-28692 Madrid, Spain\relax                                     \label{inst:0037}
\and Laboratoire d'astrophysique de Bordeaux, Universit\'{e} de Bordeaux, CNRS, B18N, all{\'e}e Geoffroy Saint-Hilaire, 33615 Pessac, France\relax                                                           \label{inst:0038}
\and Instituut voor Sterrenkunde, KU Leuven, Celestijnenlaan 200D, 3001 Leuven, Belgium\relax                                                                                                                \label{inst:0042}
\and Department of Astrophysics/IMAPP, Radboud University Nijmegen, P.O.Box 9010, 6500 GL Nijmegen, The Netherlands\relax                                                                                    \label{inst:0043}
\and Mullard Space Science Laboratory, University College London, Holmbury St Mary, Dorking, Surrey RH5 6NT, United Kingdom\relax                                                                            \label{inst:0045}
\and INAF - Osservatorio Astrofisico di Torino, via Osservatorio 20, 10025 Pino Torinese (TO), Italy\relax                                                                                                   \label{inst:0046}
\and Niels Bohr Institute, University of Copenhagen, Juliane Maries Vej 30, 2100 Copenhagen {\O}, Denmark\relax                                                                                              \label{inst:0047}
\and Centre for Electronic Imaging, Department of Physical Sciences, The Open University, Walton Hall MK7 6AA Milton Keynes, United Kingdom\relax                                                            \label{inst:0052}
\and ALTEC S.p.a, Corso Marche, 79,10146 Torino, Italy\relax                                                                                                                                                 \label{inst:0056}
\and Serco Gesti\'{o}n de Negocios for ESA/ESAC, Camino bajo del Castillo, s/n, Urbanizacion Villafranca del Castillo, Villanueva de la Ca\~{n}ada, E-28692 Madrid, Spain\relax                              \label{inst:0060}
\and STFC, Rutherford Appleton Laboratory, Harwell, Didcot, OX11 0QX, United Kingdom\relax                                                                                                                   \label{inst:0073}
\and Gaia DPAC Project Office, ESAC, Camino bajo del Castillo, s/n, Urbanizacion Villafranca del Castillo, Villanueva de la Ca\~{n}ada, E-28692 Madrid, Spain\relax                                          \label{inst:0079}
\and SYRTE, Observatoire de Paris, PSL Research University, CNRS, Sorbonne Universit{\'e}s, UPMC Univ. Paris 06, LNE, 61 avenue de l'Observatoire, 75014 Paris, France\relax                                 \label{inst:0089}
\and National Observatory of Athens, I. Metaxa and Vas. Pavlou, Palaia Penteli, 15236 Athens, Greece\relax                                                                                                   \label{inst:0092}
\and IMCCE, Observatoire de Paris, PSL Research University, CNRS, Sorbonne Universit{\'e}s, UPMC Univ. Paris 06, Univ. Lille, 77 av. Denfert-Rochereau, 75014 Paris, France\relax                            \label{inst:0094}
\and Royal Observatory of Belgium, Ringlaan 3, 1180 Brussels, Belgium\relax                                                                                                                                  \label{inst:0095}
\and Institut d'Astrophysique Spatiale, Universit\'{e} Paris XI, UMR 8617, CNRS, B\^{a}timent 121, 91405, Orsay Cedex, France\relax                                                                          \label{inst:0102}
\and Institute for Astronomy, Royal Observatory, University of Edinburgh, Blackford Hill, Edinburgh EH9 3HJ, United Kingdom\relax                                                                            \label{inst:0104}
\and HE Space Operations BV for ESA/ESAC, Camino bajo del Castillo, s/n, Urbanizacion Villafranca del Castillo, Villanueva de la Ca\~{n}ada, E-28692 Madrid, Spain\relax                                     \label{inst:0106}
\and Institut d'Astrophysique et de G\'{e}ophysique, Universit\'{e} de Li\`{e}ge, 19c, All\'{e}e du 6 Ao\^{u}t, B-4000 Li\`{e}ge, Belgium\relax                                                              \label{inst:0107}
\and \'{A}rea de Lenguajes y Sistemas Inform\'{a}ticos, Universidad Pablo de Olavide, Ctra. de Utrera, km 1. 41013, Sevilla, Spain\relax                                                                     \label{inst:0111}
\and Observatoire Astronomique de Strasbourg, Universit\'{e} de Strasbourg, CNRS, UMR 7550, 11 rue de l'Universit\'{e}, 67000 Strasbourg, France\relax                                                       \label{inst:0114}
\and Kavli Institute for Cosmology, University of Cambridge, Madingley Road, Cambride CB3 0HA, United Kingdom\relax                                                                                          \label{inst:0117}
\and Aurora Technology for ESA/ESAC, Camino bajo del Castillo, s/n, Urbanizacion Villafranca del Castillo, Villanueva de la Ca\~{n}ada, E-28692 Madrid, Spain\relax                                          \label{inst:0122}
\and Laboratoire Univers et Particules de Montpellier, Universit\'{e} Montpellier, Place Eug\`{e}ne Bataillon, CC72, 34095 Montpellier Cedex 05, France\relax                                                \label{inst:0123}
\and Department of Astrophysics, Astronomy and Mechanics, National and Kapodistrian University of Athens, Panepistimiopolis, Zografos, 15783 Athens, Greece\relax                                            \label{inst:0125}
\and Department of Physics and Astronomy, Division of Astronomy and Space Physics, Uppsala University, Box 516, 75120 Uppsala, Sweden\relax                                                                  \label{inst:0126}
\and Universit\`{a} di Catania, Dipartimento di Fisica e Astronomia, Sezione Astrofisica, Via S. Sofia 78, I-95123 Catania, Italy\relax                                                                      \label{inst:0127}
\and INAF - Osservatorio Astrofisico di Catania, via S. Sofia 78, 95123 Catania, Italy\relax                                                                                                                 \label{inst:0128}
\and Universidade da Coru\~{n}a, Facultade de Inform\'{a}tica, Campus de Elvi\~{n}a S/N, 15071, A Coru\~{n}a, Spain\relax                                                                                    \label{inst:0129}
\and CENTRA, Universidade de Lisboa, FCUL, Campo Grande, Edif. C8, 1749-016 Lisboa, Portugal\relax                                                                                                           \label{inst:0130}
\and University of Helsinki, Department of Physics, P.O. Box 64, FI-00014 University of Helsinki, Finland\relax                                                                                              \label{inst:0131}
\and Finnish Geospatial Research Institute FGI, Geodeetinrinne 2, FI-02430 Masala, Finland\relax                                                                                                             \label{inst:0132}
\and Isdefe for ESA/ESAC, Camino bajo del Castillo, s/n, Urbanizacion Villafranca del Castillo, Villanueva de la Ca\~{n}ada, E-28692 Madrid, Spain\relax                                                     \label{inst:0133}
\and ASI Science Data Center, via del Politecnico SNC, 00133 Roma, Italy\relax                                                                                                                               \label{inst:0135}
\and Institut UTINAM UMR6213, CNRS, OSU THETA Franche-Comt\'{e} Bourgogne, Universit\'{e} Bourgogne Franche-Comt\'{e}, F-25000 Besan\c{c}on, France\relax                                                    \label{inst:0137}
\and Elecnor Deimos Space for ESA/ESAC, Camino bajo del Castillo, s/n, Urbanizacion Villafranca del Castillo, Villanueva de la Ca\~{n}ada, E-28692 Madrid, Spain\relax                                       \label{inst:0148}
\and Thales Services for CNES Centre Spatial de Toulouse, 18 avenue Edouard Belin, 31401 Toulouse Cedex 9, France\relax                                                                                      \label{inst:0149}
\and EURIX S.r.l., via Carcano 26, 10153, Torino, Italy\relax                                                                                                                                                \label{inst:0152}
\newpage
\and University of Vienna, Department of Astrophysics, T\"{ u}rkenschanzstra{\ss}e 17, A1180 Vienna, Austria\relax                                                                                           \label{inst:0155}
\and ON/MCTI-BR, Rua Gal. Jos\'{e} Cristino 77, Rio de Janeiro, CEP 20921-400, RJ,  Brazil\relax                                                                                                             \label{inst:0156}
\and OV/UFRJ-BR, Ladeira Pedro Ant\^{o}nio 43, Rio de Janeiro, CEP 20080-090, RJ, Brazil\relax                                                                                                               \label{inst:0157}
\and Faculdade Ciencias, Universidade do Porto, Departamento Matematica Aplicada, Rua do Campo Alegre, 687 4169-007 Porto, Portugal\relax                                                                    \label{inst:0163}
\and Instituto de Astrof\'{\i}sica e Ci\^encias do Espa\,co, Universidade de Lisboa Faculdade de Ci\^encias, Campo Grande, PT1749-016 Lisboa, Portugal\relax                                                 \label{inst:0164}
\and Departamento de Astrof\'{i}sica, Centro de Astrobiolog\'{i}a (CSIC-INTA), ESA-ESAC. Camino Bajo del Castillo s/n. 28692 Villanueva de la Ca\~{n}ada, Madrid, Spain\relax                                \label{inst:0173}
\and Department of Physics and Astronomy, University of Leicester, University Road, Leicester LE1 7RH, United Kingdom\relax                                                                                  \label{inst:0175}
\and University of Oviedo, Campus Universitario, 33203 Gij\'{o}n, Spain\relax                                                                                                                                \label{inst:0178}
\and University of C\'{a}diz, Avd. De la universidad, Jerez de la Frontera, C\'{a}diz, Spain\relax                                                                                                           \label{inst:0181}
\and Kapteyn Astronomical Institute, University of Groningen, Landleven 12, 9747 AD Groningen, The Netherlands\relax                                                                                         \label{inst:0195}
\and Consorci de Serveis Universitaris de Catalunya, C/ Gran Capit\`{a}, 2-4 3rd floor, 08034 Barcelona, Spain\relax                                                                                         \label{inst:0205}
\and University of Turin, Department of Computer Sciences, Corso Svizzera 185, 10149 Torino, Italy\relax                                                                                                     \label{inst:0207}
\and INAF - Osservatorio Astronomico di Roma, Via di Frascati 33, 00078 Monte Porzio Catone (Roma), Italy\relax                                                                                              \label{inst:0211}
\and CRAAG - Centre de Recherche en Astronomie, Astrophysique et G\'{e}ophysique, Route de l'Observatoire Bp 63 Bouzareah 16340 Algiers, Algeria\relax                                                       \label{inst:0225}
\and Universiteit Antwerpen, Onderzoeksgroep Toegepaste Wiskunde, Middelheimlaan 1, 2020 Antwerpen, Belgium\relax                                                                                            \label{inst:0228}
\and Department of Physics and Astronomy, University of Padova, Via Marzolo 8, I-35131 Padova, Italy\relax                                                                                                   \label{inst:0230}
\and INAF - Osservatorio Astronomico di Teramo, Via Mentore Maggini, 64100 Teramo, Italy\relax                                                                                                               \label{inst:0232}
\and Instituto de Astronomia, Geof\`{i}sica e Ci\^{e}ncias Atmosf\'{e}ricas, Universidade de S\~{a}o Paulo, Rua do Mat\~{a}o, 1226, Cidade Universitaria, 05508-900 S\~{a}o Paulo, SP, Brazil\relax          \label{inst:0234}
\and Department of Geosciences, Tel Aviv University, Tel Aviv 6997801, Israel\relax                                                                                                                          \label{inst:0247}
\and Astronomical Institute Anton Pannekoek, University of Amsterdam, PO Box 94249, 1090 GE, Amsterdam, The Netherlands\relax                                                                                \label{inst:0248}
\and Leibniz Institute for Astrophysics Potsdam (AIP), An der Sternwarte 16, 14482 Potsdam, Germany\relax                                                                                                    \label{inst:0250}
\and ATOS for CNES Centre Spatial de Toulouse, 18 avenue Edouard Belin, 31401 Toulouse Cedex 9, France\relax                                                                                                 \label{inst:0253}
%\and School of Physics and Astronomy, Tel Aviv University, Tel Aviv 6997801, Israel\relax                                                                                                                    \label{inst:0256}
\and UNINOVA - CTS, Campus FCT-UNL, Monte da Caparica, 2829-516 Caparica, Portugal\relax                                                                                                                     \label{inst:0257}
\and Laboratoire G\'{e}oazur, Universit\'{e} Nice Sophia-Antipolis, UMR 7329, CNRS, Observatoire de la C\^{o}te d'Azur, 250 rue A. Einstein, F-06560 Valbonne, France\relax                                  \label{inst:0263}
\and RHEA for ESA/ESAC, Camino bajo del Castillo, s/n, Urbanizacion Villafranca del Castillo, Villanueva de la Ca\~{n}ada, E-28692 Madrid, Spain\relax                                                       \label{inst:0269}
\and Astronomical Institute, Academy of Sciences of the Czech Republic, Fri\v{c}ova 298, 25165 Ond\v{r}ejov, Czech Republic\relax                                                                            \label{inst:0271}
\and Barcelona Supercomputing Center - Centro Nacional de Supercomputaci\'{o}n, c/ Jordi Girona 29, Ed. Nexus II, 08034 Barcelona, Spain\relax                                                               \label{inst:0284}
\and Department of Mechanical Engineering, University of La Rioja, c/ San Jos\'{e} de Calasanz, 31, 26004 Logro\~{n}o, La Rioja, Spain\relax                                                                 \label{inst:0287}
\and ETSE Telecomunicaci\'{o}n, Universidade de Vigo, Campus Lagoas-Marcosende, 36310 Vigo, Galicia, Spain\relax                                                                                             \label{inst:0289}
\and SRON, Netherlands Institute for Space Research, Sorbonnelaan 2, 3584CA, Utrecht, The Netherlands\relax                                                                                                  \label{inst:0315}
\and Faculty of Mathematics and Physics, University of Ljubljana, Jadranska ulica 19, 1000 Ljubljana, Slovenia\relax                                                                                         \label{inst:0321}
\and Physics Department, University of Antwerp, Groenenborgerlaan 171, 2020 Antwerp, Belgium\relax                                                                                                           \label{inst:0323}
\and Harvard-Smithsonian Center for Astrophysics, 60 Garden Street, Cambridge MA 02138, USA\relax                                                                                                            \label{inst:0325}
\and Institut de Physique de Rennes, Universit{\'e} de Rennes 1, F-35042 Rennes, France\relax                                                                                                                \label{inst:0340}
\and Shanghai Astronomical Observatory, Chinese Academy of Sciences, 80 Nandan Rd, 200030 Shanghai, China\relax                                                                                              \label{inst:0349}
\and CSC Danmark A/S, Retortvej 8, 2500 Valby, Denmark\relax                                                                                                                                                 \label{inst:0352}
\and Las Cumbres Observatory Global Telescope Network, Inc., 6740 Cortona Drive, Suite 102, Goleta, CA  93117, USA\relax                                                                                     \label{inst:0353}
\and Astrophysics Research Institute, Liverpool John Moores University, L3 5RF, United Kingdom\relax                                                                                                         \label{inst:0365}
\and Baja Observatory of University of Szeged, Szegedi \'{u}t III/70, 6500 Baja, Hungary\relax                                                                                                               \label{inst:0371}
\and Laboratoire AIM, IRFU/Service d'Astrophysique - CEA/DSM - CNRS - Universit\'{e} Paris Diderot, B\^{a}t 709, CEA-Saclay, F-91191 Gif-sur-Yvette Cedex, France\relax                                      \label{inst:0372}
\and INAF - Osservatorio Astronomico di Trieste, Via G.B. Tiepolo 11, 34143, Trieste, Italy\relax                                                                                                            \label{inst:0384}
\and Laboratoire de l'Acc\'{e}l\'{e}rateur Lin\'{e}aire, Universit\'{e} Paris-Sud, CNRS/IN2P3, Universit\'{e} Paris-Saclay, 91898 Orsay Cedex, France\relax                                                  \label{inst:0386}
\and \'{E}cole polytechnique f\'{e}d\'{e}rale de Lausanne, SB MATHAA STAP, MA B1 473 (B\^{a}timent MA), Station 8, CH-1015 Lausanne, Switzerland\relax                                                       \label{inst:0392}
\and INAF - Istituto di Astrofisica Spaziale e Fisica Cosmica di Bologna, Via Piero Gobetti 101, 40129 Bologna, Italy\relax                                                                                  \label{inst:0398}
\and Technical University of Madrid, Jos\'{e} Guti\'{e}rrez Abascal 2, 28006 Madrid, Spain\relax                                                                                                             \label{inst:0401}
\and EQUERT International for CNES Centre Spatial de Toulouse, 18 avenue Edouard Belin, 31401 Toulouse Cedex 9, France\relax                                                                                 \label{inst:0410}
\and AKKA for CNES Centre Spatial de Toulouse, 18 avenue Edouard Belin, 31401 Toulouse Cedex 9, France\relax                                                                                                 \label{inst:0417}
\and Villanova University, Dept. of Astrophysics and Planetary Science, 800 E Lancaster Ave, Villanova PA 19085, USA\relax                                                                                   \label{inst:0418}
\and Vitrociset Belgium for ESA/ESAC, Camino bajo del Castillo, s/n, Urbanizacion Villafranca del Castillo, Villanueva de la Ca\~{n}ada, E-28692 Madrid, Spain\relax                                         \label{inst:0423}
\and Fork Research, Rua do Cruzado Osberno, Lt. 1, 9 esq., Lisboa, Portugal\relax                                                                                                                            \label{inst:0443}
\and APAVE SUDEUROPE SAS for CNES Centre Spatial de Toulouse, 18 avenue Edouard Belin, 31401 Toulouse Cedex 9, France\relax                                                                                  \label{inst:0446}
\and Spanish Virtual Observatory\relax                                                                                                                                                                       \label{inst:0453}
\and Fundaci\'{o}n Galileo Galilei - INAF, Rambla Jos\'{e} Ana Fern\'{a}ndez P\'{e}rez 7, E-38712 Bre\~{n}a Baja, Santa Cruz de Tenerife, Spain\relax                                                        \label{inst:0464}
\and INSA for ESA/ESAC, Camino bajo del Castillo, s/n, Urbanizacion Villafranca del Castillo, Villanueva de la Ca\~{n}ada, E-28692 Madrid, Spain\relax                                                       \label{inst:0469}
\and Dpto. Arquitectura de Computadores y Autom\'{a}tica, Facultad de Inform\'{a}tica, Universidad Complutense de Madrid, C/ Prof. Jos\'{e} Garc\'{i}a Santesmases s/n, 28040 Madrid, Spain\relax            \label{inst:0470}
\newpage
\and H H Wills Physics Laboratory, University of Bristol, Tyndall Avenue, Bristol BS8 1TL, United Kingdom\relax                                                                                              \label{inst:0473}
\and Stellar Astrophysics Centre, Aarhus University, Department of Physics and Astronomy, 120 Ny Munkegade, Building 1520, DK-8000 Aarhus C, Denmark\relax                                                   \label{inst:0476}
\and Applied Physics Department, University of Vigo, E-36310 Vigo, Spain\relax                                                                                                                               \label{inst:0479}
\and HE Space Operations BV for ESA/ESTEC, Keplerlaan 1, 2201AZ, Noordwijk, The Netherlands\relax                                                                                                            \label{inst:0491}
\and Warsaw University Observatory, Al. Ujazdowskie 4, 00-478 Warszawa, Poland\relax                                                                                                                         \label{inst:0500}
\and Instituto de Astrof\'{\i}sica de Canarias, E-38205 La Laguna, Tenerife, Spain\relax                                                                                                                     \label{inst:0509}
\and Universidad de La Laguna, Departamento de Astrof\'{\i}sica, E-38206 La Laguna, Tenerife, Spain\relax                                                                                                    \label{inst:0510}
\and RHEA for ESA/ESTEC, Keplerlaan 1, 2201AZ, Noordwijk, The Netherlands\relax                                                                                                                              \label{inst:0520}
\and Max Planck Institute for Solar System Research, Justus-von-Liebig-Weg 3, 37077 G\"{ o}ttingen, Germany\relax                                                                                            \label{inst:0524}
\and SISSA (Scuola Internazionale Superiore di Studi Avanzati), via Bonomea 265, 34136 Trieste, Italy\relax                                                                                                  \label{inst:0525}
\and Instituto Nacional de Pesquisas Espaciais/Minist\'{e}rio da Ciencia Tecnologia, Avenida dos Astronautas 1758, S\~{a}o Jos\'{e} Dos Campos, SP 12227-010, Brazil\relax                                   \label{inst:0534}
\and Argelander Institut f\"{ u}r Astronomie der Universit\"{ a}t Bonn, Auf dem H\"{ u}gel 71, 53121 Bonn, Germany\relax                                                                                     \label{inst:0538}
\and European Southern Observatory (ESO), Karl-Schwarzschild-Stra{\ss}e 2, 85748 Garching bei M\"{ u}nchen, Germany\relax                                                                                    \label{inst:0541}
\and Laboratory of Optics, Lasers and Systems, Faculty of Sciences, University of Lisbon, Campus do Lumiar, Estrada do Pa\c{c}o do Lumiar, 22, 1649-038 Lisboa, Portugal\relax                               \label{inst:0544}
\and Department of Physics and Astronomy, Notre Dame University, Louaize, PO Box 72, Zouk Mika\"{ e}l, Lebanon\relax                                                                                         \label{inst:0555}
\and University of Nova Gorica, Vipavska 13, 5000 Nova Gorica, Slovenia\relax                                                                                                                                \label{inst:0557}
\and Max Planck Institute for Extraterrestrial Physics, OPINAS, Gie{\ss}enbachstra{\ss}e, 85741 Garching, Germany\relax                                                                                      \label{inst:0561}
\and NASA/IPAC Infrared Science Archive, California Institute of Technology, Mail Code 100-22, 770 South Wilson Avenue, Pasadena, CA, 91125, USA\relax                                                       \label{inst:0568}
\and Center of Applied Space Technology and Microgravity (ZARM), c/o Universit\"{ a}t Bremen, Am Fallturm 1, 28359 Bremen, Germany\relax                                                                     \label{inst:0570}
\and RHEA System for ESA/ESOC, Robert Bosch Stra{\ss}e 5, 64293 Darmstadt, Germany\relax                                                                                                                     \label{inst:0571}
\and Tartu Observatory, 61602 T\~{o}ravere, Estonia\relax                                                                                                                                                    \label{inst:0577}
\and Sydney Institute for Astronomy, School of Physics A28, The University of Sydney, NSW 2006, Australia\relax                                                                                              \label{inst:0579}
\and Slovak Organisation for Space Activities, Zamocka 18, 85101 Bratislava, Slovak Republic\relax                                                                                                           \label{inst:0581}
\and National Astronomical Observatories, CAS, 100012 Beijing, China\relax                                                                                                                                   \label{inst:0585}
\and US Naval Observatory, Astrometry Department, 3450 Massachusetts Ave. NW, Washington DC 20392-5420 D.C., USA\relax                                                                                       \label{inst:0587}
\and European Southern Observatory (ESO), Alonso de C\'{o}rdova 3107, Vitacura, Casilla 19001, Santiago de Chile, Chile\relax                                                                                \label{inst:0590}
\and Airbus Defence and Space SAS, 31 Rue des Cosmonautes, 31402 Toulouse Cedex 4, France\relax                                                                                                              \label{inst:0594}
\and EJR-Quartz BV for ESA/ESTEC, Keplerlaan 1, 2201AZ, Noordwijk, The Netherlands\relax                                                                                                                     \label{inst:0601}
\and The Server Labs for ESA/ESAC, Camino bajo del Castillo, s/n, Urbanizacion Villafranca del Castillo, Villanueva de la Ca\~{n}ada, E-28692 Madrid, Spain\relax                                            \label{inst:0604}
\and Astronomical Observatory Institute, Faculty of Physics, A. Mickiewicz University, ul. S\l{}oneczna 36, 60-286 Pozna\'{n}, Poland\relax                                                                  \label{inst:0608}
\and CS Syst\`{e}mes d'Information for CNES Centre Spatial de Toulouse, 18 avenue Edouard Belin, 31401 Toulouse Cedex 9, France\relax                                                                        \label{inst:0618}
\and Directorate of Science, European Space Research and Technology Centre (ESA/ESTEC), Keplerlaan 1, 2201AZ, Noordwijk, The Netherlands\relax                                                               \label{inst:0620}
\and Praesepe BV for ESA/ESAC, Camino bajo del Castillo, s/n, Urbanizacion Villafranca del Castillo, Villanueva de la Ca\~{n}ada, E-28692 Madrid, Spain\relax                                                \label{inst:0621}
\and Sorbonne Universit\'{e}s UPMC et CNRS, UMR7095, Institut d'Astrophysique de Paris, F75014, Paris, France\relax                                                                                          \label{inst:0625}
\and GMV for ESA/ESAC, Camino bajo del Castillo, s/n, Urbanizacion Villafranca del Castillo, Villanueva de la Ca\~{n}ada, E-28692 Madrid, Spain\relax                                                        \label{inst:0631}
\and Institute of Theoretical Physics and Astronomy, Vilnius University, Sauletekio al. 3, Vilnius, LT-10222, Lithuania\relax                                                                                \label{inst:0640}
%\newpage
\and S[\&]T Corporation, PO Box 608, 2600 AP, Delft, The Netherlands\relax                                                                                                                                   \label{inst:0642}
\and Department of Space Studies, Southwest Research Institute (SwRI), 1050 Walnut Street, Suite 300, Boulder, Colorado 80302, USA\relax                                                                     \label{inst:0643}
\and Deutsches Zentrum f\"{ u}r Luft- und Raumfahrt, Institute of Space Systems, Am Fallturm 1, D-28359 Bremen, Germany\relax                                                                                \label{inst:0644}
\and University of Applied Sciences Munich, Karlstr. 6, 80333 Munich, Germany\relax                                                                                                                          \label{inst:0646}
\and Dipartimento di Fisica, Universit\`{a} di Roma Tor Vergata, via della Ricerca Scientifica 1, 00133 Rome, Italy\relax                                                                                    \label{inst:0648}
\and Department of Physics and Astronomy, University of the Western Cape, Robert Sobukwe Road, 7535 Bellville, Cape Town, South Africa\relax                                                                 \label{inst:0651}
\and INAF - Istituto di Radioastronomia, via Piero Gobetti 101, 40129 Bologna, Italy\relax                                                                                                                   \label{inst:0652}
\and Department of Physics, Florida International University, 11200 SW 8th Street, Miami, FL 33199, USA\relax                                                                                                \label{inst:0655}
\and Hamburger Sternwarte, Gojenbergsweg 112, D-21029 Hamburg, Germany\relax                                                                                                                                 \label{inst:0661}
}

%% file: ack.tex
\begin{acknowledgements}

This work has made use of results from the European Space Agency (ESA) space mission {\it Gaia}, the data from which were processed by the {\it Gaia} Data Processing and Analysis Consortium (DPAC). Funding for the DPAC has been provided by national institutions, in particular the institutions participating in the {\it Gaia} Multilateral Agreement. The {\it Gaia} mission website is \url{http://www.cosmos.esa.int/gaia}. The authors are current or past members of the ESA and Airbus DS {\it Gaia} mission teams and of the {\it Gaia} DPAC.
This research has made use of the SIMBAD database, operated at CDS, Strasbourg, France.
 We thank the referee, Pierre Kervella, for his detailed comments and suggestions that have helped to improve
the paper analysis and presentation.
This work has financially been supported by:
%the Agenzia Spaziale Italiana (ASI) through grants ASI I/058/10/0 and ASI;
the Agenzia Spaziale Italiana (ASI) through grants I/037/08/0, I/058/10/0, 2014-025-R.0, and 2014-025-R.1.2015 to INAF and contracts I/008/10/0 and 2013/030/I.0 to ALTEC S.p.A.;
the Algerian Centre de Recherche en Astronomie, Astrophysique et G\'{e}ophysique of Bouzareah Observatory;
the Austrian FWF Hertha Firnberg Programme through grants T359, P20046, and P23737;
the BELgian federal Science Policy Office (BELSPO) through various PROgramme de D\'eveloppement d'Exp\'eriences scientifiques (PRODEX) grants;
the Brazil-France exchange programmes FAPESP-COFECUB and CAPES-COFECUB;
the Chinese National Science Foundation through grant NSFC 11573054;
the Czech-Republic Ministry of Education, Youth, and Sports through grant LG 15010;
the Danish Ministry of Science;
the Estonian Ministry of Education and Research through grant IUT40-1;
the European Commission’s Sixth Framework Programme through the European Leadership in Space Astrometry (ELSA) Marie Curie Research Training Network (MRTN-CT-2006-033481), through Marie Curie project PIOF-GA-2009-255267 (SAS-RRL), and through a Marie Curie Transfer-of-Knowledge (ToK) fellowship (MTKD-CT-2004-014188); the European Commission's Seventh Framework Programme through grant FP7-606740 (FP7-SPACE-2013-1) for the {\it Gaia} European Network for Improved data User Services (GENIUS) and through grant 264895 for the {\it Gaia} Research for European Astronomy Training (GREAT-ITN) network;
the European Research Council (ERC) through grant 320360 and through the European Union’s Horizon 2020 research and innovation programme through grant agreement 670519 (Mixing and Angular Momentum tranSport of massIvE stars -- MAMSIE);
the European Science Foundation (ESF), in the framework of the {\it Gaia} Research for European Astronomy Training Research Network Programme (GREAT-ESF);
the European Space Agency in the framework of the {\it Gaia} project;
the European Space Agency Plan for European Cooperating States (PECS) programme through grants for Slovenia; the Czech Space Office through ESA PECS contract 98058;
the Academy of Finland; the Magnus Ehrnrooth Foundation;
the French Centre National de la Recherche Scientifique (CNRS) through action `D\'efi MASTODONS';
the French Centre National d'Etudes Spatiales (CNES);
the French L'Agence Nationale de la Recherche (ANR) `investissements d'avenir' Initiatives D’EXcellence (IDEX) programme PSL$\ast$ through grant ANR-10-IDEX-0001-02;
the R\'egion Aquitaine;
the Universit\'e de Bordeaux;
the French Utinam Institute of the Universit\'e de Franche-Comt\'e, supported by the R\'egion de Franche-Comt\'e and the Institut des Sciences de l'Univers (INSU);
the German Aerospace Agency (Deutsches Zentrum f\"{u}r Luft- und Raumfahrt e.V., DLR) through grants 50QG0501, 50QG0601, 50QG0602, 50QG0701, 50QG0901, 50QG1001, 50QG1101, 50QG140, 50QG1401, 50QG1402, and 50QG1404;
the Hungarian Academy of Sciences through Lend\"ulet Programme LP2014-17;
the Hungarian National Research, Development, and Innovation Office through grants NKFIH K-115709, K-119517 and PD-116175;
the Israel Ministry of Science and Technology through grant 3-9082;
the Italian Istituto Nazionale di Astrofisica (INAF);
the Netherlands Organisation for Scientific Research (NWO) through grant NWO-M-614.061.414 and through a VICI grant to A.~Helmi;
the Netherlands Research School for Astronomy (NOVA);
the Polish National Science Centre through HARMONIA grant 2015/18/M/ST9/00544;
the Portugese Funda\c{c}\~ao para a Ci\^{e}ncia e a Tecnologia (FCT) through grants PTDC/CTE-SPA/118692/2010, PDCTE/CTE-AST/81711/2003, and SFRH/BPD/74697/2010; the Strategic Programmes PEst-OE/AMB/UI4006/2011 for SIM, UID/FIS/00099/2013 for CENTRA, and UID/EEA/00066/2013 for UNINOVA;
the Slovenian Research Agency;
the Spanish Ministry of Economy MINECO-FEDER through grants AyA2014-55216, AyA2011-24052, ESP2013-48318-C2-R, and ESP2014-55996-C2-R and MDM-2014-0369 of ICCUB (Unidad de Excelencia `Mar\'{\i}a de Maeztu);
the Swedish National Space Board (SNSB/Rymdstyrelsen);
%\change{
the Swiss State Secretariat for Education, Research, and Innovation through the ESA PRODEX programme;
the Swiss Mesures d’Accompagnement;
the Swiss Activit\'es Nationales Compl\'ementaires;
the Swiss National Science Foundation, including an Early Postdoc.Mobility fellowship;
%}
the United Kingdom Rutherford Appleton Laboratory;
the United Kingdom Science and Technology Facilities Council (STFC) through grants PP/C506756/1 and ST/I00047X/1; and
the United Kingdom Space Agency (UKSA) through grants ST/K000578/1 and ST/N000978/1.

\end{acknowledgements}

%% file: Clementini-AA-2016-29925-accepted.bbl
\begin{thebibliography}{}
\bibitem[Anderson \& Darling(1952)]{anderson1952} Anderson, T. W. \& Darling, D. A., 1952, Ann.~Math.~Statist., 2, 193
\bibitem[Anderson et al.(2016)]{anderson16} Anderson, R.I., Casertano, S., Riess, A.G., et al.\ 2016,  \apjs, 226, 18
\bibitem[Arenou \& Luri(1999)]{Arenou1999} Arenou, F., \& Luri, X. 1999, ASPC, 167, 13
\bibitem[Arenou et al.(2017)]{Arenou2017} Arenou, F., Luri, X., Babusiaux, S., et al.  2017, A\&A, 599, A50
\bibitem[Bailer-Jones(2015)]{2015PASP..127..994B} Bailer-Jones, C.~A.~L. 2015, PASP, 129, 994
\bibitem[Beaton et al. (2016)]{beaton2016} Beaton, R.~L., Freedman, W.~L., Madore, B.~F., et al. 2016, \apj, 832, 210
 \bibitem[Benedict et al.(2002)]{Ben2002} Benedict, G.~F., McArthur, B.~E., Fredrick, L.~W., et al. 2002, \aj, 123, 473
   \bibitem[Benedict et al.(2007)]{ben07} Benedict, G.~F., McArthur, B.~E., Feast, M.~W., et al.\ 2007, \aj, 133, 1810
 \bibitem[Benedict et al.(2011)]{Benedict2011} Benedict, G. F., McArthur, B. E., Feast, M. W., et al. 2011, AJ, 142,187
 \bibitem[Berdnikov(2008)]{berdni08} Berdnikov, L.~N.\ 2008, VizieR Online Data Catalog, 2285
\bibitem[Berdnikov et al.(2000)]{berdni00} Berdnikov , L.~N., Dambis, A.~K., \& Vozyakova, O.~V.\ 2000, A\&AS, 143, 211
\bibitem[Bono et al.(2000)]{bcm00} Bono, G., Castellani, V., \& Marconi, M.\ 2000, \apjl, 532, L129 
\bibitem[Bono et al.(2002)]{bcm02} Bono, G., Castellani, V., \& Marconi, M.\ 2002, \apjl, 565, L83 
\bibitem[Bono, Caputo \& Santolamazza(1997)]{Bono1997} Bono, G., Caputo, F., \& Santolamazza, P. 1997, A\&A, 317, 171
\bibitem[Bono et al.(2001)]{bono01} Bono, G., Caputo, F., Castellani, V., Marconi, M., \& Storm, J.\ 2001, \mnras, 326, 1183 
\bibitem[Bono et al.(2003)]{Bono2003} Bono, G., Caputo, F., Castellani, V., et al. 2003, MNRAS, 344, 1097 
\bibitem[Bono et al.(2010)]{bono10} Bono, G., Caputo, F., Marconi, M., \& Musella, I.\ 2010, \apj, 715, 277 
\bibitem[Borissova et al.(2009)]{Bor2009} Borissova, J., Rejkuba, M., Minniti, D., Catelan, M. and Ivanov, V.~D. 2009, A\&A, 502, 505
\bibitem[Breitfelder et  al.(2016)]{breitfelder16} Breitfelder, J., M\'erand, A., Kervella, P.,  et al. \ 2016,  \aap, 587, A117
 \bibitem[Cacciari \& Clementini (2003)]{cacciari03} Cacciari, C.,  \& Clementini, G. 2003, in Lecture Notes in Physics,
Berlin Springer Verlag, Vol. 635, Stellar Candles for the
Extragalactic Distance Scale, ed. D. Alloin \& W. Gieren, 105-122
 \bibitem[Cacciari et al. (1992)]{cacciari92} Cacciari, C., Clementini, G., \& Fernley, J. A. 1992, \apj, 396, 219
\bibitem[Caputo et al.(2000b)]{capu00} Caputo, F., Castellani, V., Marconi, M., \& Ripepi, V.\ 2000b, \mnras, 316, 819 
\bibitem[Caputo et al.(2000a)]{cmm00} Caputo, F., Marconi, M., \& Musella, I.\ 2000a, \aap, 354, 610 
\bibitem[Cardelli et al.(1989)]{Cardelli1989} Cardelli, J. A., Clayton, C., \& Mathis, J. S. 1989, AJ, 345, 245
  \bibitem[Casertano et al.(2016)]{cas16} Casertano, S., Riess, A.~G., Anderson, J., et al.\ 2016, \apj, 825, 11 
  \bibitem[Casertano et al.(2017)]{cas17} Casertano, S., Riess, A.~G., Bucciarelli, B., \& Lattanzi, M.~G. \ 2017, A\&A, 599, 67
 \bibitem[Cassisi et al.(1998)]{cassisi98} Cassisi, S., Castellani, V., degl'Innocenti, S., \& Weiss, A.\ 1998, \aaps, 129, 267 
\bibitem[Catelan et al.(2004)]{Cat2004} Catelan, M.,  Pritzl, B.~J. \& Smith, H.~A.~E. 2004,   ApJS, 154, 633 
\bibitem[Catelan \& Cortes(2008)]{Cat2008} Catelan, M., \& Cortes, C. 2008, ApJ, 676, L135
\bibitem[Cioni et al.(2011)]{Cioni2011} Cioni, M.-R. L., Clementini, G., Girardi, L., et al. 2011, A\&A, 527, A116
\bibitem[Clementini et al.(2003)]{Clementini2003} Clementini, G., Gratton, R.~G., Bragaglia, A., et al. 2003, AJ, 125, 1309 
\bibitem[Clementini et al.(2016)]{clementini16} Clementini, G., Ripepi, V., Leccia, S., et al.  2016, A\&A, 595, A133
 \bibitem[Coppola et al.(2011)]{Cop2011} Coppola, G., Dall'Ora, M., Ripepi, V., et al. 2011, MNRAS, 416, 1056 
\bibitem[Cutri et al. (2003)]{Cutri2003} Cutri, R., et al. 2003, University of Massachusetts and Infrared Processing and Analysis Centre (IPAC/California Institute of Technology)
\bibitem[Dall'Ora et al.(2004)]{DO2004} Dall'Ora, M., Storm, J., Bono, G., et al. 2004, ApJ, 610, 269
\bibitem[Dambis et al.(2013)]{Dambis2013} Dambis, A. K., Berdnikov, L. N., Kniazev, A. Y., et al. 2013, \mnras, 435, 3206
\bibitem[Dambis et al.(2014)]{Dambis2014} Dambis, A.~K., Rastorguev, A.~S., \& Zabolotskikh,~M.~V. 2014, MNRAS, 439, 3765
\bibitem[de Bruijne et~al.(2005)]{Bruijneetal} de Bruijne, J., Perryman, M., Lindegren, L., Jordi, C., H{\o}g, E. Katz, D., \& Cropper, M., 2005, {\it Gaia} Data
  Processing and Analysis Consortium (DPAC) technical note GAIA-JDB-022
\bibitem[de Grijs et al.(2014)]{deGr2014} De Grijs, R., Wicker J. E., \& Bono, G., 2014, AJ, 147, 122
\bibitem[Del Principe et al.(2006)]{DelP2006} Del Principe, M., Piersimoni, A. M., Storm, J., et al. 2006, AJ, 652, 362
\bibitem[Di Fabrizio et al.(2002)]{difa02} Di Fabrizio, L., Clementini, G., Marconi, M., et al.\ 2002, \mnras, 336, 841 
  \bibitem[Drimmel et al.(2003)]{Drimmel2003} Drimmel, R., Cabrera-Lavers, A.,  \& Lo\'pez-Corredoira, M. 2003, A\&A, 409, 205
\bibitem[ESA (1997)]{esa97} ESA, ed. 1997, ESA Special Publication, Vol. 1200, The HIPPARCOS and TYCHO
catalogues. Astrometric and photometric star catalogues derived from
the ESA HIPPARCOS Space Astrometry Mission
\bibitem[Evans(1994)]{evans1994} Evans, N.~R.\ 1994, \apj, 436, 273 
 \bibitem[Feast et al.(2008)]{Feast2008} Feast, M.~W., Clifton, D.~L., Kinman, T.~D., van Leeuwen, F. \& Whitelock, P.~A. 2008, \mnras, 386, 2115
  \bibitem[Fernie(2000)]{fernie2000} Fernie, J.~D.\ 2000, \aj, 120, 978 
 \bibitem[Fernie et al. (1995)]{ddo} Fernie, J.D., Beattie, B., Evans, N.R., \& Seager, S. 1995, IBVS No. 4148
  \bibitem[Fernley(1994)]{fernley1994} Fernley, J.~A.\ 1994, \aap, 284, L16 
\bibitem[Fiorentino et al.(2013)]{fio13} Fiorentino, G., Musella, I.,
  \& Marconi, M.\ 2013, \mnras, 434, 2866 
\bibitem[Fouqu{\'e} et al.(2007)]{fou2007}Fouqu{\'e}, P., Arriagada, P., Storm, J., et al.\ 2007, \aap, 476, 73
\bibitem[Freedman et al.(2001)]{free01} Freedman, W.~L., Madore, B.~F., Gibson, B.~K., et al.\ 2001, \apj, 553, 47 
 \bibitem[{\it Gaia} Collaboration et al.(2016a)]{gaiacol-brown} {\it Gaia} Collaboration, Brown, A.G.A., Vallenari, A., Prusti, T., et al.\ 2016a, A\&A, 595, A2
 \bibitem[{\it Gaia} Collaboration et al.(2016b)]{gaiacol-prusti} {\it Gaia} Collaboration, Prusti, T., de Bruijne, J.H.J., et al.\ 2016b,   A\&A, 595, A1
    \bibitem[Gallenne et al.(2012)]{gallenne2012} Gallenne, A., Kervella, P., M\'erand, A., et al. 2012,  \aap, 541, A87
  \bibitem[Genovali et al.(2014)]{gen14} Genovali, K., Lemasle, B., Bono, G., et al.\ 2014, \aap, 566, A37
  \bibitem[Gibson (2000)]{Gib2000} Gibson, B. K. 2000, MmSAIt, 71, 693
  \bibitem[Gieren et al.(2013)]{Gieren2013} Gieren, W., G{\'o}rski, M., Pietrzy{\'n}ski, G., et al.\ 2013, \apj, 773, 69
  \bibitem[Gingold(1985)]{Gingold1985} Gingold, R. A. 1985, Mem. Soc. Astron. Ital., 56, 169
 \bibitem[Gratton et al.(2004)] {Grat2004} Gratton, R. G., Bragaglia, A., Clementini, G., et al. 2004, A\&A, 421, 937
 \bibitem[Groenewegen(1999)]{groe1999} Groenewegen, M.~A.~T.\ 1999, A\&AS, 139, 245
 \bibitem[Groenewegen et al.(2008)]{Groenewegen2008} Groenewegen, M.~A.~T., Udalski, A., \& Bono, G.\ 2008, A\&A, 481, 441 
\bibitem[Harris(1985)]{har85} Harris, H.~C.\ 1985, \aj, 90, 756
\bibitem[H{\o}g et al.(2000)]{hog00} H{\o}g, E., Fabricius, C., Makarov, V.~V., et al.\ 2000, \aap, 355, L27
\bibitem[Inno et al.(2013)]{inno13} Inno, L., Matsunaga, N., Bono, G., et al.\ 2013, \apj, 764, 84 
\bibitem[Inno et al.(2015)]{Inno15} Inno, L., Matsunaga, N., Romaniello, M., et al.\ 2015, \aap, 576, A30
 \bibitem[Jordi et al. (2010)]{jordi10} Jordi, C., Gebran, M., Carrasco, J. M., et al. 2010, A\&A, 523, 48 %RD13
\bibitem[Keller \& Wood(2002)]{kw02} Keller, S.~C., \& Wood, P.~R.\ 2002, \apj, 578, 144 
\bibitem[\protect\citeauthoryear{Keller \& Wood}{2006}]{kw06} Keller, S.~C., \& Wood, P.~R.\ 2006, ApJ, 642, 834 
\bibitem[Kervella et al.(2008)]{kerve08} Kervella, P.,  M{\'e}rand, A., Szabados, L., et al.\ 2008, \aap, 480, 167 
\bibitem[Kervella et al.(2014)]{kervella14} Kervella, P., Bond, H.~E.,  Cracraft, M., et al.\ 2014, \aap, 572, A7 
\bibitem[Klagyivik \& Szabados(2009)]{Klag2009} Klagyivik, P., \& Szabados, L.\ 2009, \aap, 504, 959
\bibitem[Klein et al.(2014)]{Klein2014} Klein,~C.~R., Richards,~J.~W., Butler,~N.~R., \& Bloom~J.~S., 2014, MNRAS, 440, L96
\bibitem[Kubiak \& Udalski(2003)]{Kubiak2003} Kubiak, M., \& Udalski, A.\ 2003, Acta Astronomica, 53, 117 
\bibitem[Laney \& Stobie(1992)]{ls92} Laney, C.~D., \& Stobie, R.~S.\ 1992, \aaps, 93, 93
\bibitem[Lee et al.(1990)]{lee90} Lee, Y.-W., Demarque, P., \& Zinn, R.\ 1990, \apj, 350, 155  
\bibitem[Lemasle et al.(2007)]{lem07} Lemasle, B., Fran{\c c}ois, P., Bono, G., et al.\ 2007, \aap, 467, 283
\bibitem[Lemasle et al.(2008)]{lem08} Lemasle, B., Fran{\c c}ois, P., Piersimoni, A., et al.\ 2008, \aap, 490, 613
\bibitem[Lemasle et al.(2015)]{lem15} Lemasle, B., Kovtyukh, V., Bono, G., et al.\ 2015, \aap, 579, A47
\bibitem[Lindegren et al.(2016)]{lindegren16} Lindegren, L., Lammers, U., Bastian, U., et al.\ 2016, A\&A, 595, A4
\bibitem[Longmore et al.(1986)]{long86} Longmore, A.~J., Fernley, J.~A., \& Jameson, R.~F.\ 1986, \mnras, 220, 279 
\bibitem[Longmore et al.(1990)]{long90} Longmore, A.~J., Dixon, R., Skillen, I., Jameson, R.~F., \& Fernley, J.~A.\ 1990, \mnras, 247, 684 
\bibitem[Luck et al.(2011)]{luc11} Luck, R.~E., Andrievsky, S.~M., Kovtyukh, V.~V., Gieren, W., \& Graczyk, D.\ 2011, \aj, 142, 51
\bibitem[Luck \& Lambert(2011)]{lul11} Luck, R.~E., \& Lambert, D.~L.\ 2011, \aj, 142, 136
\bibitem[Lutz \& Kelker(1973)]{Lutz-Kelker} Lutz, T.~E. \& Kelker, D.~H. 1973, \pasp, 85, 573
\bibitem[Madore(1982)]{mdr82} Madore, B.~F.\ 1982, \apj, 253, 575 
\bibitem[Madore \& Freedman(1991)]{madore91} Madore, B.~F., \& Freedman, W.~L.\ 1991, \pasp, 103, 933 
\bibitem[Madore et al.(2013)] {Madore2013} Madore, B.~F., Hoffman D., Freedman W.~L., et al. 2013, ApJ, 776, 135
\bibitem[Maintz(2005)]{Maintz2005} Maintz, G. 2005, A\&A, 442, 381
\bibitem[Malmquist(1936)]{Malmquist1936} Malmquist, K.~G. 1936, Stockholms Ob. Medd., 26
 \bibitem[Marconi(2015)]{marconi15} Marconi, M.\ 2015, \memsai, 86, 190 
\bibitem[Marconi \& Clementini(2005)]{mc05} Marconi, M., \& Clementini, G.\ 2005, \aj, 129, 2257 
\bibitem[Marconi \& Degl'Innocenti(2007)]{md07} Marconi, M., \& Degl'Innocenti, S.\ 2007, \aap, 474, 557
\bibitem[Marconi \& Di Criscienzo(2007)]{Marconi2007} Marconi, M., \& Di Criscienzo, M.\ 2007, A\&A, 467, 223 
\bibitem[Marconi et al.(2013a)]{marconi13a} Marconi, M., Molinaro, R., Bono, G., et al.\ 2013a, \apj,  768, L6 
\bibitem[Marconi et al.(2013b)]{marconi13b} Marconi, M., Molinaro, R., Ripepi, V., Musella, I., \& Brocato, E.\ 2013b, \mnras, 428, 2185 
\bibitem[Marconi et al. (2005)]{marconi05} Marconi, M., Musella, I., \& Fiorentino, G. 2005, \apj, 632, 590
\bibitem[Matsunaga et al.(2011)]{Matsunaga2011} Matsunaga, N., Feast, M.~W., \& Soszy{\'n}ski, I.\ 2011, MNRAS, 413, 223 
\bibitem[Matsunaga et al.(2006)]{Matsunaga2006} Matsunaga, N., Fukushi, H., Nakada, Y., et al.\ 2006, MNRAS, 370, 1979
\bibitem[McNamara et al.(2007)]{mcna07} McNamara, D.~H., Clementini, G., \& Marconi, M.\ 2007, \aj, 133, 2752
\bibitem[M\'erand et  al.(2015)]{merand15} M\'erand, A., Kervella, P., Breitfelder, J., et al. \ 2015,  \aap, 584, A80
\bibitem[Michalik et al. (2015)]{michalik15} Michalik, D., Lindegren, L., \& Hobbs, D., 2015, A\&A, 574, A115
\bibitem[Moln{\'a}r \& Szabados(2014)]{Molnar2014} Moln{\'a}r, L., \& Szabados, L.\ 2014, \mnras, 442, 3222 
\bibitem[Monson \& Pierce(2011)]{mp11} Monson, A.~J., \& Pierce, M.~J.\ 2011, \apjs, 193, 12 
\bibitem[Muraveva et al.(2015)]{Muraveva2015} Muraveva, T.,  Palmer, M.,  Clementini, G., et al. 2015, \apj, 807, 127
\bibitem[Natale et al.(2008)]{natale08} Natale, G., Marconi, M., \& Bono, G.\ 2008, \apj, 674, L93 
 \bibitem[Neeley et al.(2015)] {Neeley2015} Neeley, J.~R., Marengo, M., Bono, G., et al. 2015, ApJ, 808, 11
\bibitem[Nemec et al.(1994)]{Nemec1994} Nemec, J.~M., Nemec, A.~F.~L., \& Lutz, T.~E.\ 1994, AJ, 108, 222 
\bibitem[Ngeow \& Kanbur(2006)]{nk06} Ngeow, C., \& Kanbur, S.~M.\ 2006, \apj, 650, 180 
\bibitem[Ngeow(2012)]{ngeow2012} Ngeow, C.-C.\ 2012, \apj, 747, 50
\bibitem[Ngeow et al.(2012)]{ngeow12} Ngeow, C.-C., Kanbur, S.~M., Bellinger, E.~P., et al.\ 2012, \apss, 341, 105 
 \bibitem[Pejcha \& Kochanek(2012)]{Pejcha2012} Pejcha, O., \& Kochanek, C.~S.\ 2012, \apj, 748, 107
 \bibitem[Pietrzynski et al. (2013)]{pietrz13} Pietrzynski, G., Graczyk, D., Gieren, W.,  et al. 2013, Nature, 495, 76
  \bibitem[Pojmanski(2002)]{Pojmanski2002} Pojmanski, G. 2002, Acta Astron., 52, 397
  \bibitem[Pritzl et al.(2003)]{Pritzl2003} Pritzl, B.~J., Smith, H.~A., Stetson, P.~B., et al.\ 2003, AJ, 126, 1381 
\bibitem[Riess et al.(2011)]{riess11} Riess, A.~G., Macri, L., Casertano, S., et al.\ 2011, \apj, 730, 119 
\bibitem[Riess et al.(2014)]{riess14}  Riess, A.~G., Casertano, S., Anderson, J., MacKenty, J., \& Filippenko, A.~V.\ 2014, \apj, 785, 161
\bibitem[Riess et al.(2016)]{riess16} Riess, A.~G.,  Macri, L.~M., Hoffmann, S.~L., et al. \ 2016, \apj, 826, 56
\bibitem[Ripepi et al.(2012)]{ripe12} Ripepi, V., Moretti, M.~I., Marconi, M., et al.\ 2012, \mnras, 424, 1807 
\bibitem[Ripepi et al.(2015)]{Ripepi2015} Ripepi, V., Moretti, M.-I., Marconi, M., et al. 2015, MNRAS, 446, 3034
\bibitem[Ripepi et al.(2016)]{ripe16} Ripepi, V., Marconi, M., Moretti, M.~I., et al.\ 2016, \apjs, 224, 21 
\bibitem[Romaniello et al.(2008)]{roma08} Romaniello, M., Primas, F., Mottini, M., et al.\ 2008, \aap, 488, 731 
\bibitem[Saha et al.(2006)]{saha06} Saha, A., Thim, F., Tammann, G.~A., Reindl, B., \& Sandage, A.\ 2006, \apjs, 165, 108 
\bibitem[Samus et al. (2007-2015)]{gcvs} Samus, N.N., Durlevich O.V., Goranskij V.P., Kazarovets E. V., Kireeva N.N., Pastukhova E.N., Zharova A.V., General Catalogue of Variable Stars (Samus+ 2007-2015)
\bibitem[Samus et al. (2017)]{samus17}  Samus', N.~N., Kazarovets, E.~V., Durlevich, O.~V., Kireeva, N.~N., \& Pastukhova, E.~N.\ 2017, Astronomy Reports, 61, 80
 \bibitem[Schaefer (2008)]{Sch2008} Schaefer, B. E. 2008, AJ, 135, 112
 \bibitem[Schlafly \& Finkbeiner (2011)]{schl2011} Schlafly, E.~F. \& Finkbeiner~D.~P.\ 2011, \apj, 737, 103
  \bibitem[Sesar et al.(2016)]{Sesar2016} Sesar, B., Fouesneau, M., Price-Whelan, A.~M., Bailer-Jones, C.~A.~L., Gould, A., \& Rix, H.-W.\ 2016, arXiv:1611.07035
 \bibitem[Skillen et al.(1993)]{Skillen1993} Skillen, I., Fernley, J. A., Stobie, R. S., \& Jameson, R. F.\ 1993, \mnras, 265, 301
 \bibitem[Sollima et al.(2006)]{Sol2006} Sollima, A., Cacciari, C. \& Valenti, E., 2006, MNRAS, 372, 1675
\bibitem[Sollima et al.(2008)]{Sol2008} Sollima, A., Cacciari, C., Arkharov, A. A. H., et al. 2008, MNRAS, 384, 1583
\bibitem[Soszy{\'n}ski et al.(2008)]{Sos2008} Soszy{\'n}ski, I., Udalski, A., Szyma{\'n}ski, M.~K., et al.\ 2008, Acta Astronomica, 58, 293 
\bibitem[Szabados (1997)]{szabados97} Szabados, L. 1997, in `Hipparcos - Venice '97', ESA SP-402, p. 657
\bibitem[Szabados (2003)]{szabados03} Szabados, L. 2003, IBVS, 5394 %RD4
\bibitem[Szil{\'a}di et al.(2007)]{szi07} Szil{\'a}di, K., Vink{\'o}, J., Poretti, E., Szabados, L., \& Kun, M.\ 2007, \aap, 473, 579 
\bibitem[Tammann et al.(2003)]{ta03} Tammann, G.~A., Sandage, A., \& Reindl, B.\ 2003, \aap, 404, 423 
\bibitem[Turner et al.(2001)]{Turner2001}Turner, D.~G., Billings, G.~W., \& Berdnikov, L.~N.\ 2001, \pasp, 113, 715
\bibitem[Udalski et al.(1992)]{Udalski1992} Udalski, A., M., Kaluzny, J., Kubiak, M., \& Mateo, M.\ 1992, Acta Astronomica, 42, 253 
\bibitem[Udalski et al.(1999)]{Udalski1999} Udalski, A., Soszynski, I., Szymanski, M., et al.\ 1999, \actaa, 49, 45 
\bibitem[van Leeuwen(2007a)]{van07a} van Leeuwen, F., ed. 2007a, Astrophysics and Space Science Library, Vol. 350,
Hipparcos, the New Reduction of the Raw Data
\bibitem[van Leeuwen(2007b)] {van07b} van Leeuwen, F.\ 2007b, \aap, 474, 653
\bibitem[Wade et al. (1999)]{wade99} Wade,  R. A., Donley, J., Fried, R.,  White, R. E., \& Saha, A. 1999, \aj, 118, 2442
\bibitem[Wallerstein \& Cox(1984)]{Wallerstein1984} Wallerstein, G., \& Cox, A. N. 1984, PASP, 96, 677
\bibitem[Wallerstein(2002)]{Wallerstein2002} Wallerstein, G. 2002, PASP, 114, 689
\bibitem[Welch et al.(1984)]{w84} Welch, D.~L., Wieland, F., McAlary, C.~W., et al.\ 1984, \apjs, 54, 547 
\bibitem[Wenger et al.(2000)]{Wenger2000} Wenger, M., Ochsenbein, F., Egret, D., et al. \ 2000,  A\&AS, 143, 9
\bibitem[Wood et al.(1997)]{was97} Wood, P.~R., S.~Arnold, A., \& Sebo, K.~M.\ 1997, \apjl, 485, L25
\bibitem[Zinn \& West(1984)]{ZW1984} Zinn, R., \& West, M. J. 1984, \apjs, 55, 45

\end{thebibliography}
